\documentclass[fleqn,usenatbib]{mnras}
\usepackage{newtxtext,newtxmath}
\usepackage[T1]{fontenc}
\usepackage{ae,aecompl}
\usepackage{graphicx}	
\usepackage{lscape}
\usepackage{caption}
\def\lea{\mathrel{<\kern-1.0em\lower0.9ex\hbox{$\sim$}}}
\def\gea{\mathrel{>\kern-1.0em\lower0.9ex\hbox{$\sim$}}}
\title[Star Cluster Classification in the PHANGS-HST Survey]{Star Cluster Classification in the PHANGS-HST Survey: Comparison between Human and 
Machine Learning Approaches}  
\author[Whitmore et al.]{Bradley C. Whitmore,$^{1}$\thanks{E-mail: whitmore@stsci.edu}
Janice~C.~Lee,$^{2,3}$
Rupali~Chandar,$^{4}$
David~A.~Thilker,$^{5}$
\newauthor
Stephen~Hannon,$^{6}$
Wei, Wei,$^{7}$ 
E. A. Huerta,$^{7}$
Frank Bigiel,$^{8}$
\newauthor
M\'ed\'eric~Boquien,$^{9}$
M\'elanie Chevance,$^{10}$
Daniel~A.~Dale,$^{11}$
Sinan~Deger,$^{12}$
\newauthor
Kathryn~Grasha,$^{13}$
Ralf~S.~Klessen,$^{14,15}$
J.~M.~Diederik Kruijssen,$^{10}$
Kirsten~L.~Larson,$^{3}$
\newauthor
Angus~Mok,$^{4}$
Erik~Rosolowsky,$^{16}$
Eva~Schinnerer,$^{17}$
Andreas~Schruba,$^{17}$
\newauthor
Leonardo~Ubeda,$^{1}$
Schuyler~D.~Van~Dyk,$^{3}$
Elizabeth~Watkins$^{10}$
Thomas Williams$^{17}$
\\\\
\parbox{\textwidth}{$^{1}$Space Telescope Science Institute, 3700 San Martin Drive, Baltimore, MD, 21218, USA\\
$^{2}$Gemini Observatory/NSF's NOIRLab, 950 N. Cherry Ave, Tucson, AZ 85719, USA\\
$^{3}$Caltech/IPAC, 1200 California Blvd, Pasadena, CA, 91125, USA\\
$^{4}$Department of Physics \& Astronomy, University of Toledo, Toledo, OH, 43606, USA\\
$^{5}$Department of Physics and Astronomy, The Johns Hopkins University, Baltimore, MD, 21218 USA\\
$^{6}$Department of Physics and Astronomy, University of California, Riverside, CA, 92521 USA\\
$^{7}$NCSA \& Center for Artificial Intelligence Innovation, University of Illinois at Urbana-Champaign, Urbana, Illinois 61801, USA\\
$^{8}$Argelander-Institut f\"ur Astronomie, Universit\"at Bonn, Auf dem H\"ugel 71, 53121 Bonn, Germany\\
$^{9}$Centro de Astronom\'ia, (CITEVA), Universidad de Antofagasta,  Avenida Angamos 601, Antofagasta 1270300, Chile\\
$^{10}$Astronomisches Rechen-Institut, Zentrum f\" ur Astronomie der Universit\"at Heidelberg, M\"onchhofstra\ss e 12-14, D-69120 Heidelberg, Germany\\
$^{11}$Department of Physics and Astronomy, University of Wyoming, Laramie, WY 82071, USA\\
$^{12}$TAPIR, California Institute of Technology, Pasadena, CA, 91125 USA\\
$^{13}$International Centre for Radio Astronomy Research, Australian National University, Canberra, ACT 2611, Australia\\
$^{14}$Universit\"at Heidelberg, Zentrum f\"ur Astronomie, Institut f\"ur Theoretische Astrophysik, Heidelberg, Germany\\
$^{15}$Universit\"at Heidelberg, Interdisziplin\"ares Zentrum f\"ur Wissenschaftliches Rechnen, Heidelberg, Germany\\
$^{16}$Department of Physics, University of Alberta, Edmonton, AB T6G 2E1, Canada\\
$^{17}$Max Planck Institut f\"ur Astronomie, K\"onigstuhl 17, 69117 Heidelberg, Germany\\
}}

\date{Accepted XXX. Received YYY; in original form ZZZ}

\pubyear{2015}

\begin{document}
\label{firstpage}
\pagerange{\pageref{firstpage}--\pageref{lastpage}}
\maketitle

\begin{abstract}

When completed, the PHANGS-HST project will provide a census of roughly 50,000 compact star clusters and associations, as well as human morphological classifications for roughly 20,000 of those objects. 
These large numbers motivated the development of a more objective and repeatable method to help perform source classifications. 
In this paper we consider the results for five PHANGS-HST galaxies (NGC 628, NGC 1433, NGC 1566, NGC 3351, NGC 3627)  using classifications from two convolutional neural network architectures (RESNET and VGG) trained using deep transfer learning techniques. The results are
compared to classifications performed by humans. 
The primary result is that the neural network classifications are comparable in  quality to the human classifications with typical agreement around 70 to 80  \% for Class 1 clusters (symmetric, centrally concentrated) and  
 40 to 70 \% for Class 2 clusters (asymmetric, centrally concentrated). 
 If Class 1 and 2 are considered together the agreement is 82 $\pm$ 3 \%.
 Dependencies on magnitudes, crowding, and background surface brightness are examined. A detailed description of the criteria and methodology used for the human classifications is included along with an examination of systematic differences between PHANGS-HST and LEGUS.
The distribution of data points in a colour-colour diagram is used as a ``figure of merit" to further test the relative performances of the different methods.
The effects on science results (e.g., determinations of mass and age functions) of using different cluster classification methods are examined and found to be minimal.

\bigskip

\end{abstract}

\clearpage

\bigskip

\begin{keywords}
galaxies: star clusters, catalogues, software:public release 
\end{keywords}




\bigskip
\bigskip

\section{Introduction}
\label{sec:intro}

The identification of star clusters in external galaxies is useful for a variety of purposes. Besides providing insight into the transition from giant molecular clouds to  stars and star clusters, they are also useful as ``clocks" to time this evolution since they can be effectively characterized as single-aged stellar populations. Perhaps the most important timescale for understanding the evolution of galaxies is the rate at which feedback removes the gas from the region around forming stars, limiting the rate at which the gas is used and stars are formed in the galaxy.

Early efforts to classify star clusters in external galaxies relied on visual inspection of photographic plates of nearby galaxies 
(e.g., \citealt{Hodge81}).
This resulted in  catalogues with typically a few hundred objects for roughly a dozen galaxies (e.g., M31, M33, M81, LMC, SMC).
The availability of the Hubble Space Telescope (HST), with roughly a factor of ten improvement in spatial resolution,   acted as a major catalyst for the study of extragalactic star clusters (e.g., \citealt{holtzman92}, \citealt{meurer95},  \citealt{Whitmore95}). With the  10-fold improvement in  spatial resolution came 
a 1000-fold increase in the volume that could be searched. Rather than a few hundred clusters in each galaxy the accompanying improvement in detection level led to typical numbers of clusters for each galaxy in the thousands.

The HST archives contain a few hundred nearby galaxies with data sets that could support detailed studies of clusters (i.e., with three or more bands to enable age-dating through SED modeling). The increase in the number of clusters that might be classified  compared to the earlier pre-HST era is therefore a factor of a few hundred [i.e. (300 galaxies $\times$ 3,000 clusters) / (10 galaxies  $\times$ 300 clusters)], with a total number of clusters that could be classified around 1 million. Classification of this many objects represents a limiting constraint for the study of clusters in nearby galaxies, and was the primary reason for development of automated, neural network methods for cluster classifications (\citealt{messa18}, \citealt{Grasha19}, \citealt{bia2019}, \citealt{wei20}, \citealt{perez21}).

 Using neural networks not only accelerates the process of classifications, which has been a limiting step in the production of star cluster  catalogues, it also improves the consistency of the classifications by reducing both random and systematic errors introduced by the subjective nature of human classifications (i.e., the same person may give different classifications for the same object on different occasions). 
We thus developed a new automated approach  to star cluster classifications using neural networks \citep{wei20} as part of the Physics at High Angular Resolution in Nearby Galaxies with the Hubble Space Telescope project (PHANGS-HST); PI: J.C. Lee, GO-15654)
(J. C. Lee et al., in preparation).
\footnote{\url{https://archive.stsci.edu/hlsp/phangs-hst}}.  PHANGS-HST  is a Cycle 26 Treasury program obtaining 5-band UV-optical, F275W (NUV), F336W (U), F438W (B), F555W (V), F814W (I), WFC3 (or ACS in some cases with existing data) imaging for 38 nearby spiral galaxies with previous 
CO(2-1) observations from the PHANGS-ALMA large program
 \citep{leroy21}. \footnote{\url{https://sites.google.com/view/phangs/home}}.  
 
 An early attempt to classify clusters using quantitative morphological parameters is described in \citet{whitmore14}. Using only a simple concentration index (CI = difference in photometry in circular apertures with radii of 1 and 3 pixel), they were able to work down to $M_V$ =  --8 mag (Vega magnitude system) without degrading the sample with a large fraction of blended stars being misidentified as clusters. This study provided cluster luminosity functions for 20 galaxies, with typically a few hundred clusters in each galaxy.

A citizen science approach to cluster classification was used for the PHAT (Panchromatic Hubble Andromeda Treasury) project (\citealt{johnson12}, \citealt{Johnson15}) 
to accelerate classification. While this works well for nearby well-resolved clusters, exploratory efforts for the more distant galaxies in the  Legacy  ExtraGalactic  Ultra-violet  Survey - (\citealt{calzetti15} - LEGUS) project (out to $\sim$11 Mpc)
were unsuccessful due to the more subtle differences between clusters, associations, stars, and interlopers due to the decreased physical resolution, as reported in \citet{perez21}.

Another pioneering study was
performed using a machine learning approach developed as part of the LEGUS project.
While this was successful for nearby well resolved clusters (i.e., recovery rates on the order 60 to 70 \% between human and machine learning approaches), exploratory efforts for the more distant galaxies, and for the compact associations,  showed  lower recovery rates.
Subsequently, \citet{perez21}   developed multiscale convolutional neural network models for the LEGUS project, with agreement fractions on par with the performance of our models in \citet{wei20}.
An additional contribution is \citet{bia2019}, who used neural network classifcations based on simulations for resolved star clusters in M31.

Many  of the most promising recent machine learning  approaches make use of neural networks.
Given the small sizes of existing, human-labelled HST star cluster samples (approximately 2000 objects per class spread out over 39 fields) relative to the samples needed for the robust training of neural networks, we decided to use deep transfer learning techniques as described in detail in \citet{wei20}.  That is, a
neural network model, pre-trained on images of everyday objects from the \texttt{ImageNet} dataset \citep{imagenet_cvpr09}, is fine-tuned using HST image data of clusters with human classifications, rather than training all of the layers in the network from scratch.  The results of our proof-of-concept experiment in \citet{wei20} were encouraging, as the prediction accuracies based on testing with the first PHANGS-HST galaxy observed, NGC~1559, were found to be competitive with the consistency between different human classifications.  In this paper, we take an expanded look at the performance of the neural network models presented in \citet{wei20} by applying the models to star clusters in five additional galaxies, and examining the accuracies as functions of various properties including magnitudes, crowding, background surface brightness, and colours.

We expect our neural network models and cluster catalogs to evolve and improve with time.
The cluster catalogs used for the current paper are from version 0.9 of the PHANGS-HST pipeline 
(D. Thilker et al., in preparation)
and neural network models from \citet{wei20}.  
As new datasets and algorithms are produced they will be made available on the PHANGS-HST website at 
\url{https://archive.stsci.edu/hlsp/phangs-hst}, and will be evaluated in future papers from our team.   

The paper is  organised as follows. In Section
\ref{sec:data} we describe the data set and how the machine learning classifications were performed. In Section \ref{sec:human} we describe the methodology for the human cluster classifications and in Section \ref{sec:compare} we compare the agreement fraction for human classifications with the machine learning classifications.
In Section \ref{sec:cc} we develop several ``figures of merit" based on the distribution of data points in the U-B vs. V-I  colour-colour diagram, in an attempt to determine which classifications provide the best results. 
Section \ref{sec:complete} investigates issues related to completeness while
the dependence of various science results on different methods of classifying clusters is examined in Section \ref{sec:sci}. 
Section \ref{sec:improve}  describes plans to improve the machine learning classifications in the future.
A summary and conclusions are presented in Section \ref{sec:summary}. 
An appendix provides a basic description of the convolutional neural network models described in \citet{wei20} and a step-by-step tutorial on using the models to classify cluster candidates with five band HST imaging.

\section{Sample, Data, and Machine Learning Classification}
\label{sec:data}

In this paper we use five galaxies from the PHANGS-HST sample (NGC 628, NGC 1433, NGC 1566, NGC 3351 and NGC 3627) to evaluate the agreement fractions between classifications made by humans and the convolutional neural network (i.e.,  machine learning)  models of \citet{wei20}.  These particular galaxies are chosen because cluster catalogs with human classifications have been released by the LEGUS program: \url{https://archive.stsci.edu/prepds/legus/dataproducts-public.html} \citep{adamo17}, which provide a  point of comparison for both the PHANGS-HST human and machine learning classifications.   

HST imaging in five bands, F275W (NUV), F336W (U), F438W (B), F555W (V), F814W (I), WFC3 (or ACS in some cases with existing data), were obtained by LEGUS for all five galaxies.  PHANGS-HST obtained imaging in an additional WFC3 pointing for two of the galaxies, NGC 3351 and NGC 3627, to complete coverage of the area of disks mapped in CO(2-1), and to support joint ALMA-HST analysis (J. C. Lee et al., in preparation).

With these data, V-band selected catalogs of compact star clusters and associations, which include five-band aperture photometry, human classifications, and ages, masses, and reddenings derived from SED fitting, were produced by both programs.
We have performed a cross-match of the catalogs.  For this paper, we primarily study the objects in common between the two catalogs and only use the photometry and physical properties resulting from the PHANGS-HST pipeline.
This allows  our analysis to focus on differences resulting from classification methodology rather than from detection, selection, and basic photometric procedures such as aperture corrections (e.g., see 
D. Thilker et al., in preparation
and S. Deger et al., in preparation
for discussion of these properties).

We have incorporated the neural network models of \cite{wei20} into the PHANGS-HST pipeline and use them to produce classifications for all of these objects.  As mentioned above, a deep transfer learning approach was used to train models with two different architectures, \texttt{ResNet18}~\citep{he2016deep} (= RESNET hereafter) and \texttt{VGG19}  \citep{simonyan2014very} with batch normalization \texttt{VGG19-BN} (= VGG hereafter).  Briefly, ResNet18 is a convolutional neural network that is 18 layers deep. This architecture introduced a number of innovations, skip connections and batch normalization, that enabled the training of very deep neural networks (hundreds of layers).  VGG is a convolutional neural network with a depth of 19 layers. Both networks are open source, and thus their weights may be readily adjusted and tuned through transfer learning to be used for a variety of image recognition tasks.

Two different training sets were used in \cite{wei20}: 1) ``3-person consensus'' classifications for clusters in 29 LEGUS galaxies (11268 objects), based on the mode of classifications made by three people as published in the LEGUS cluster catalogs, and 2) single-person classifications for 10 LEGUS galaxies (5488 objects) performed by BCW, the first author of the current paper, who also is performing the human classifications for the PHANGS-HST project. 
Models based on the two different training sets have similar performance.  
Here, we use the models trained using the BCW-only classifications to perform the PHANGS-HST classifications, and examine results from both the RESNET and VGG architectures.  Hence, for all objects in common between the PHANGS-HST and LEGUS cluster catalogs, we compare classifications from four different sources: PHANGS-HST human (BCW-only), LEGUS human (3-person consensus for NGC 628, NGC 1433, and NGC 1566; and BCW-only for  NGC 3351 and NGC 3627), RESNET and VGG. We note that two of the ten galaxies in the \cite{wei20} BCW-only training set are two of the program galaxies in the present study, hence the two samples are not completely independent. This is probably why the performance for these two galaxies is roughly 10 \% better than for the other three, as shown in Table 1 and  discussed later in the text.

Human classifications are determined for sources as faint as $m_V$ = 22.5 to 24~mag in the Vega magnitude system depending on the number of candidate clusters in the galaxy (i.e., brighter limits are used for  the richer galaxies since the numbers to visually examine can become prohibitive, i.e., more than 10,000 in a single galaxy).  We also determine neural network classifications for sources up to about one magnitude fainter, and examine the performance of the models for these faint sources in this paper.

\section{Human Classification}
\label{sec:human}

Human classification of star clusters during the selection process 
has been an important step in most studies of extra-galactic star cluster systems, as reviewed by \citealt{adamo21}).
The methods developed and lessons learned lay the foundation for the development of automated, reliable cluster classification methods.

\subsection{Background \& History}
\label{sec:background}

In this section we describe the approach used for human classification of clusters in the 38 PHANGS-HST galaxies.
The methodology and criteria are very similar
to those previously applied to the LEGUS cluster candidate catalogs \citep{adamo17}, which were used as training sets for the RESNET and VGG neural network models developed by \cite{wei20}.

Based on experience, we have previously found that roughly 50 \% of clusters observed in nearby ($<$ 10~Mpc) spiral galaxies with HST can be reliably classified in just a glance; perhaps 25 \% can be reliably classified after more careful study; and the remaining $\sim25$ \% are often challenging, with properties that make it difficult to confidently establish their classification, for example whether a source is a single star or a very compact cluster.
This level of classification 
is sufficient to determine 
clear correlations between the classes and  various physical properties of the cluster populations, such as their  colour, mass and age distributions. This demonstrates the utility of human classifications, even if they are subjective.  We will examine our classification accuracy in Section \ref{sec:cc}, where we use the location of different classes of clusters in a  colour-colour diagram to test the quality of the classifications, and will assess the effect different methods of classification have on the determination of the age and mass distributions  in Section \ref{sec:sci}.

Previous works have used different numbers of people to perform the human classifications, from a single person up to large numbers from a citizen science approach (e.g., the PHAT survey; \citealt{Johnson15}).
Sampling statistics suggest that a larger number of classifiers should result in more robust results, but this assumes that each human classifier uses similar definitions and internal weighting systems. In practice, most studies to date have used either a single person (e.g., \citealt{chandar10b}, \citealt{bastian14}, \citealt{silva14}, ...),  a few people, or as many as eight people (\citealt{johnson12}). For LEGUS, three different people from a pool of roughly a dozen  classified each object  for most of the galaxies, as described in \citet{adamo17} (see also \citealt{perez21}, and H. Kim et al., in preparation), while 10 galaxies were classified by just one person, BCW.

\begin{figure}
\begin{center}

\includegraphics[width =3.3in , angle= 0]{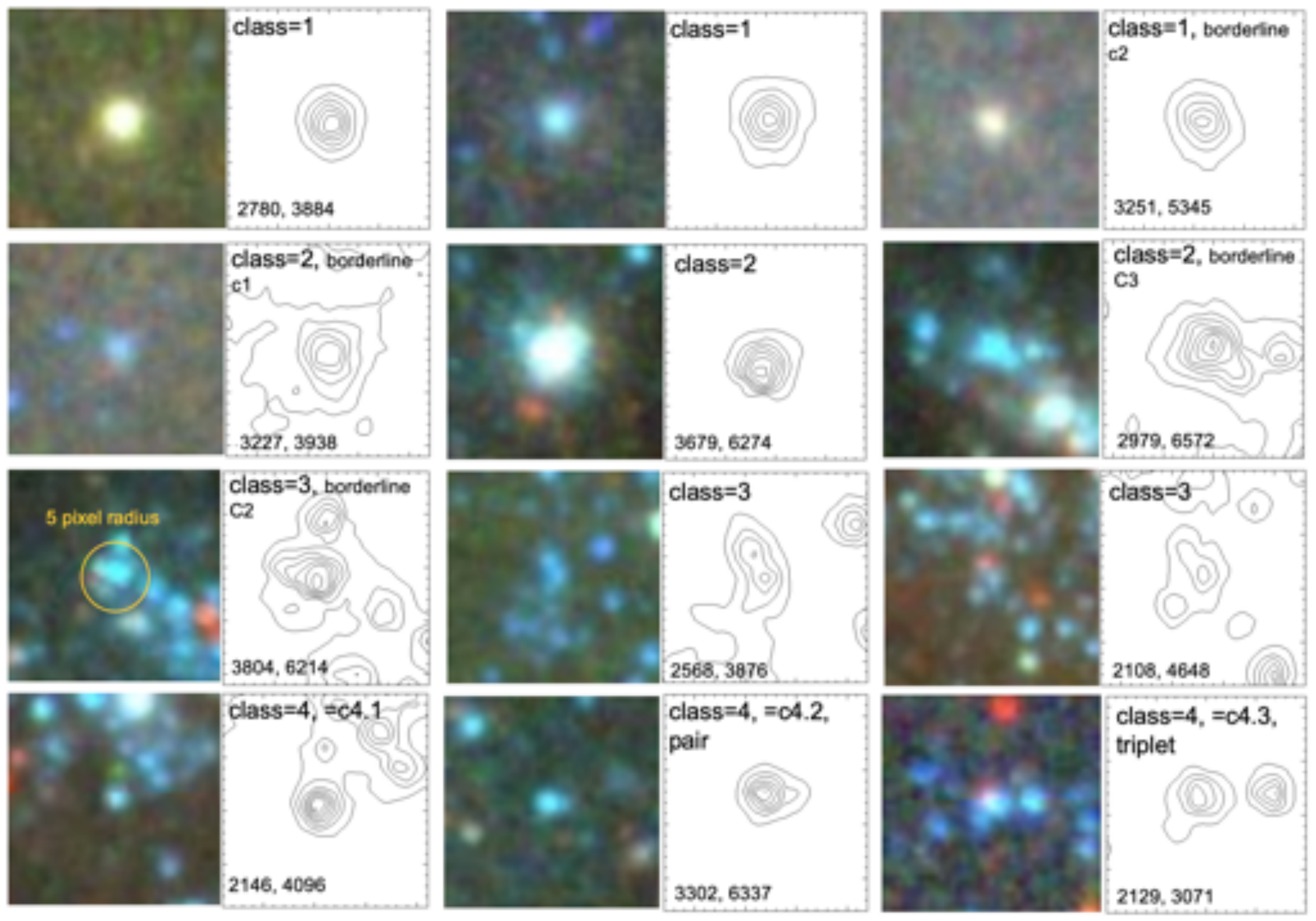}

\end{center}
\caption{Illustration of the four cluster classification types from the galaxy NGC 628.
The color images (using F814W, F555W and F336W) show a 40 by 40 pixel field while the contour plots show a 20 by 20 pixel field from the F555W image. Coordinates in the reference frame
of the PHANGS-HST images available at \url{https://archive.stsci.edu/hlsp/phangs-hst}
are included in the bottom left corner of the contour plots. An example of a circle with a 10 pixel diameter (19 pc at the distance of NGC 628) is include for one object to show the scale. Several borderline cases between two classes are included. }
\label{fig:clust_illus}
\end{figure}

While automated, algorithmically based approaches to classification might be considered objective in one  sense, since they are repeatable, it is important to keep in mind that they are based on subjectively determined training sets.
Hence we cannot  characterize them as fully objective. We note that we are  pursuing two approaches that would be more objective, as discussed in Section \ref{sec:improv_obj}.

\subsection{Procedure}
\label{sec:procedure}

As discussed in \citet{wei20}, 
most early cluster studies in external galaxies provided a single classification class. An exception was the study of \citet{schweizer96}
who defined nine object types and grouped them into  two  classes:  candidate  globular  clusters  and  extended stellar  associations. \citet{bastian12a}
classified  star  clusters  in M83 as  either  symmetric  or  asymmetric clusters, and argued that the difference between the results of their study and \citet{chandar10b}
(who used both human and automated cluster  catalogues) is largely due to the inclusion of asymmetric clusters. 
Following this work, many studies in the field, including LEGUS \citep{calzetti15},
began differentiating candidate clusters into two or three different categories, so that they could be studied separately or together depending on the goals of the project.
It appears that this dichotomy, which has been characterized as 
``exclusive'' (symmetric clusters) and ``inclusive'' (symmetric and asymmetric clusters and in some cases also small associations) by \citet{krumholz19}, can explain much of the difference in the slopes of the age distributions in different studies, 
as first suggested by \citet{bastian12a} (also see \citet{adamo21} for a recent review).
We address this issue by presenting age distribution results from different combinations of classes (see below) in Section \ref{sec:sci}.

In LEGUS, cluster candidates are sorted into four classes based on their morphological appearance as follows (\citealt{adamo17}; \citealt{cook19})

    Class 1: compact, symmetric, single central peak, radial profile extended relative to point source

Class 2:  asymmetric,  single  central peak,  radial  profile extended relative to point source

Class 3: asymmetric, multiple peaks

Class 4: not a star cluster (image artifacts, background galaxies, pairs and multiple stars in crowded regions, individual stars.)

\noindent We adopt the same general classification system for this paper. We refer to Classes 1, 2, and 3 as  symmetric clusters, asymmetric clusters, and compact associations, respectively.

Our primary focus is to identify clusters and groups of stars that are likely to have formed together. Class 1 and 2 objects are  referred to as 
``clusters'', simply based on their centrally peaked profiles, and Class 3 objects are referred to as ``associations'' because they have multiple peaks.
We note that the topic of  associations has largely been superseded in the PHANGS-HST project by the study of K. Larson et al. (in preparation),
who use a more uniform hierarchical approach to finding multi-scale associations. 

These morphological classifications do not provide an unambiguous way of assessing whether or not clusters are bound (have negative energy) or unbound (positive energy), but the bound fraction should depend at least somewhat on the class (e.g., \citealt{2012MNRAS.419..841K,2016A&A...595A..27G,2020arXiv200804453G}). In general, centrally concentrated clusters that have survived for many crossing times (i.e. are older than  10~Myr - see \citealt{gieles2011}) are likely to be gravitationally bound, while the candidates younger than this contain an unknown mix of bound and unbound clusters, where the bound fraction is predicted to increase towards lower classes.

In the youngest ($<10$~Myr) clusters, individual bright, massive stars can lead to an asymmetric appearance, regardless of the spatial distribution of the more numerous, lower-mass stars or the internal energy state of a cluster.  Clusters naturally become smoother in appearance over time as these massive stars die off. In bound clusters, the distribution becomes smoother due to dynamical interactions between the stars, causing them to relax \citep[e.g.][]{2012MNRAS.420.3264G,2012MNRAS.427..637P,2014MNRAS.438..620P}. In unbound cluster candidates, the distribution becomes smoother by ballistic dispersal
\citep[e.g.][]{baumgardt07,2018MNRAS.475.5659W,2020MNRAS.495..663W,2020NewAR..9001549W}.  In general, all clusters lose mass continuously starting shortly after they form.  Therefore, the evolution of clusters should naturally result in some correlation between age and class, where we expect a larger fraction of symmetric clusters (Class 1) at old ages rather than at young ages. Future simulations of evolving clusters which include both gas and stars, and which mimic real observations like the ones made here, would be very helpful to establish how well we can assess the internal energy of the youngest clusters based on their morphologies. 

The criteria for Class 1 and 2 are essentially identical for PHANGS-HST and LEGUS. For Class 3, PHANGS-HST uses a more specific definition than LEGUS, namely that at least four stars are detected within a five pixel radius. This is to avoid stellar pairs and triplets which are sometimes included by LEGUS as Class 2 and 3 objects, as will be discussed in Section \ref{sec:phangs_vs_legus}.  The primary rationale for eliminating pairs and triplets is that these have a much higher probability of being chance super-positions in crowded regions than groups/clusters of stars that formed together.
Examples of objects in each of the four classes are shown in Figure \ref{fig:clust_illus}.

\begin{figure}
\begin{center}

\includegraphics[width =3.2in, angle= 0]{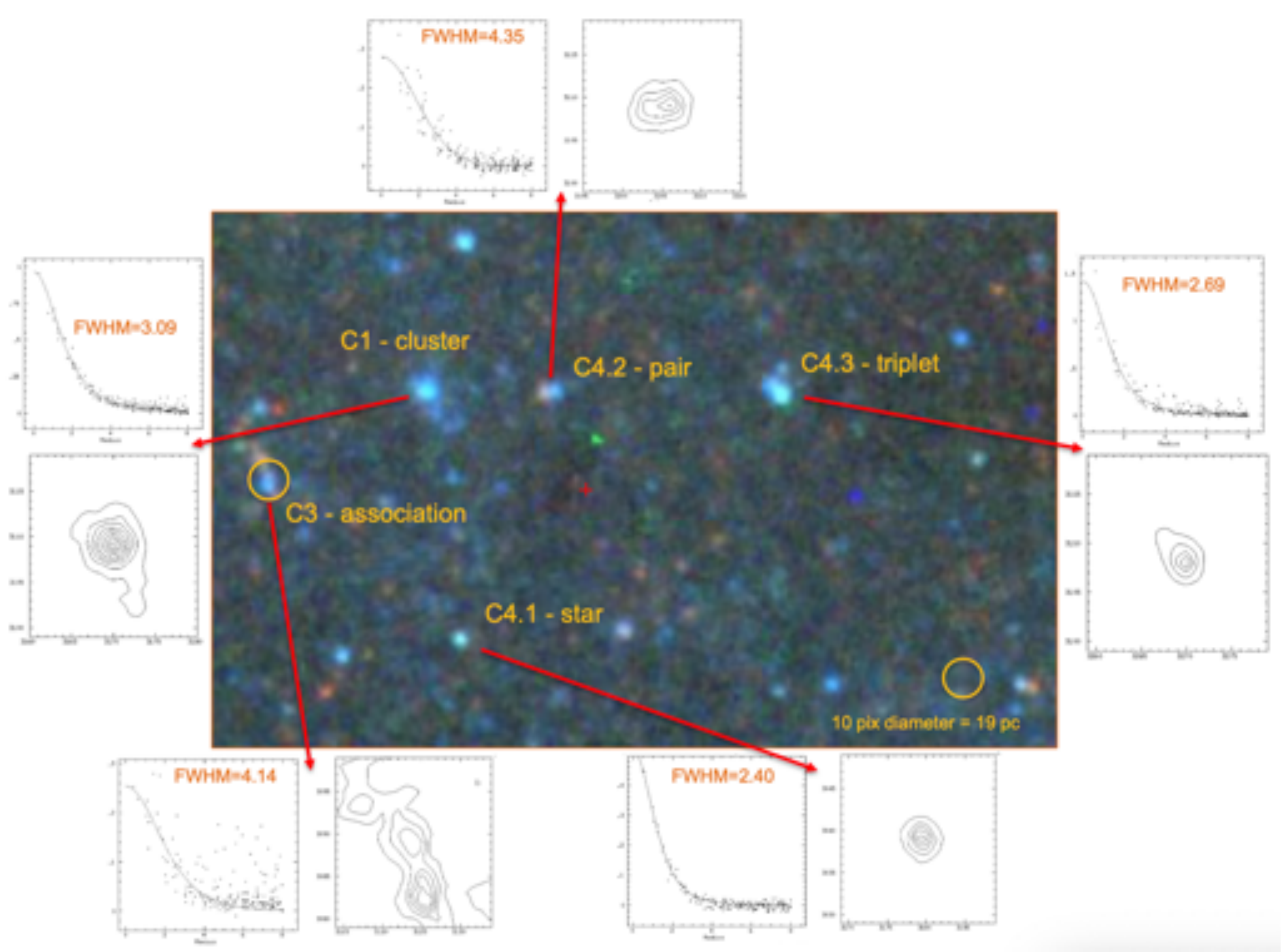}

\end{center}
\caption{Colour image of a field in NGC 628 illustrating profiles, contour plots and appearances for five objects. A circle with a diameter of 10 pixels = 19 pc is shown to provide scale. Note that using the  colour image helps distinguish pairs (i.e., C4.2 = pair) when the stars have different  colours. 
}
\label{fig:profiles}
\end{figure}

Another difference from LEGUS is that Class 4 (artifacts) has been broken up into 12 subclasses, including C4.1 = single star, C4.2 = pair, C4.3 = triplet, C4.4 = saturated star, ... to C4.12 = bad pixels).
The full set of subclasses are listed in Table \ref{tab:table3}.
One of the reasons for this approach was to be able to test whether machine learning could be better trained if the artifacts were divided into more similar morphologies (see Section \ref{sec:improv_train}). Another reason was to allow more flexibility if some users wanted to include some objects into Class 3 (e.g., C4.3 = triplets), or wanted to examine the properties of a specific subsample (e.g., C4.6 nucleus of the program galaxy or C4.7 = background galaxies). 
In most of what follows, the 12 subclasses
will be rolled into a single Class 4, to provide more direct comparisons with LEGUS.

In Figure \ref{fig:profiles} we examine a  colour image for five objects. Radial profiles and contour plots for various objects are also included. Note that the  colour image is especially useful for identifying pairs and triplets when the stars have different  colours.
The division between Class 2 and 3 is sometimes difficult to make. The primary rule is that Class 2 should be centrally concentrated and relatively circular in the contour profiles (i.e., not several objects in a line). The stars in Class 3 objects can generally be seen as separate objects, but in some cases are only visible as strong spurs on the contours.

\begin{figure}
\begin{center}

\includegraphics[width =3.2in, angle= 0]{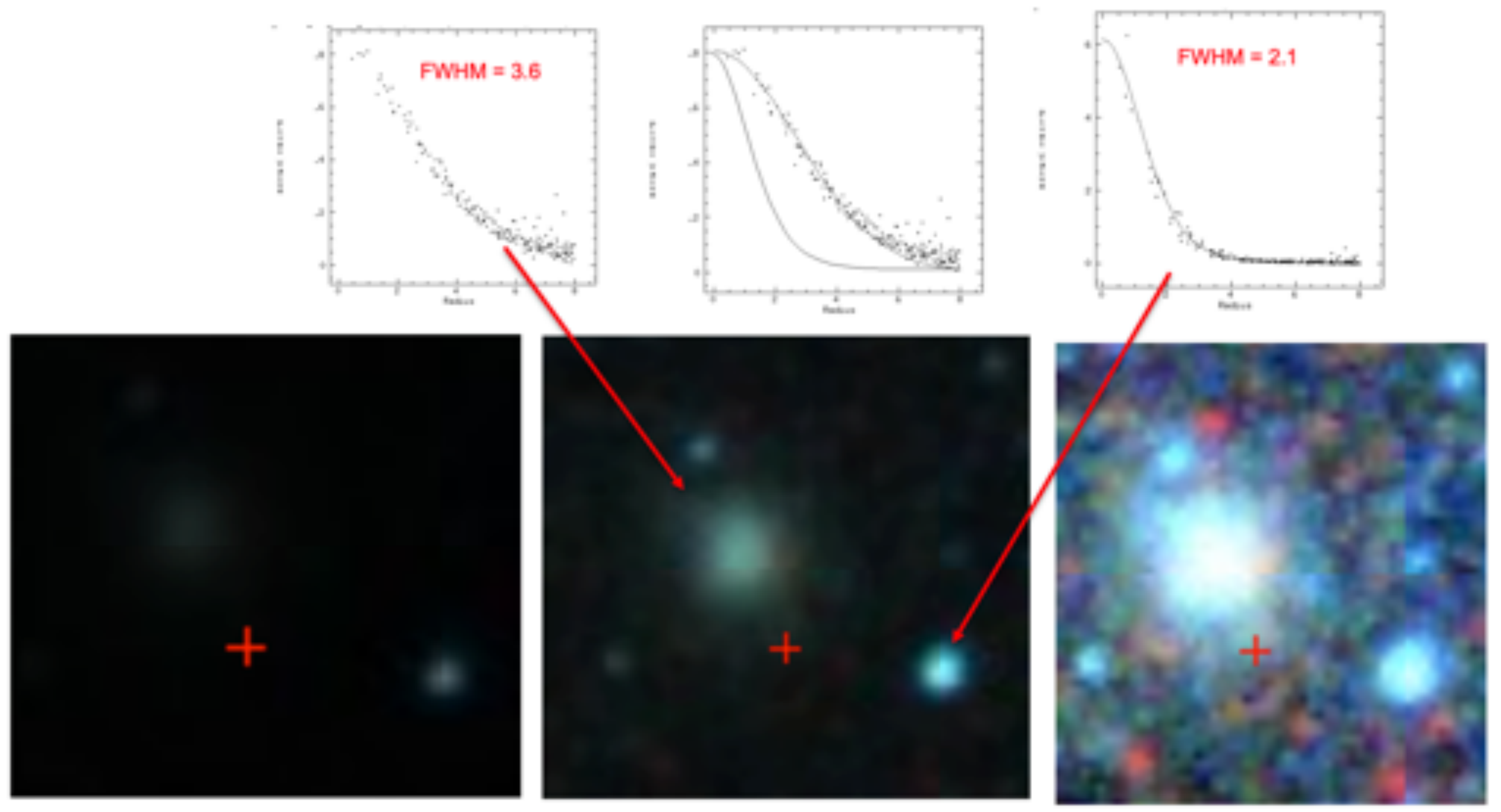}

\end{center}
\caption{Figure showing how a diffuse Class 1 cluster and nearby star are easily distinguished using different contrast levels in the DS9 display tool (bottom panels - Step 2 in the classification procedure) and the spatial profile using the {\tt imexamine} task in IRAF (top panels - Step 3 in the procedure). The upper central panel shows both the cluster and stellar profiles for comparison, with FWHM values of 3.6 pixels (cluster) and 2.1 (star) for these particular objects. The dividing line between stars and clusters is generally around FWHM = 2.4 pixels for uncrowded objects with small scatter in the profile.  }
\label{fig:contrast}
\end{figure}

Distinguishing a bright, well-resolved, isolated star cluster from a single star is generally an easy task for our sample of PHANGS-HST galaxies, as illustrated in Figure 
\ref{fig:contrast}.
Using a combination
of image examination, contrast control, and surface brightness profile review, agreement fractions of over 90 \% can be obtained between human classifiers, as will be demonstrated in Section \ref{sec:compare}.

Of course, it is not always this easy.  
In most cases the candidate clusters are fainter, less clearly resolved, and in crowded or high background regions. Below we describe in detail how the human classifications have been done for the PHANGS-HST galaxies. 

\begin{figure}
\begin{center}

\includegraphics[width =3.2in, angle= 0]{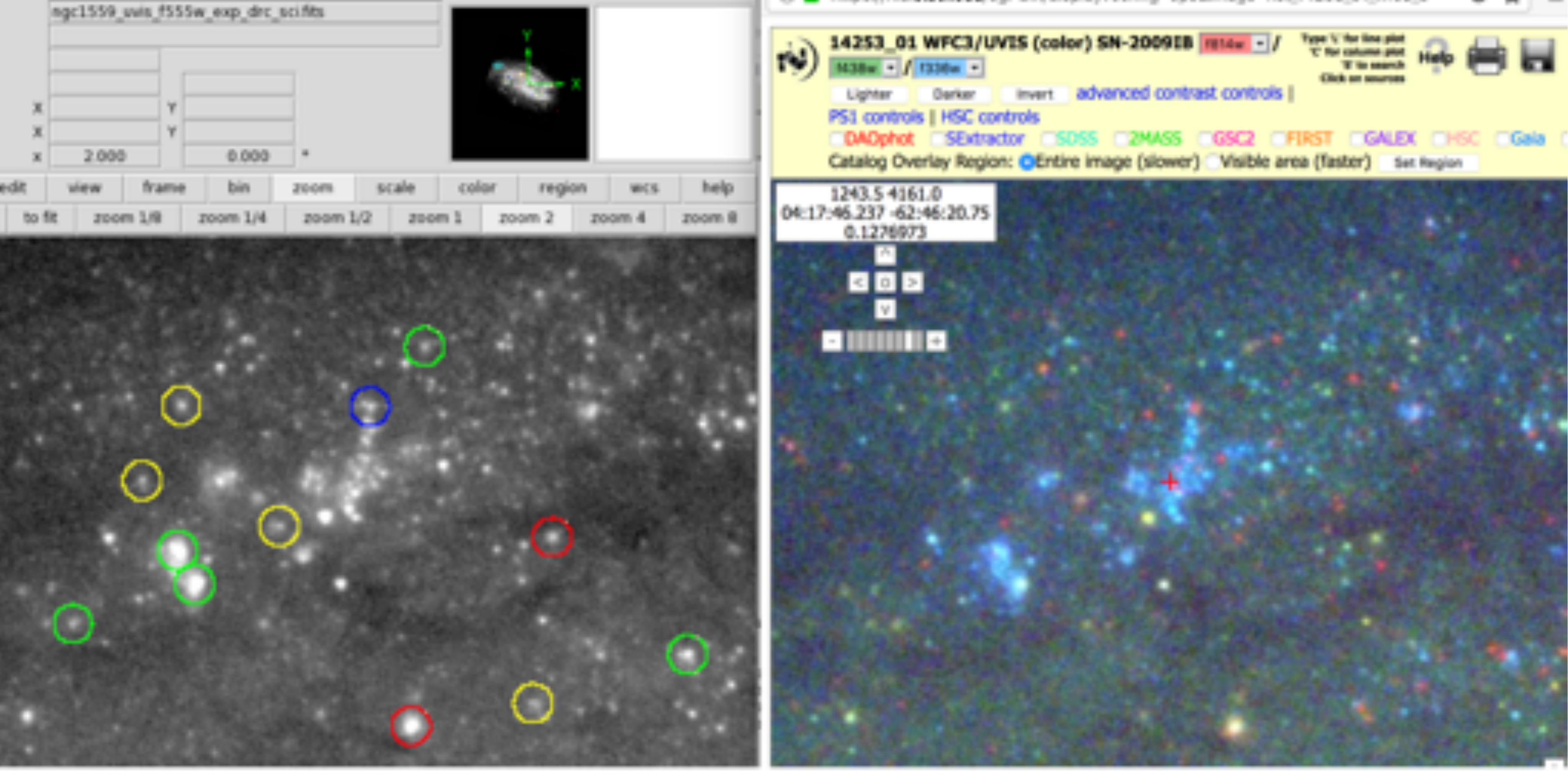}

\end{center}
\caption{Figure showing how the side-by-side SAOImage DS9 F555W image (left) and HLA  colour image (right) are used to make the classifications; Class 1 = red, Class 2 = green, Class 3 = blue, and Class 4 = yellow.   }
\label{fig:ds9}
\end{figure}

\begin{itemize}

\item STEP 1 - IDENTIFY AN INITIAL SAMPLE OF CLUSTER CANDIDATES: A subset of likely clusters are automatically identified in the F555W ($V$) filter image, using the multiple concentration index approach (MCI - see 
D. Thilker et al., in preparation)
Very briefly, the MCI approach uses a combination of  photometric measurements with seven apertures ranging from 1 to 5 pixels in radius to distinguish  resolved clusters from unresolved stars. The more standard CI-based approach relies on just two apertures (e.g., 1 and 3 pixels \citealt{whitmore99}, 
\citealt{adamo17}).
The MCI approach typically reduces the source sample from a few hundred thousand point-like objects (mainly stars or close blends) to a few hundred or few thousand cluster candidates in each galaxy. The F555W  filter has been selected as a  compromise between bluer (e.g., F275W) and redder (e.g., F814W) filters, since both young and old clusters are  generally reasonably bright in this filter. 
We note that all five filters are actually used in the machine learning determinations \citep{wei20}.
\item STEP 2 - VISUALLY INSPECT IMAGES OF CANDIDATE CLUSTERS:
Candidate clusters selected in step 1 are examined using SAOImage DS9, zoomed in by a factor of 2. In many cases a mere glance at the image will reveal whether the object is a cluster (fuzzy, soft edges) or a star (sharp edges). If not, the contrast is adjusted to compare the candidate cluster with nearby stars of roughly the same brightness. As illustrated in Figure \ref{fig:contrast}, a star of comparable luminosity will show up first because of its bright core. As the contrast is increased, the cluster will grow more rapidly and eventually will be larger than the star because of its flatter profile. Even in cases where subsequent tests are not definitive (e.g., determining the FWHM for pairs), this contrast test generally works fairly well. 

\item STEP 3 - MEASUREMENT OF FWHM AND EXAMINATION OF CONTOURS: 
The IRAF task {\tt imexamine} is used to measure the FWHM of the cluster candidate.
Stars typically have FWHM in the range 1.8 to 2.4 pixels in the PHANGS-HST data, while clusters have values in the range 2.4 to 5 (or more) pixels, depending somewhat on the distance of the host galaxy.
These correspond to radii for clusters from a few to about 20 pc.
If the candidate is fairly spherical, and is in an uncrowded region, the scatter around the best fit profile is small and the classification as a ``cluster'' is fairly secure (e.g., see Figure \ref{fig:profiles}). However, if the object is elongated, or is in a crowded region, the scatter can be   large and this particular test does not help with classification. For example, a close pair of stars  can have a large FWHM and a flat looking profile, but the large scatter indicates that such a source is not a cluster (e.g., objects  labelled C3 - association, C4.2 - pair, and C4.3 triplet objects in Figure \ref{fig:profiles}).  Contour plots are then used to help determine if a cluster is 
symmetric (Class 1) or asymmetric (Class 2), and can also help identify stellar pairs (or triplets) and elongated clusters.

Figure \ref{fig:profiles} shows several illustrative examples of different contour plots. 
While the pair (C4.2) in Figure \ref{fig:profiles} is obvious, primarily because one star is red and one is blue in the  colour image, closer pairs (and triplets) can be  challenging to differentiate from slightly extended clusters. In questionable cases, it is often useful to go back and try the contrast test (e.g., does an object grow like two stars, or like an extended cluster).

\item STEP 4 - SOURCE MORPHOLOGY IN COLOUR IMAGE:
 A  colour image of each cluster candidate is examined concurrently with steps 2 and 3.  This image is produced by the Hubble Legacy Archive software and is displayed in its interactive display tool 
 \url{https://archive.stsci.edu/hlsp/phangs-hst}
  (HLA - \citealt{whitmore16}). The HLA software has been incorporated into the PHANGS project by coauthor R. White.  
The  colour image is generally created using images in the F336W, F555W and  F814W filters, if available, although in certain cases substituting a different filter (e.g., using the F435W rather than the F336W filter) provides better contrast.

Figure \ref{fig:ds9} shows an example of using the DS9 image along side the  colour image  when making human classifications. The resulting classifications are  colour coded: Class 1 = red, Class 2 = green, Class 3 = blue, and Class 4 = yellow. 
\end{itemize}

\section{Comparison of Results from Different Classification Methods}
\label{sec:compare}

As described in \citet{wei20}, two different deep transfer algorithms (RESNET and VGG) were used to train and test machine learning classifications against both the LEGUS (39 fields in 32 galaxies) and one PHANGS-HST  (NGC 1559) human classifications. As briefly  summarised in the Introduction, they found
prediction accuracies for Classes 1, 2, 3, 4 at roughly
the 70, 40, 45, 60 \% level, respectively.
This level of agreement is similar to that found between human-vs-human classification in NGC 4656 (see \citealt{wei20} for details). 

In this section we quantify how source luminosity, crowding and background affects the agreement between different classification methods. 
As discussed in Section \ref{sec:data}, the models
trained using the BCW-only classifications 
(i.e. Table 1 from \citealt{wei20}) were used to perform the RESNET and VGG classifications.

A few words are in order concerning how we calculate ``agreement fractions''. The first step is to match the two  catalogues being compared, so that we are working with the intersection of the two studies rather than the union. A matching radius of two pixels is used. 
We determine the fraction of exact matches for each class. For that we calculate the ratio of number of matched objects divided by (a) the total number of objects from one study and (b) the total number of objects in the other study. We adopt the mean of these two ratios as the agreement fraction. We also note that the LEGUS field of view in NGC 3351 is roughly 30 \% smaller than the PHANGS-HST field of view. Only the region of overlap has been used in the calculation of the agreement fraction and the colour-colour statistics discussed in Section \ref{sec:cc}.

\subsection{Agreement as a Function of Cluster Class}
\label{sec:agree_class}

\begin{figure}
\begin{center}
\includegraphics[width =3.2in, angle= 0]{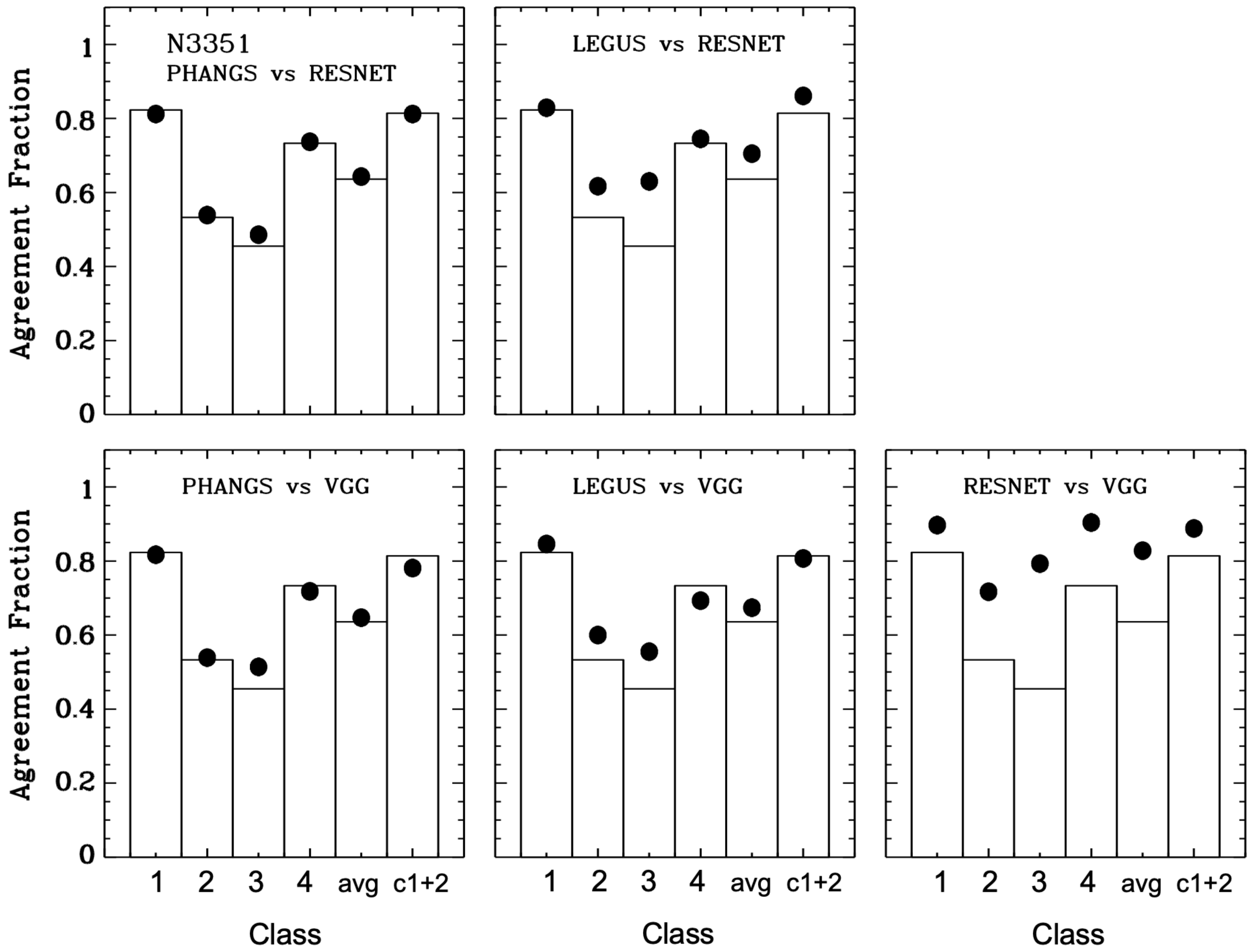}

\end{center}
\caption{Comparison between classification for PHANGS-HST (human), LEGUS (human), RESNET and VGG for NGC 3351, as described in text. The bars in the histogram, which are identical in all five panels, show the human-to-human baseline defined by comparing human classifications from PHANGS-HST and LEGUS. The filled circles show the agreement factions based on the comparison between the different studies, as denoted at the top of each panel. The fifth column shows the average of columns 1 to 4, while the column  labelled c1+2 shows the results when Classes 1 and 2 are considered as a single bin. Note that the filled circles,  which include comparisons involving machine learning classifications, are generally the same as or above the histograms representing the PHANGS-HST vs. LEGUS human-to-human comparison. } 
\label{fig:agree_class}
\end{figure}

We begin by comparing agreement fractions in the galaxy NGC 3351, as shown in Figure \ref{fig:agree_class}. The results for the other galaxies are similar, and are included in Table \ref{tab:table1}.
The  comparison between human classifications, made as part of the PHANGS-HST and the LEGUS studies, will be used as the human-to-human baseline in this figure. 
In both  cases sources were classified by the same person, co-author BCW, 
but several years apart.
The agreement fractions for Classes 1, 2, 3, and 4 are represented by the identical histogram bars in all five panels of Figure \ref{fig:agree_class}. {\it The agreement  for the human-to-human comparison between PHANGS-HST and LEGUS in NGC 3351 are 82, 53, 46, 73 \%, with a mean value for the four classes of 64 \% (shown as the  5th histogram bar)}.

These numbers for the human-to-human comparison in NGC 3351 are in somewhat better agreement (i.e., by 13 \% in the mean) than the human-to-human comparison quoted by \citet{wei20}, who compared agreement in source classifications between LEGUS-BCW and the LEGUS 3-person-consensus for the galaxy NGC 4656. The slightly higher values we find here likely result from the same person classifying sources in NGC~3351, which reduces inherent bias which may exist between human classifiers. Our experience, both here and later in Table \ref{tab:table1}, is that this is  typically a $\sim$10 \% effect. 

In Figure \ref{fig:agree_class}, we also compare both the PHANGS-HST and LEGUS human source classifications in NGC 3351  with the machine learning classifications from RESNET and VGG. The agreement between the five different combinations which include at least one human classification (i.e., PHANGS-HST vs. LEGUS, PHANGS-HST vs. RESNET, PHANGS-HST vs. VGG, LEGUS vs. RESNET, and LEGUS vs. VGG) are listed in Table \ref{tab:table1} for NGC 3351. 
We find that in nearly all cases, the comparisons between the human and machine learning classifications (i.e., the solid circles in Figure  \ref{fig:agree_class}) are as good or better than the human-to-human classifications (i.e., the histogram). 
We find the same general trends as in the human-to-human classifications, with the best agreement for Class 1 and the worst for Class 3.
The mean for all five comparisons involving human classifications (i.e., leaving out the RESNET vs. VGG classifications which are always higher)  is 66 \%, i.e., essentially the same as the mean for the human-to-human comparison in NGC 3351 being used as the benchmark (i.e., 64 \%).

An additional bar ( labelled ``C1+2") is included in Figure \ref{fig:agree_class} for the case where the Class 1 and Class 2 objects are combined into a single bin, a common  procedure in many studies. 
We find an agreement fraction of 81 \% for the baseline human-to-human (i.e., PHANGS-HST versus LEGUS) classifications for Class 1 + 2 in NGC 3351, and similar values for all four of the other combinations in Figure \ref{fig:agree_class}. These numbers are very similar to the agreement fractions for Class 1 alone.

The C1+2 sample will be considered the ``standard" sample in many aspects of the discussion throughout this paper. However, 
our general advice is to use both Class 1$+$2 and Class 1 alone to see how they affect your science results, as we have done in Section \ref{sec:sci}. 

Finally, we note the very high agreement fractions when the two machine learning algorithms (RESNET vs. VGG) are compared (e.g., in the bottom right panel in Figure  \ref{fig:agree_class} for NGC 3351).
We interpret this to be due to the repeatability when computer classification algorithms are used. However, there is no guarantee that the machine learning classifications are actually ``better" or ``correct", only that they are more repeatable. We shall revisit this point in Section \ref{sec:cc} where we examine ``figures of merit".

Table \ref{tab:table1} also includes the agreement fractions in the four classes for all five galaxies studied here.
We find that for all galaxies except NGC~628, the mean agreement fractions including the machine learning algorithms are higher than the PHANGS-HST vs. LEGUS (human-to-human) ones.  {\it This, and the similar result for NGC 3351 shown in Figure \ref{fig:agree_class}, are the primary reasons for our statement that the machine  learning classifications are as good or slightly better than human classifications.}

The two highest {\em mean} agreement fractions in Table \ref{tab:table1} are for NGC 3627 (74 \%) and NGC 3351 (66 \%), which are the two galaxies classified by BCW in both PHANGS-HST and LEGUS. The slightly higher agreement, $\sim10$\%, is likely for the same reason described earlier, that this reduces the systematic human-to-human differences in classification methodology in this cases since the same human is involved, and the BCW LEGUS training set was used for RESNET and VGG classifications.

In general, there is no one study that appears to be much better than the others. For example, when considering the mean values from column 5 of Table \ref{tab:table1}  (excluding the RESNET vs. VGG comparison), three different classifier combinations have the highest values for a given galaxy (i.e., PHANGS-HST vs. VGG for NGC 628 and NGC 1566; PHANGS-HST vs. RESNET for NGC 1433; and LEGUS vs. RESNET for  NGC 3351).

To summarize, we find that comparisons between all four classification methods give fairly similar results; all of them appear to provide source classifications of comparable quality.

\subsection{Agreement as a Function of Source Brightness}
\label{sec:agree_bright}

\begin{figure}
\begin{center}
\includegraphics[width =3.3in, angle= 0]{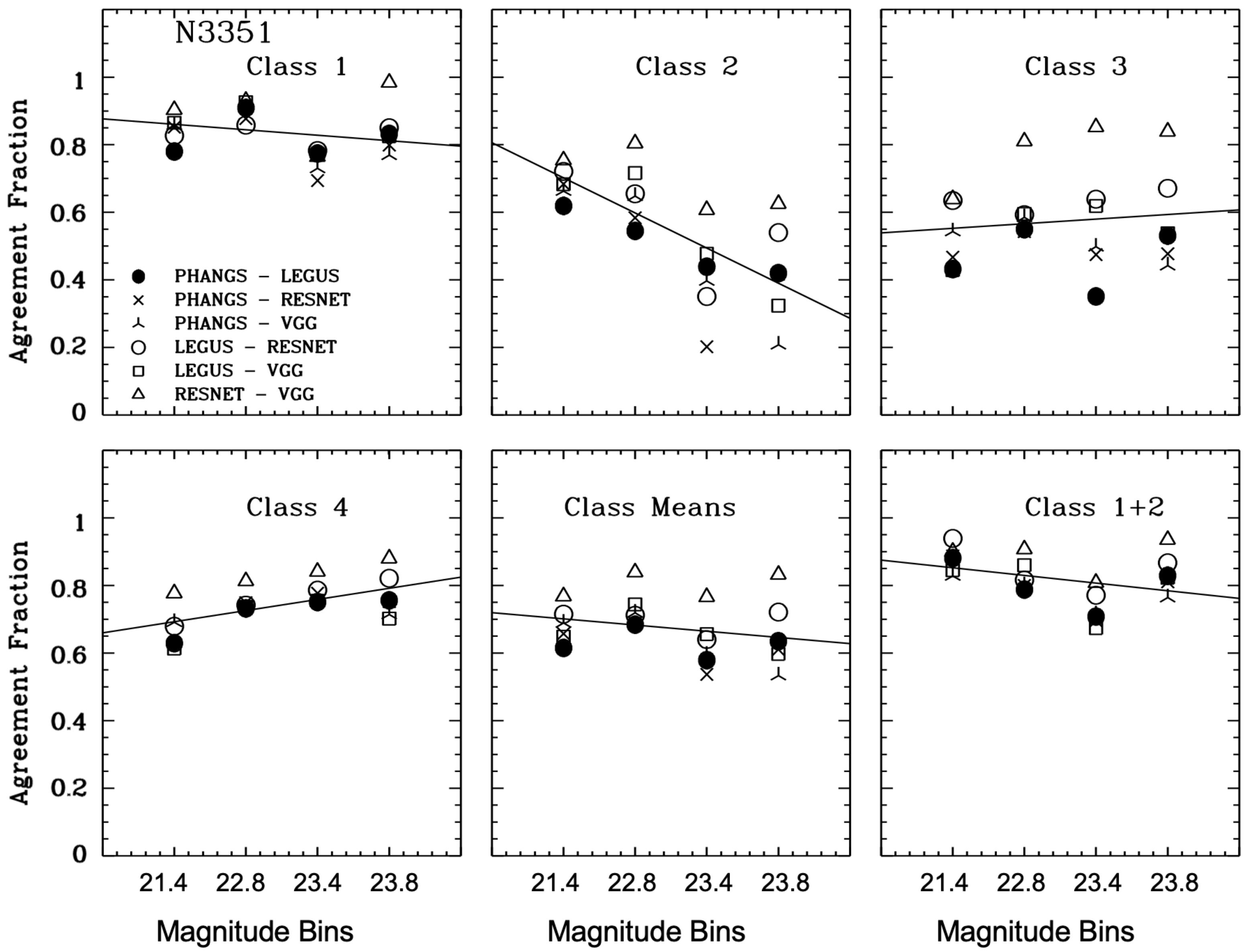}
\end{center}
\caption{Agreement fractions as a function of magnitude bins  for Classes 1 to 4, the means of Classes 1 to 4 added together (bottom middle), and the values if Class 1 + 2 are considered as one bin (bottom right). The two studies being compared in each case are plotted with different symbols. Least-squares fits to all the data are shown by the lines. We find that while many of the dependencies are relatively flat (i.e., Class 1, Class 1 + 2) the agreement fractions for Class 2 falls rapidly as a  function of magnitude.  }
\label{fig:agree_mag}
\end{figure}

In Figure \ref{fig:agree_mag} we examine how the results from different classifiers change with brightness, comparing the agreement fraction of sources in NGC~3351 in different magnitude bins. Classes 1, 2, 3, and 4 are shown in different panels.
The symbols represent the different combinations of classifiers (PHANGS-HST, LEGUS, RESNET, VGG).  The magnitude bins have been selected to contain roughly equal numbers of sources (of all classes), which leads to non-uniform size bins because there are many more faint clusters than bright ones.
The medians of the bins are $m_V$ = 21.4, 22.8, 23.4, and 23.8   ~mag.

Probably the most important conclusion in this subsection is that the agreement fractions for Class 1, and also for Class 1 + 2 when considered one bin (i.e., the standard sample),
are similar regardless of source brightness, with values around 85 \% over the full magnitude range. This is reassuring since finding Class 1 and Class 2 clusters 
is the focus of our cluster classification effort. It also implies that we may be able to use the machine learning algorithms to push down to fainter magnitudes for these types of clusters. This is discussed in more detail in Section \ref{sec:cc_faint} and in
D. Thilker et al. (in preparation).

Class 2 clusters alone, on the other hand, show a strong decreasing agreement fraction towards fainter magnitudes, starting around 70 \% for bright clusters and dropping to about 40 \% (with large scatter) at the faint end. This may reflect the fact that most bright Class 2 clusters are similar to Class 1, with some relatively minor asymmetries, while many of the fainter Class 2 clusters are more difficult to discriminate from Class 3 objects.

Note that the agreement fractions are generally higher when combining the Class 1 and 2 clusters into a single bin compared to averaging them together.
This is because a frequent difference in classification is to interchange the two classes; i.e., to draw the line between symmetric and asymmetric slightly differently. Hence, some of the shortcomings discussed in this section and elsewhere in the paper for the classification of the Class 2 clusters alone are not as serious when combining the two into one bin.

The agreement fractions for Class 3 (compact associations) are essentially flat as a function of magnitude, albeit with a large scatter, while the agreement fractions for Class 4 (artifacts) actually increase at fainter magnitudes. The latter effect likely results because the fraction of contaminants in the form of individual and pairs of stars, which are relatively easy to classify,  increases at fainter magnitudes.

\subsection{Agreement as a Function of Crowding}
\label{sec:agree_crowd}

\begin{figure}
\begin{center}
\includegraphics[width =3.3in , angle= 0]{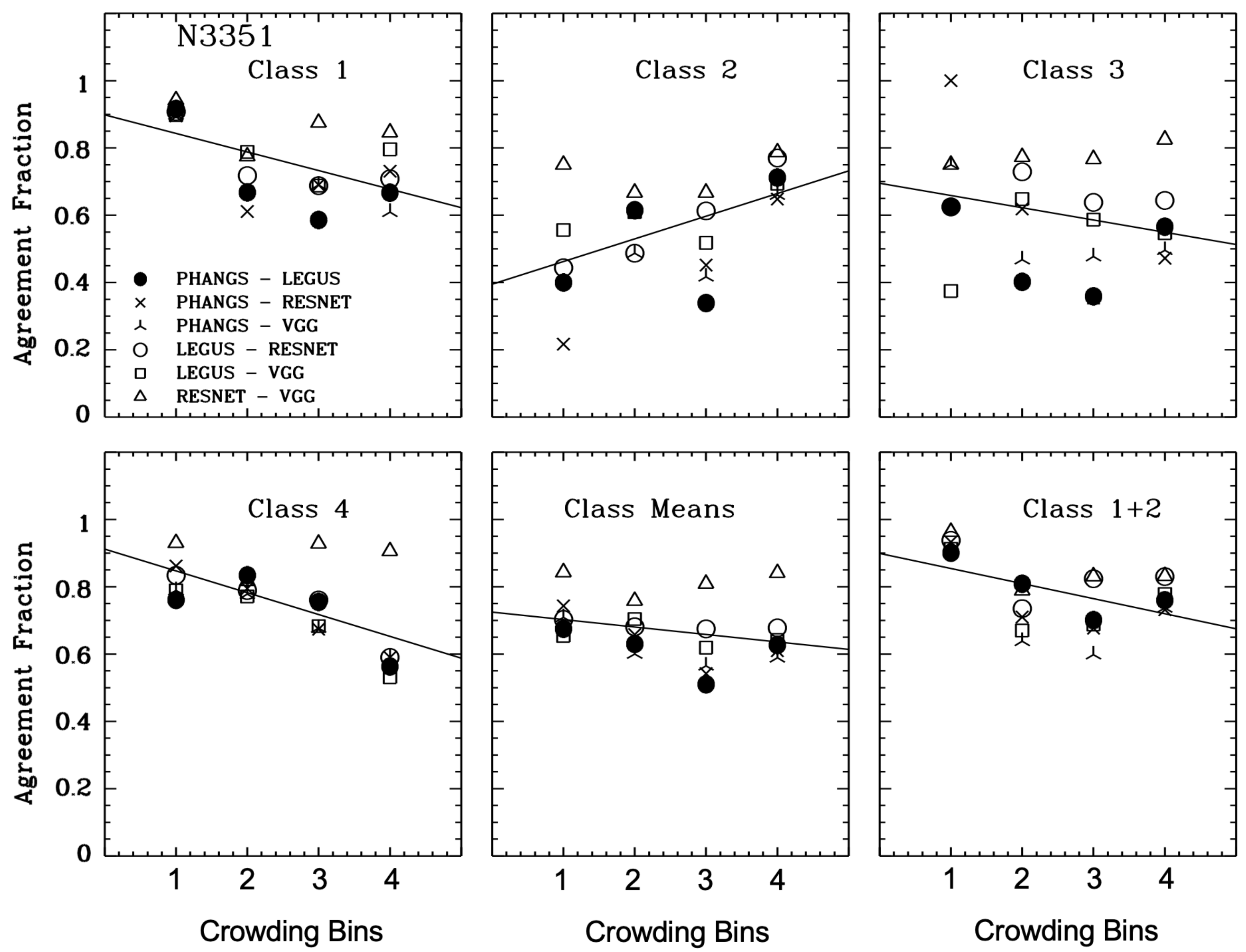}
\end{center}
\caption{Agreement fractions as a function of crowding bins (bin 1 = isolated, bin 4 = most crowded) for Classes 1  to 4, the means of Classes 1 to 4, and Class 1 +  2 considered as one bin. The two studies being compared in each case are plotted with different symbols. We find that most of the dependencies are steeper than versus magnitudes. }
\label{fig:agree_crowd}
\end{figure}

In Figure \ref{fig:agree_crowd} we make comparisons for the agreement fractions as a function of source crowding. The methodology is essentially the same as for our comparisons with magnitudes shown in Figure \ref{fig:agree_mag}. The crowding parameter employed is from the  DOLPHOT V-band ``crowd' parameter.

The  catalogue is  broken into four subsamples, with crowding bin = 1 for isolated objects (e.g., old isolated clusters in the  smooth bulge regions outside of the central starburst ring), to crowding bin = 4 for very crowded regions (e.g., large associations and the central starburst ring). 

The trends with crowding are as strong or stronger than those found with brightness. The results for each object class is shown in its own panel, as  labelled in Figure~ \ref{fig:agree_crowd}.  We note the excellent classification agreement (mean = 92 \%) for isolated Class 1 objects (crowding bin = 1), shown as the upper left points in the upper left panel. These are primarily old clusters in the bulge which have had ample time to separate from their birth clouds and hence are generally isolated. The agreement falls to about 70 \% for Class 1 for the more crowded bins. The reverse trend is seen for Class 2 objects, with the highest agreement fraction for the most crowded regions. This reflects the fact that crowded regions will often introduce apparent asymmetries in the outskirts of clusters, causing them to be classified as Class 2 (asymmetric clusters) rather than Class 1 (symmetric clusters). We note, however, that the relation for the  standard Class 1 + 2 sample, and  the means, are relatively flat, though not as flat as versus magnitude. This indicates that crowding has a stronger impact than brightness on the classification of Class 1 and Class 2 objects, the main focus of this paper.
\citet{perez21} report similar issues with Class 2 and Class 3 classifications, hence this effect appears to be inherent in the difficulty of classifying these objects in crowded regions rather than in the particular machine learning classification method employed.

The scatter between the four different classification methods for some of the source types (e.g., Class 2 and Class 3) is large, presumably reflecting differences in how the different methods perform the classifications.

\subsection{Agreement as a Function of Local Background}
\label{sec:agree_background}

\begin{figure}
\begin{center}
\includegraphics[width = 3.3in, angle= 0]{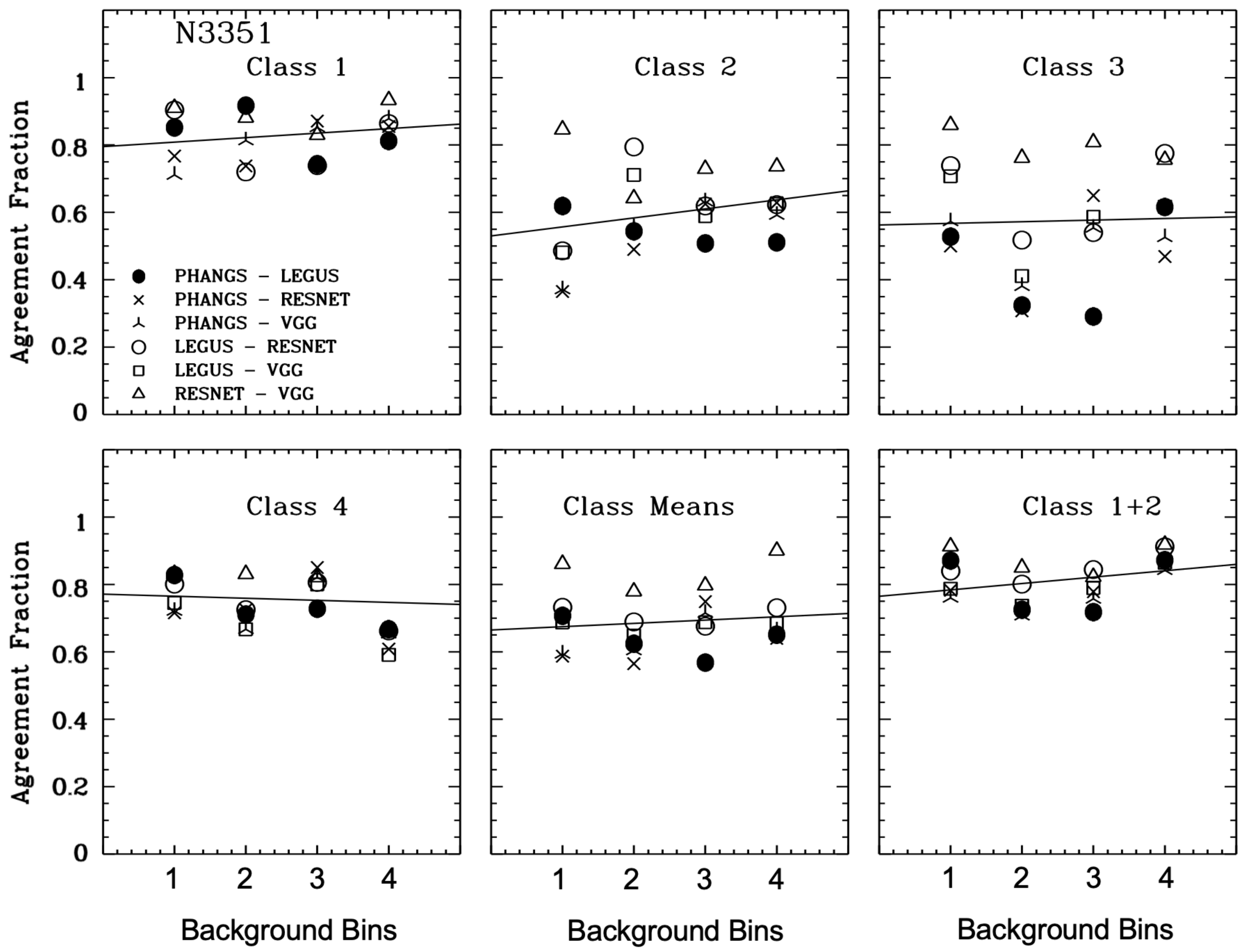}
\end{center}
\caption{Agreement fractions as a function of local background bins (bin 1 = low background, bin 4 = highest background) for Classes 1 to 4, the means of Classes 1 to 4, and Classes 1 and  2 together. The two studies being compared in each case are plotted with different symbols. We find that  most of the dependencies are relatively flat.   }
\label{fig:agree_back}
\end{figure}

In Figure \ref{fig:agree_back} we assess how much the background surface brightness  affects the classification results, where the background is defined to be the median flux in the background annulus used for photometry (i.e., an annulus between 7 and 8 pixels in radius).
In general, the agreement fractions do not vary significantly with background level, unlike the situation with  magnitude and crowding. Part of the reason for this might be that some sources in regions of high background are found in the central region around the chaotic star formation ring. The agreement fractions are low here. Other sources with high background are in the smooth bulge component just outside of this region. The agreement fractions are actually high here. Hence, the results for these regions tend to balance each other out, resulting in relatively flat correlations.

\subsection{Agreement as a Function of Spatial Resolution}
\label{sec:agree_resol}

The agreement fractions are likely to be poorer for more distant galaxies, as it becomes more difficult to distinguish clusters from individual stars. However, it is difficult to test this hypothesis when we have only five galaxies in our sample. While there does appear to be a weak correlation in the expected sense (see Table \ref{tab:table1}), a larger sample will be required to make this determination in the future.

To summarize subsections \ref{sec:agree_class} to \ref{sec:agree_resol}, crowding appears  to introduce the largest uncertainties in the classifications, more so  than faint magnitudes or high background. Hence, classification might be expected to be more problematic for galaxies with high star formation rates, producing large numbers of very young stars, which tend to both cluster more strongly resulting in higher crowding, and produce higher backgrounds.

\bigskip

\section{Classification Accuracy Using Colour-Colour Diagrams}
\label{sec:cc}

While we make relative rather than absolute comparisons throughout most of this paper, we 
also develop ``figures of merit" to help guide us toward the best methods of classification.
One straightforward approach is to use locations 
of measured clusters in  colour-colour plots.

\begin{figure}
\begin{center}
\includegraphics[width =3.3in, angle= 0]{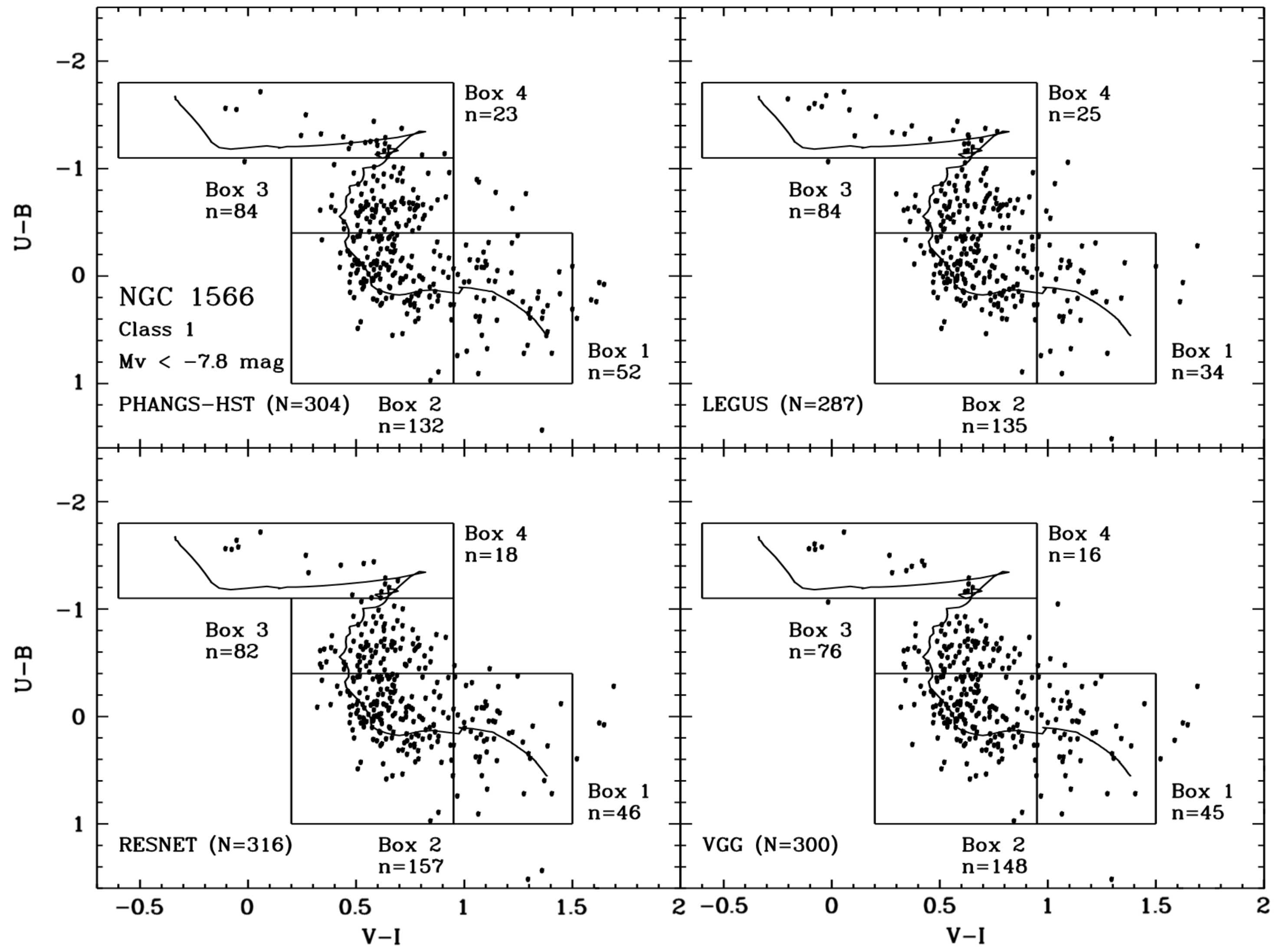}
\end{center}
\caption{U-B vs. V-I  colour-colour diagrams for Class~1 objects for each of the four classification methods for $M_V < -7.8$ mag. Four boxes corresponding to different age clusters are included, along with the number of objects in each box. A solar metallicity isochrone from \citet{bruzual03} is  included, running from 1 Myr in the upper left to 10 Gyr in the lower right. Note how similar the distributions of points are for all four of the classification methods. 
}
 
\label{fig:cc_c1}
\end{figure}

\begin{figure}
\begin{center}
\includegraphics[width =3.2in, angle= 0]{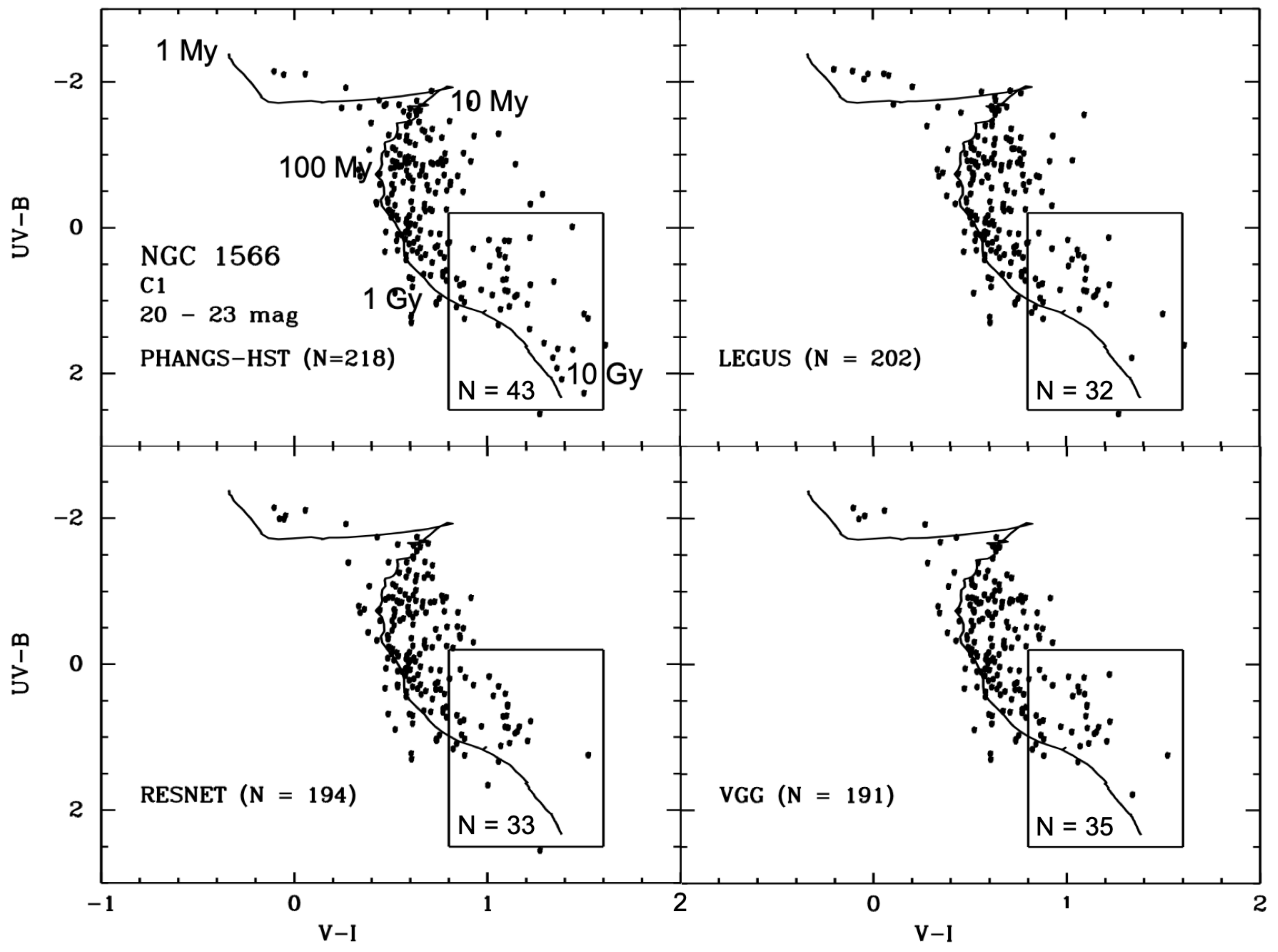}
\end{center}
\caption{UV-B vs V-I  colour-colour diagrams for Class~1 objects for each of the four classification methods for $m_V$ = 20 - 23 mag. Ages for positions along the isochrone have been added in the upper left panel. A box appropriate for old globular clusters is included.  }
\label{fig:cc_uv}
\end{figure}

Figure \ref{fig:cc_c1} shows the U-B vs. V-I diagram in NGC 1566 based on the four different classification methods.  Only Class~1 objects are shown.  This  colour-colour plot will be used throughout Section \ref{sec:cc}. Figure \ref{fig:cc_uv} shows another version of the  colour-colour plot, but substitutes the UV filter for the U band. We focus on NGC 1566 as an illustrative example; results for all five galaxies are included in Table \ref{tab:table2}. The results for all the galaxies are similar, as will be shown in Section \ref{sec:cc_merit}.

In these colour-colour diagrams, measured cluster  colours (data points) are compared with predictions from the solar metallicity \citep{bruzual03} cluster evolution models (lines).  The  colours of clusters change with age, as the most massive (generally blue) stars die off.  This predicted evolution through  colour-colour space is highlighted in the upper left panel of Figure \ref{fig:cc_uv}, where the predicted ages are shown, starting at 1 Myr in the upper left-part of the diagram and ending at 10 Gyr in the lower-right.

\begin{figure}
\begin{center}
\includegraphics[width =3.3in, angle= 0]{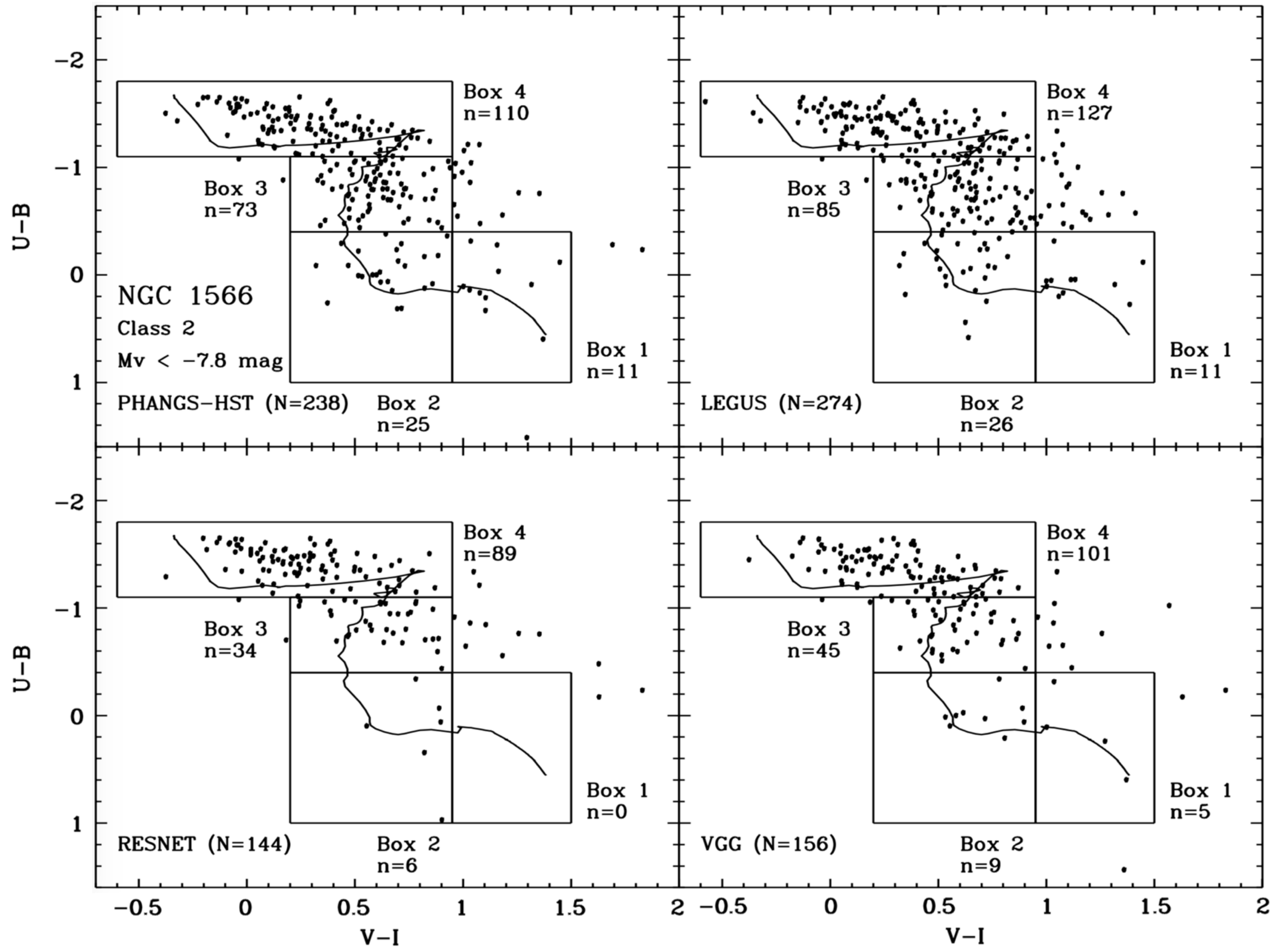}

\end{center}
\caption{Same as Figure \ref{fig:cc_c1}, but for Class~2 objects. Note the lower numbers of objects in Box 3 for RESNET (34) and VGG  (45) relative to  PHANGS-HST (73) and LEGUS (85). }
\label{fig:cc_c2}
\end{figure}

Several papers demonstrate how the  colour distributions of the sources change with morphological type  (e.g., \citealt{adamo17}, \citealt{whitmore20}, \citealt{turner21}), 
 since older clusters tend to be both redder and  more symmetric.  This is clearly observed in Figure~\ref{fig:cc_c1} by the lack of clusters along the youngest portion of the cluster evolutionary track. 
 Not surprisingly, the same lack of very young clusters is found in Figure \ref{fig:cc_uv}.
Class~2 clusters are asymmetric by definition, typically because they are in more crowded regions or because the cluster has not had sufficient time to dynamically relax, and we therefore expect them to be younger than Class~1 objects, i.e. to have  colours indicative of younger ages.
Figure \ref{fig:cc_c2} shows the distributions for Class~2 objects.
A significant fraction of these objects 
have  colours suggesting they are $10-100$~Myr old.
There are also a relatively large number of Class~2 objects with  colours indicating they are younger than 10 Myr (i.e., with U-B  colours bluer than U-B = -1.0).
Therefore, this correlation  provides a way to  test between Class~1 vs. Class~2 clusters, i.e. most old, red clusters should be identified as Class~1, while few old, red clusters should be identified as Class~2.

\begin{figure}
\begin{center}
\includegraphics[width = 3.3in, angle= 0]{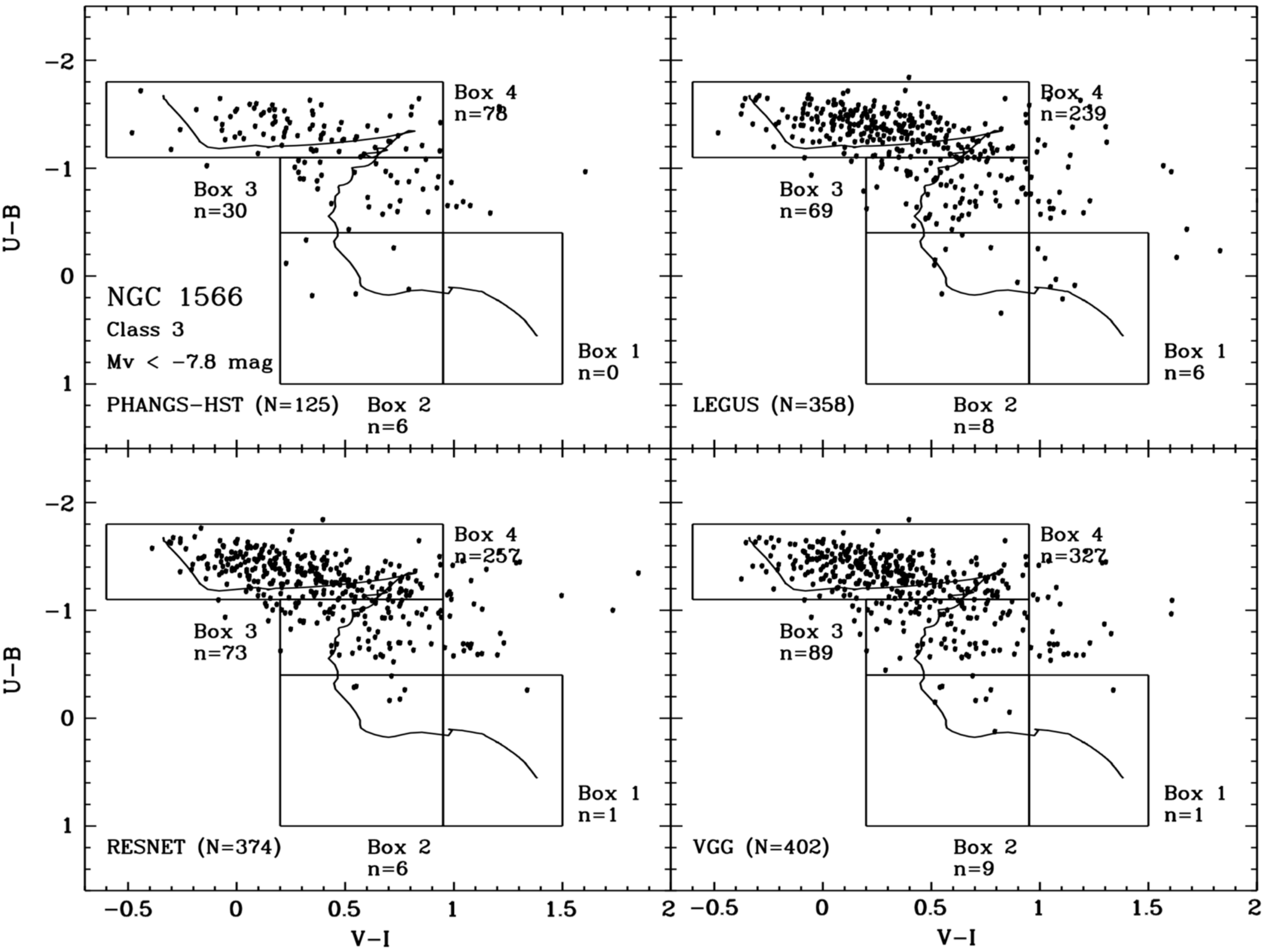}

\end{center}
\caption{Same as Figure~\ref{fig:cc_c1}, but for Class~3 objects. The lower numbers of PHANGS-HST Class~3 clusters is expected since the MCI method of selecting candidate clusters is designed to reduce the number of Class~3 (compact associations) and Class~4 (artifacts) (see D. Thilker et al., in preparation).}

\label{fig:cc_c3}
\end{figure}

Figure \ref{fig:cc_c3} shows the distributions for Class~3 objects, which are generally found to be bluer and hence have younger ages (i.e., < 10 Myr),  with typical values of V-I  between -0.6 and 1.0, and U-B less than -1.0. There are also a sprinkling of Class~3 objects that have redder  colours, likely due to a combination of age, reddening, and stochasticity effects (see \citealt{maiz09}, \citealt{fouesneau12}, \citealt{hannon20} and \citealt{whitmore20}).  
As discussed in these papers, stochasticity is primarily an issue for lower mass clusters, i.e. with log mass < 3.5 solar masses.

Hence, it appears that the distribution of data points in the colour-colour diagram can be used as a  ``figure of merit" for our cluster classifications.
Below we make quantitative checks to test the quality and uniformity of the different classification methods.

\begin{figure}
\begin{center}
\includegraphics[width = 3.3in, angle= 0]{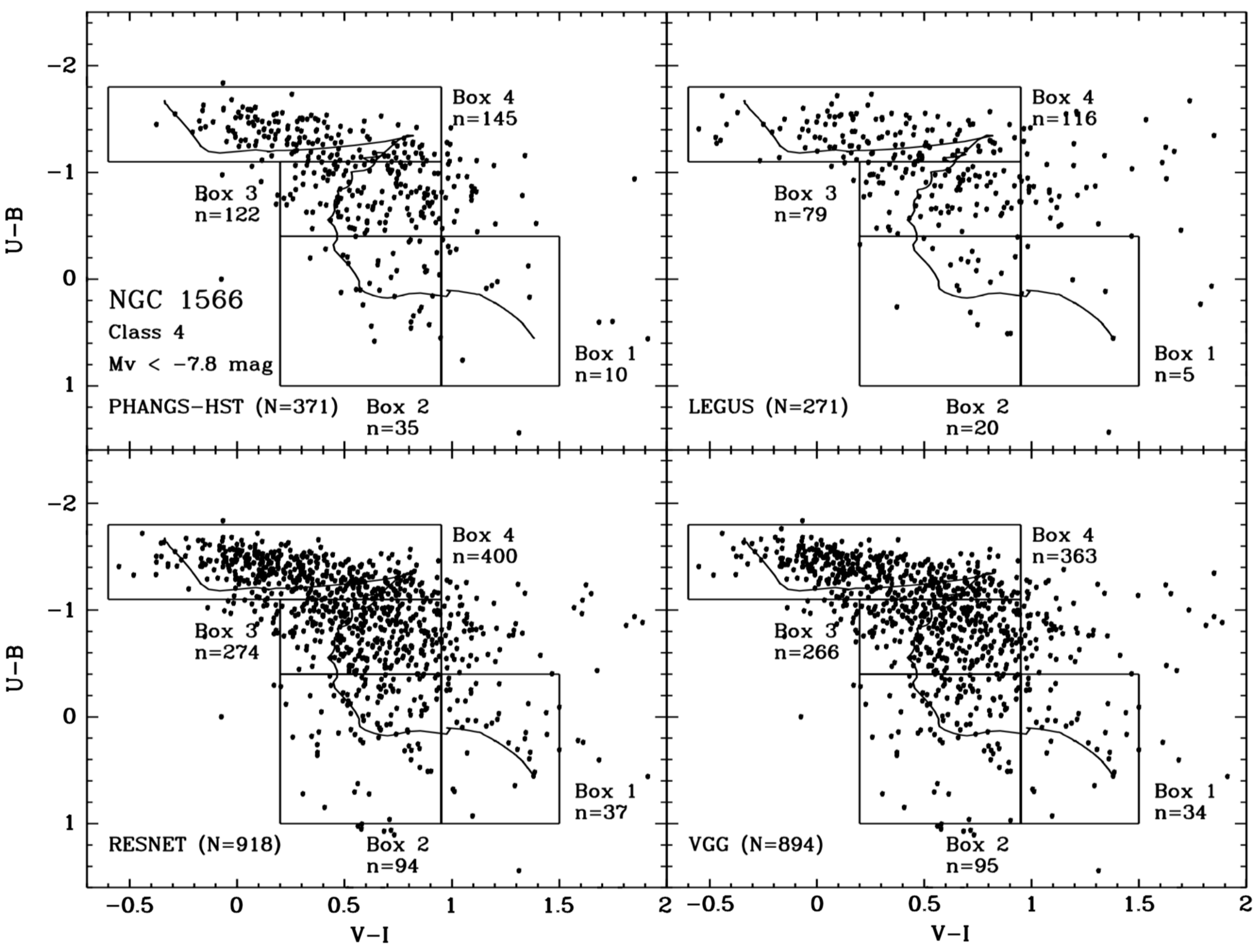}
\end{center}
\caption{Same as Figure~\ref{fig:cc_c1}, but for Class~4 objects. The larger numbers of RESNET and VGG Class~4 objects (artifacts) is primarily due to the use of the 
larger model selection region in the MCI plane for the two machine learning methods, rather than the smaller empirical polygon selection region used to identify candidates for the human PHANGS-HST classification
(see D. Thilker et al., in preparation, 
for details). This demonstrates that a much higher percentage of the objects in regions outside the polygon are artifacts, as expected.}
\label{fig:cc_c4}
\end{figure}

Figures \ref{fig:cc_c1}, \ref{fig:cc_c2},  \ref{fig:cc_c3} and \ref{fig:cc_c4} also
include four boxes drawn to roughly demarcate different ages in the U-B vs. V-I  colour-colour diagram, as defined below.
This procedure is similar to an approach originally used in \citet{whitmore10} to divide the  colour-colour plane into regions for the purpose of separating stars and clusters in the Antennae galaxies (also see \citet{chandar10b} for a similar treatment in M83).

The regions are defined as:

\bigskip

{\bf Box~1}: Old  clusters (> 1 Gyr): 0.95 < V-I < 1.5 and -0.4 < U-B < 1.0

\bigskip

{\bf Box~2}: Intermediate-age clusters (0.1 to 1.0 Gyr): 0.2 < V-I < 0.95 and -0.4 < U-B < 1.0

\bigskip

{\bf Box~3}: Young clusters (10 - 100 Myr): 0.2 < V-I < 0.95 and -1.1 < U-B < -0.4

\bigskip

{\bf Box~4}: Very young clusters  (<  10 Myr): -0.6 < V-I < 0.95 and -1.8 < U-B < -1.1

\bigskip

We note that when mapping the regions of the  colour-colour diagram onto apparent ages we  assume no reddening. This is a reasonable assumption for objects older than 10 Myr (see \citealt{whitmore20}), but it should be kept in mind that these age estimates are approximate. Values for the number of clusters in each box for each classification method are included in Table \ref{tab:table2}.

Examining Figure \ref{fig:cc_c1}, we first note that the total number of Class~1 objects (symmetric clusters) classified by the four methods are similar, with 304, 287, 316, and 300 objects from box 1 to box 4. {\it We note the strong similarity in the overall  colour distributions for Class~1 objects for all four methods}, each hugging the right side of the stellar isochrone in the region from 10 Myr to about 1 Gyr,
and then broadly following the isochrone itself in the region beyond 1 Gyr.  Figure \ref{fig:cc_uv} shows the same using the UV instead of the U filter, although the old globular clusters 
tend to be somewhat high. This is because we show a solar metallicity isochrone which is appropriate for the young  but not the oldest clusters (see \citealt{turner21} for a discussion).
{\it The strong similarity between the four panels in Figures \ref{fig:cc_c1} and \ref{fig:cc_uv} indicates that all four of the methods are finding similar Class 1 objects, which is very reassuring.}

\begin{figure}
\begin{center}
\includegraphics[width = 3.3in, angle= 0]{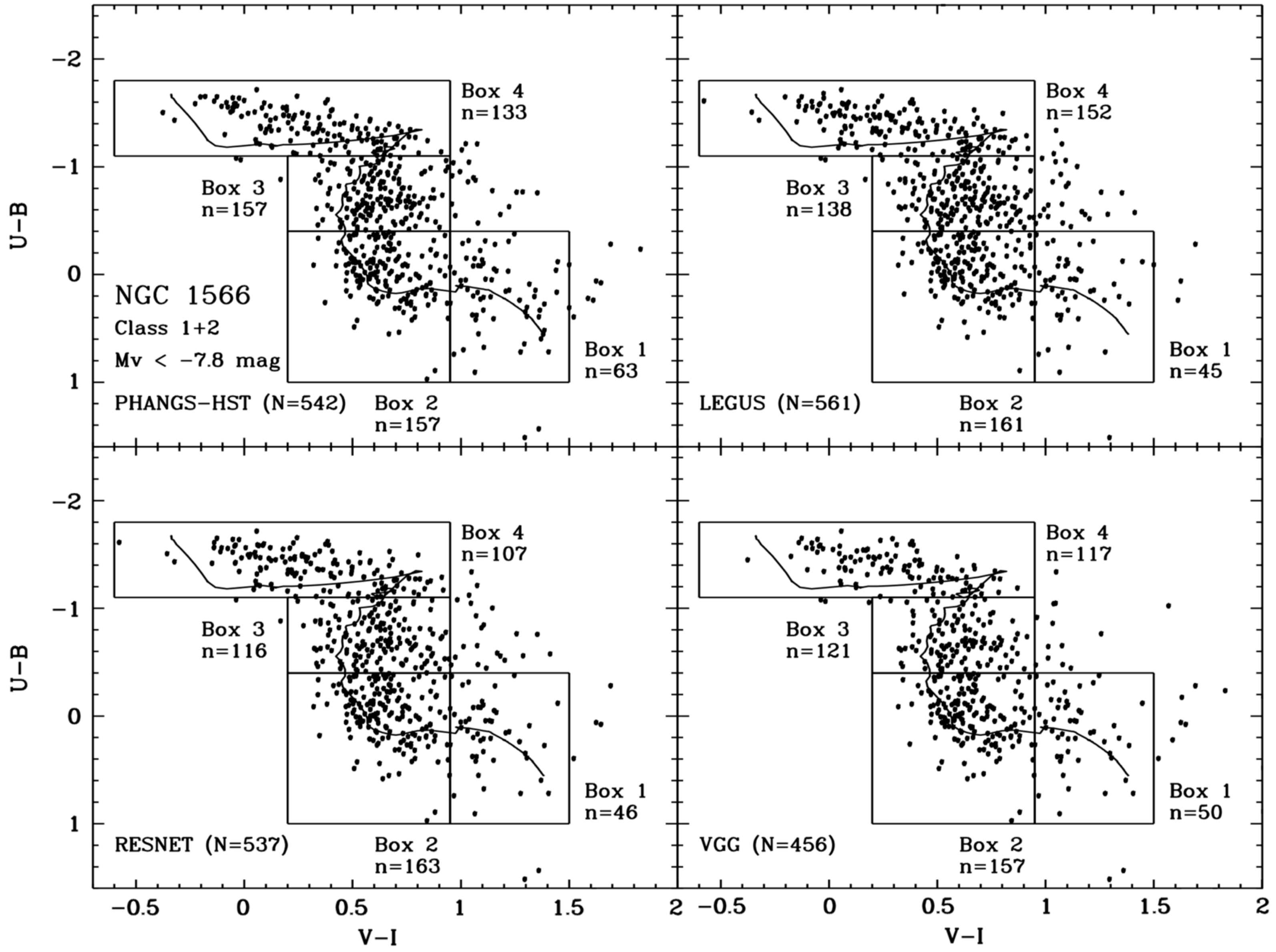}
\end{center}
\caption{Same as Figure~\ref{fig:cc_c1}, but for Class~1+2 objects (i.e., the "standard" sample.)
Note how similar the distributions of points are for all four of the classification methods, similar to the results for Class~1 in Figure~\ref{fig:cc_c1}, but with a larger number of clusters. 
}
\label{fig:cc_c12alone}
\end{figure}

Figure \ref{fig:cc_c2} includes the relative distributions within the four boxes for Class~2 objects (asymmetric clusters). The distributions of cluster  colours in these four panels look less similar than for Class~1, with a clear deficit in Box 3 based on RESNET and VGG  classifications (34 and 45) relative to the PHANGS-HST and LEGUS classifications (73 and 85).  This suggests that the machine learning algorithms do not identify as many  young objects (Box 3) as Class~2 when compared with PHANGS-HST and LEGUS.

Figure \ref{fig:cc_c3} shows the results for Class~3 sources (compact associations). The most obvious difference is the much smaller number of classified objects in PHANGS-HST. 
This is largely by design, since the MCI method discussed in Section \ref{sec:procedure} was introduced to minimize the number of Class~3 (compact associations) and Class~4 (artifacts) objects. In addition, 
Class~4 has been broken into several subclasses (see Table \ref{tab:table3}), in an attempt to allow the machine learning algorithms to better classify (and remove as artifacts) similar objects in future treatments. Hence, pairs and triplets are included as Class~4 objects (artifacts). Since LEGUS used a somewhat looser definition for its Class~3, it includes a fair fraction of pairs and triplets, most of which would be included as part of the expanded Class~4 in PHANGS-HST rather than as Class~3 objects. 
If we add the Class~4.2 (pairs) and Class~4.3 (triplets) objects to Class~3 for PHANGS-HST, the number would increase from 125 to 345, roughly the same as found for LEGUS (i.e., 358).

Figure \ref{fig:cc_c4} shows the results for Class~4 sources. The most obvious difference is the larger number of classified objects for RESNET and VGG compared with human classification. 
This is primarily due to the selection of candidates from the
larger region in the MCI plane for the two machine learning methods, rather than the smaller empirical polygon selection region used to identify candidates for the human PHANGS-HST classifications. See 
D. Thilker et al. (in preparation)
for a description of the candidate selection procedure for PHANGS-HST.
The larger number of Class 4 objects for RESNET and VGG demonstrates that a much higher percentage of objects in regions outside the polygon in the MCI plane are artifacts, as expected. 

The distribution of Class~4 points in the colour-colour plots is most similar to the Class~2 objects, but with a wider spread due to the stochasticity imposed by the low number of stars in the objects (i.e., c4.2 = pairs and c4.3 = triplets are the most populated subclasses). We also  note a small increase in the percentage of objects in Box 1 (old clusters) for RESNET and VGG (3.7 \% and 3.6 \% respectively) compared to PHANGS and LEGUS (2.5 \% and 2.1 \% respectively). This suggests that the machine learning algorithms misclassify a slightly larger fraction of old globular clusters as artifacts. 

Figure \ref{fig:cc_c12alone} shows the results for Class~1+2 sources included together. As discussed in Section \ref{sec:agree_class}, combining these  two classes (where the primary criteria is central concentration) is a common practice, although we also suggest experimenting with Class 1 alone to see how the science results of a given project might be affected. 
In Section \ref{sec:comp_matrix} we will find that Class 1 and Class 2 are frequently interchanged,
due to slightly different estimates of the degree of asymmetry. Hence combining the two classes results in agreement fractions and other properties that are roughly as good or slightly better than  Class 1 alone in most cases. For these reasons we call Class 1$+$2 the ``standard'' cluster sample in this paper.

We first note that like Class 1, the distributions are very similar for all four classification methods in NGC 1566. Hence all the methods find similar objects. While Class 1 alone has very few young objects, Class 1 + 2 has a more even distribution of ages, which can be a useful characteristic for many science projects. VGG has marginally fewer Class 1 + 2 clusters than the other three methods, with a slight tendency to find more Class 3 objects, as can be seen in Figure  \ref{fig:cc_c3}. A similar but weaker tendency is seen for the other galaxies, as shown in Table \ref{tab:table2}.

\begin{figure}
\begin{center}
\includegraphics[width =3.6in, angle= 0]{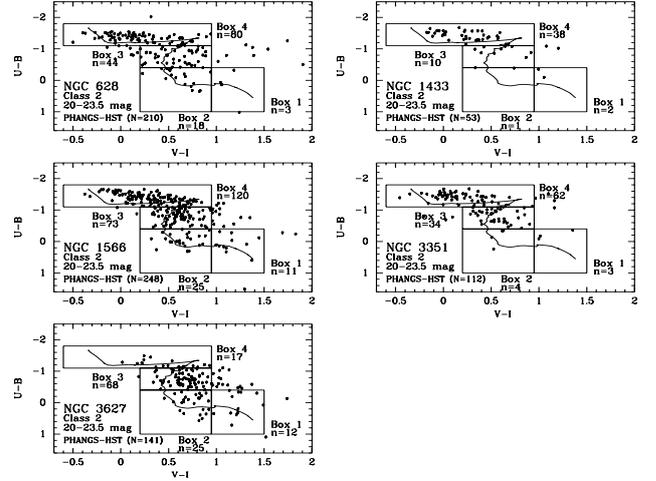}

\end{center}
\caption{Same as Figure~\ref{fig:cc_c1} for Class 2 clusters for all five program galaxies in range $m_v$ = 20 - 23.5 mag. We find a wide variety of different cluster formation histories from NGC~1433, with 72~\% of the objects in the youngest box, to NGC~3627, with 12~\% of the objects in the youngest box.}  
\label{fig:5plot}
\end{figure}

One of the primary results based on examining Figures~\ref{fig:cc_c1} to \ref{fig:cc_c4} is that all four methods of classification give fairly similar results for NGC~1566. This will be addressed in more detail in Section~\ref{sec:cc_merit}. However, one surprise is that the distribution of Class 2 objects looks more like the Class 3 objects than the Class 1 objects in NGC~1566, unlike previous results such as for NGC~4449 (\citealt{whitmore20}). One possible  explanation is that the galaxies have different cluster formation histories. 

In Figure \ref{fig:5plot} we show the Class 2 colour-colour diagrams  for all five galaxies. We find a fairly wide range in cluster distributions from NGC~1433 with 72~\% of the points in the youngest box to NGC~3627 with only 12~\% in the youngest box. NGC~1566 is intermediate with 48~\%. Hence, NGC~3627 looks more similar to NGC~4449, and we conclude that the degree of similarity between Class 2 and 3 is largely dependent on the cluster formation history. Table~\ref{tab:table2} provides similar statistics for all four classes and all five program galaxies.   
We now perform four specific tests, as motivated above.

\subsection{Old Clusters ($>1$~Gyr)}
\label{sec:cc_old}

For this test we ask how many Class~1 objects are found in Box~1  for each classification approach.

Figure \ref{fig:cc_c1} shows that using the galaxy NGC 1566 as our test case, the U-B vs. V-I diagram, and $M_V < $ -7.8 mag,  PHANGS-HST finds the most Class~1 objects in Box~1 (old clusters), with 52, compared to 34, 46, and 45 for LEGUS, RESNET, VGG respectively. If the UV-B vs. V-I diagram is used instead, and a range from $m_V$ from 20 to 23 as in Figure \ref{fig:cc_uv}, the numbers are 43, 32, 33, 35 respectively. 
We conclude that all four methods are able to find old clusters at similar levels.

\begin{figure}
\begin{center}

\includegraphics[width =3.2in, angle= 0]{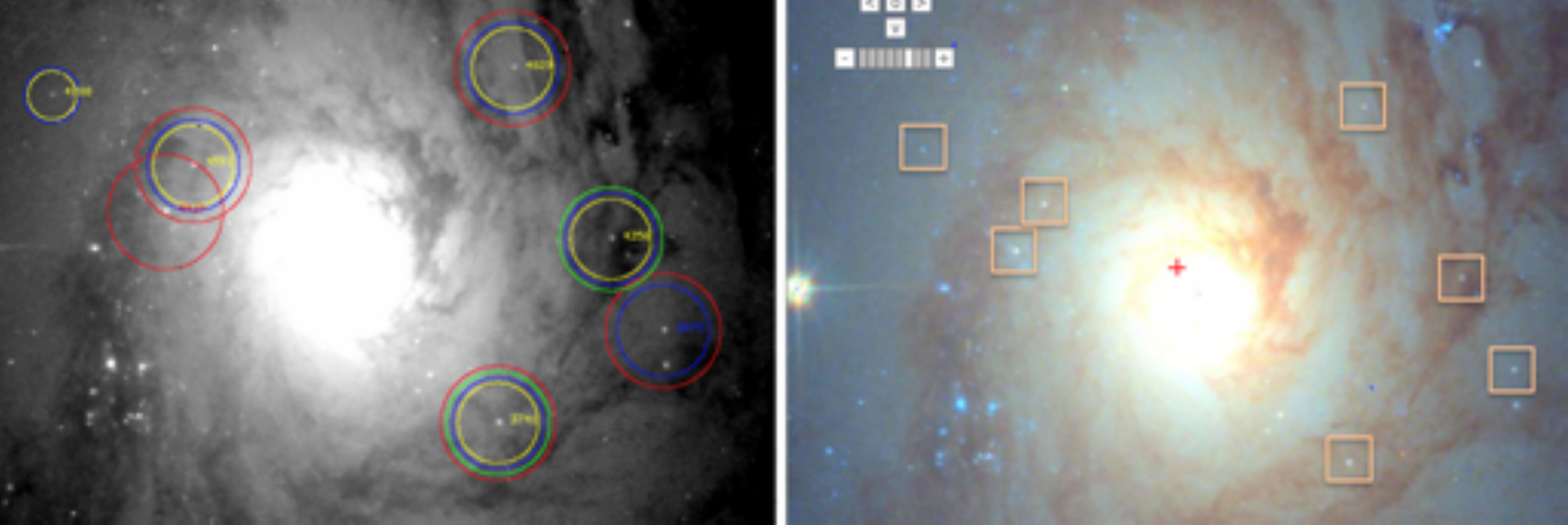}

\end{center}
\caption{Left panel shows the F555W image in the inner region of NGC 1566, with all the Class~1 clusters from Box~1 in this field of view 
identified. Red = PHANGS-HST, green = LEGUS, blue = RESNET, yellow = VGG. The right panel shows a  colour image using F814W, F555W, and F438W, with squares around the locations of the clusters.} 
\label{fig:img_1}
\end{figure}

Figure \ref{fig:img_1} shows an image of the  old cluster candidates (i.e., Class~1 objects in Box~1) in the nuclear region of NGC 1566.
This bulge region has the highest density of Class~1 objects from Box~1, as would be expected if they are old globular clusters.
Red is used for PHANGS-HST, green is used for LEGUS, blue is used for RESNET and yellow for VGG. 
The larger circles show sources in the range $m_V$ = 20 to 23.5 mag while the smaller circles represent sources in the range $m_V$ = 23.5 to 24.5 mag.
One of the objects (3741) is found by all four of the classifications. Three more of the objects are found by three of the four methods.

A visual examination of the right panel of Figure \ref{fig:img_1} 
shows 
the objects all have fairly uniform  yellowish colours and symmetric morphologies.  
Hence these are all good candidate old globular clusters, demonstrating that all four methods are able to identify this type of object fairly successfully.

We also  note from Figure \ref{fig:cc_c2} that Class~2 includes very few objects in Box~1 for all four classification approaches (i.e., 11, 11, 0, 5 for PHANGS-HST, LEGUS, RESNET, VGG respectively), and from Figure \ref{fig:cc_c3} we find even fewer (0, 6, 1, 1) in Box~1 for Class~3. 
Hence the rejection of the oldest clusters for  Classes~2 and 3 is  also good.

\subsection{Intermediate Age Clusters ($0.1-1$~Gyr)}
\label{sec:cc_interm}

For this test we ask how many Class~1 objects are found in Box~2 of  colour-colour space for all four approaches.

From Figure \ref{fig:cc_c1} we find that the  RESNET and VGG classification methods identify somewhat more symmetric, intermediate-age clusters (i.e., 157 and 148 objects in Box~2) than PHANGS-HST and LEGUS (132, and 135 in Box~2). 
We note, however, that most of this shortfall for Class~1 objects in Box~2 shows up as higher numbers in Class~2 for PHANGS-HST and LEGUS (i.e., Figure \ref{fig:cc_c2}, with 25 and 26 for PHANGS-HST and LEGUS, respectively, compared with 6 and 9 for RESNET, and VGG respectively).

\subsection{Young Clusters ($10-100$~Myr)}
\label{sec:cc_young}

A similar approach can be used for the objects in Box~3, but here we add the results from Classes~1, 2 and 3 since they are spread out more in the different classes.
We find  that LEGUS and VGG find the most young objects (238 and 210) in Box~3, while RESNET has 189.  The largest numbers of objects in Box~3 are found in Class~1, with a small spread ranging from 76 to 84 for all four methods (see Figure \ref{fig:cc_c1}).

\subsection{Very Young Clusters ($1-10$~Myr)}
\label{sec:cc_very_young}

For our Box~4 comparison we only use Class 2 clusters since there are only a handful of very young clusters in Class 1 for all four methods and the rejection of most of the Class 3 objects using the MCI method makes inclusion of PHANGS-HST  problematic.
The agreement between the four methods of classification is fairly good, with values of 110 and 127 for  PHANGS-HST and LEGUS and slightly lower values of 89 and 101  for RESNET and VGG. Hence,  the machine learning algorithms and human classifications are finding fairly similar objects for the very young objects in Box~4. 

Although the focus in this paper is on the clusters, and to a lesser degree on compact associations (see 
K. Larson et al., in preparation, for the preferred approach to working with associations in PHANGS-HST), we note from Figure \ref{fig:cc_c4} that the most populated box for Class 4 (artifacts - i.e., single stars, pairs, triplets etc) is Box 4, indicating that individual stars as well as young clusters can be found in Box 4. Indeed, the left half of Box 4 was called 'star/cluster space' in \citet{chandar10a}.

\begin{figure}
\begin{center}

\includegraphics[width = 3.3in, angle= 0]{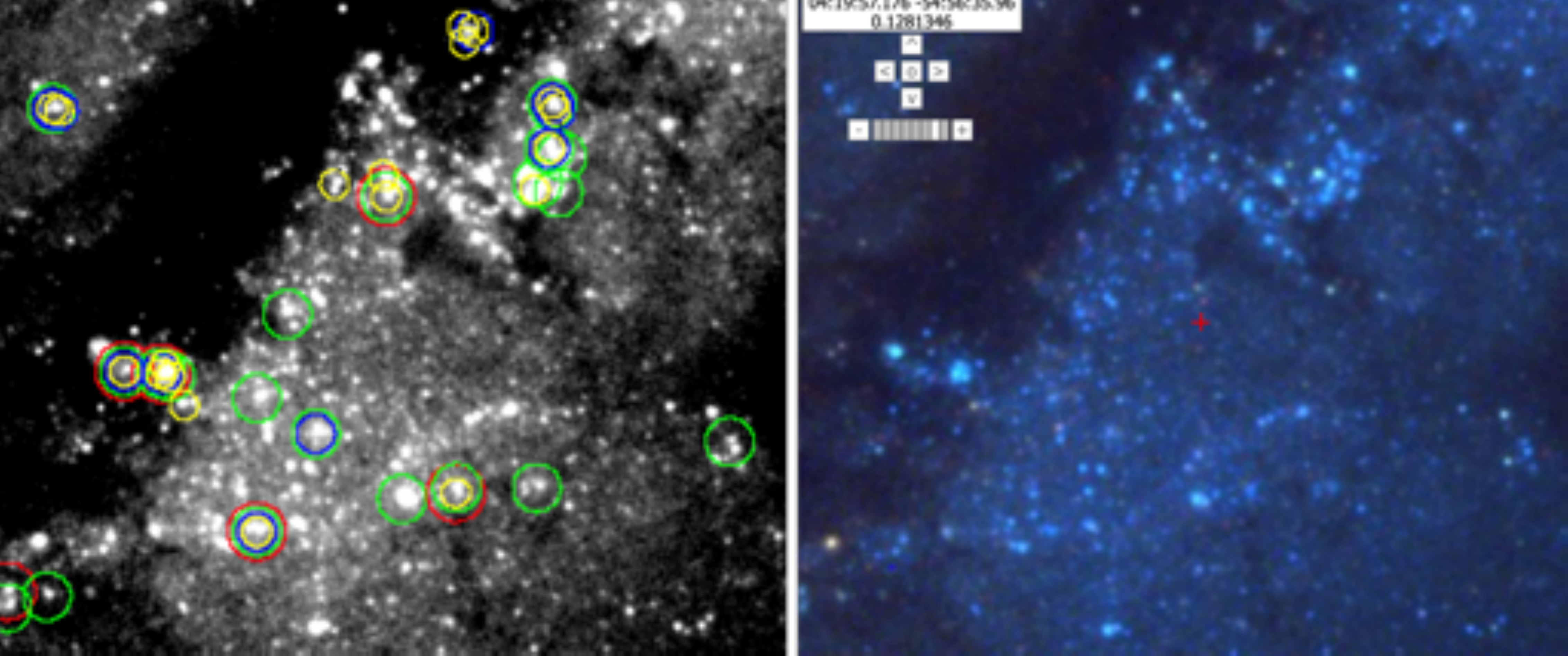}

\end{center}
\caption{Same as Figure \ref{fig:img_1}, but for Class~3 compact associations. Note that nearly all of the Class~3 objects, selected based on their morphology, are bluish in  colour as expected for very young objects. There are fewer PHANGS-HST (red) Class~3 objects, as discussed in the text.}
\label{fig:img_3}
\end{figure}

We conclude that all four of the classification methods result in fairly similar distributions in the U-B vs. V-I  colour-colour diagrams. However, there are also some important second-order differences that should be kept in mind, for example, RESNET and VGG find fewer young (Box~3) clusters in Class~2 than PHANGS-HST and LEGUS.

\subsection{Using Colour-Colour Statistics as a Figure of Merit}
\label{sec:cc_merit}

Similarities in the agreement fractions between the four classification methods, as discussed in Section \ref{sec:agree_class}, provides evidence that the machine and human classifications are similar in performance.  In this section we take a more quantitative look by comparing the colour-colour statistics for all four methods and all five galaxies. 
This ``figure-of-merit" provides both absolute  and relative comparisons of how well the different methods are able to identify different age clusters.  

Table \ref{tab:table2} includes the number of objects in each of the four boxes in the  colour-colour diagram for all five galaxies 
A limiting  magnitude of $M_v$ = -7.8 mag is used to normalize the sample. This is the magnitude cutoff of $m_v$ = 22.5 for the PHANGS-HST human selected sample for NGC 3627, which represents the brightest limiting magnitude  of the five galaxies. A value of $m_v$ = 22.5 mag provides a signal-to-noise of $\sim$100, and represents a very conservative limit (see S. Deger et al., in preparation) for classification.
A magnitude cutoff of $m_v$ = 23.5 corresponds to a range of $M_V$ = -6.5 to -7.8 mag for our galaxies, and represents a more standard limit. Pushing to $m_v$ = 24.5 mag, as we will in Section \ref{sec:cc_faint}, corresponds to a signal-to-noise around 10, and should be considered the practical limit.
When all four methods have 12 or more sources in Table \ref{tab:table2} we check to see how constant the values are, i.e., do the different methods find similar numbers of clusters in the same regions of the  colour-colour diagram?

To make the comparison we use the ratio  between the standard deviation of the four measurements and the square root of the mean number of objects in the box (i.e., the predicted standard deviation assuming Poisson statistics), which we define as the Quality Ratio = QR. These numbers are included at the end of the rows that qualify for analysis  in Table \ref{tab:table2}. Measurements with values 1.0 or lower (i.e., fairly constant values) are shown in blue, while measurements with values greater than 1.0 (i.e, discrepant values) are shown in red in Table \ref{tab:table2}. 

Class 4 objects (artifacts) are not included in the analysis since the goal of the PHANGS-HST and LEGUS studies was to eliminate as many artifacts as possible using constraints on either the concentration index (LEGUS) or on multiple concentration indices (PHANGS-HST). In addition, the machine learning algorithms used much broader constraints since the time required to make the classification is not a factor, unlike for the human classifications. These two effects  result in much larger numbers of Class 4 artifacts for RESNET and VGG, which skew the QR statistic.  This increase in the number of Class 4 objects for the machine learning algorithms can be seen in Figure \ref{fig:cc_c4}. The result is that all Class 4 QR values are very high, by design, but this is not relevant for a comparison of the goodness of the different classification methods. 
Similar statements could be made for the Class 3 objects, as shown in Figure \ref{fig:cc_c3}, since the PHANGS-HST candidate selection is designed to eliminated as many of these objects as possible.
However, since the discrepancies are smaller in most cases, and there is more interest in Class 3 (compact associations) than Class 4 (artifacts) objects, they have been included in the analysis for the purpose of comparison. We find, as expected, that most of the Class 3 objects have QR values greater than 1.0 in Table \ref{tab:table2}.

\begin{figure}
\begin{center}
\includegraphics[width = 3.3in, angle= 0]{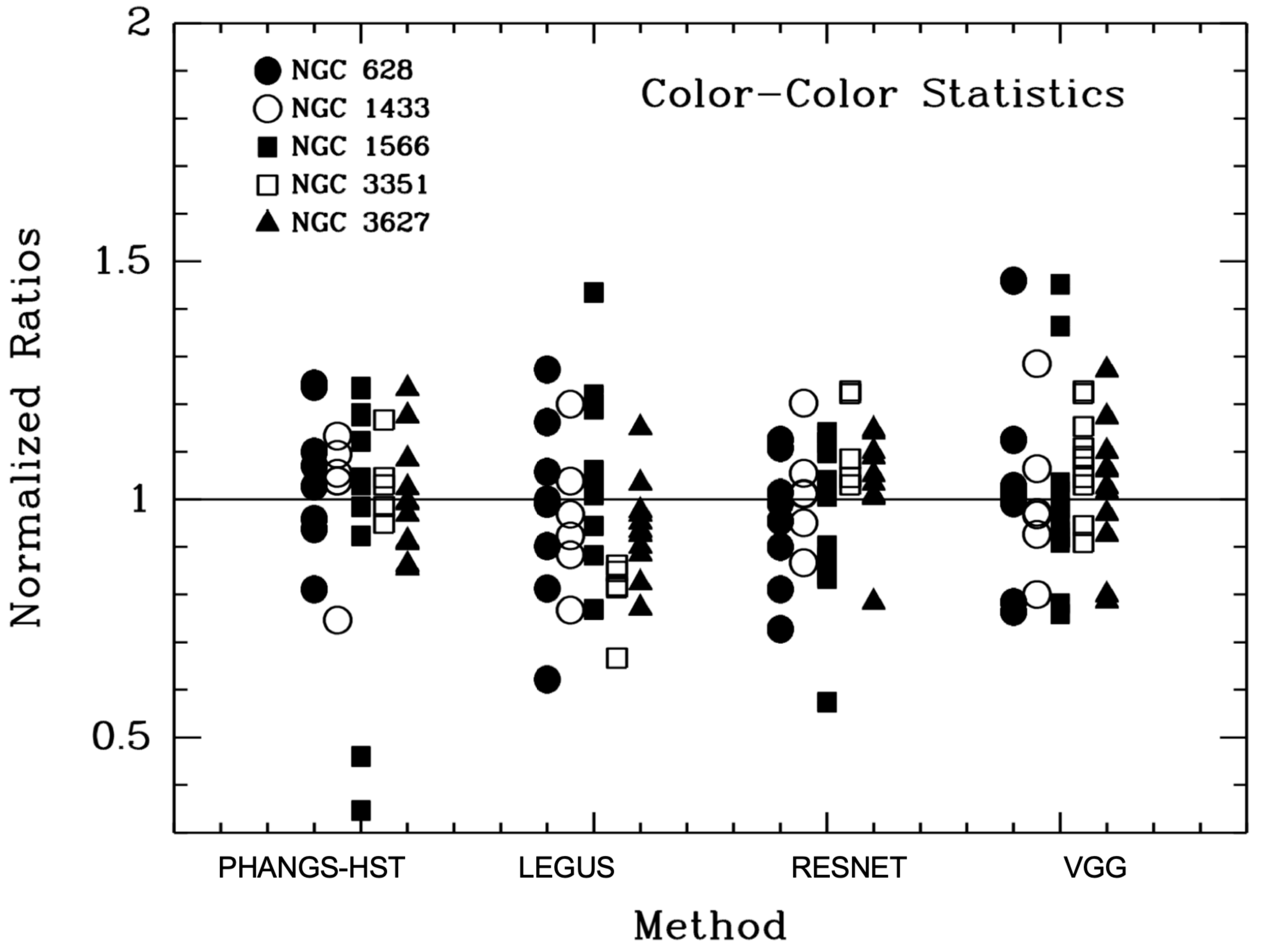}

\end{center}
\caption{Figure of merit measurements using the number of objects in  colour-colour diagram boxes from Table 2. Only cases where all four methods have at least 12 objects within the box are included. Values are normalized by the mean for the four methods. Class~4 comparisons are not included.}
\label{fig:cc_stats}
\end{figure}

The results of the figure of merit analysis are shown in Figure \ref{fig:cc_stats}. We find that all four methods are successful in finding similar numbers of objects based on the colour-colour statistics, with mean values and standard deviation values for PHANGS-HST = 1.01 $\pm$ 0.18, LEGUS = 0.97 $\pm$ 0.16, RESNET = 1.00 $\pm$ 0.14 and VGG = 1.02 $\pm$ 0.17. Similarly, values for the different galaxies are roughly the same, with no clear outliers in the mean.

Twenty-eight rows in Table \ref{tab:table2}  have values of QR = 1.0 or lower   (i.e., good agreement, blue) while fifteen rows have values greater than QR = 1 (i.e., poor agreement, red). This shows that while the four methods find similar objects in the majority of cases, there are also a number of cases where systematic differences are present. A close look at Table \ref{tab:table2} 
shows that the primary causes for large values of QR are either smaller number of Class 3 associations, as designed into the PHANGS-HST selection criteria, or the smaller number of young (Box 3 and Box 4) Class 2 objects found by RESNET and VGG, as shown in  Figure \ref{fig:cc_c2} and discussed in Section \ref{sec:cc}.

\subsection{Results for Faint Objects using Machine Learning Classifications}
\label{sec:cc_faint}

An important advantage of using machine learning classifications is the opportunity to provide cluster  catalogues that include fainter objects, since classifying large numbers of sources is no longer prohibitive. Below we investigate how accurately fainter clusters can be classified by comparing the numbers of sources found in the different regions of  colour-colour space with their brighter counterparts.

\begin{figure}
\begin{center}
\includegraphics[width =3.6in, angle= 0]{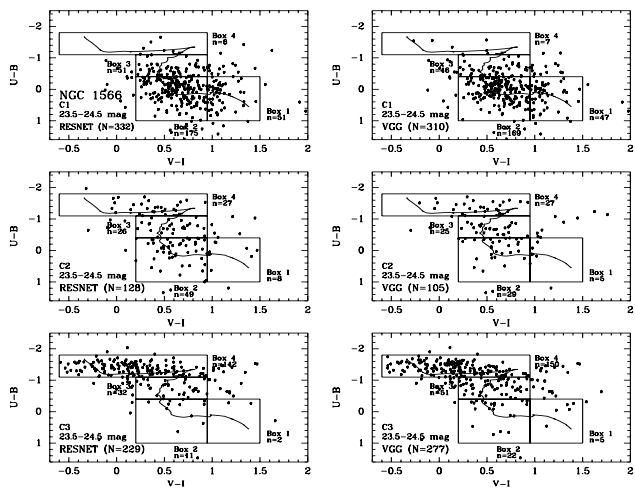}
\end{center}
\caption{U-B vs. V-I  colour-colour diagrams for Classes~1 (top), 2 (middle), and 3 (bottom) objects for the RESNET (left) and VGG (right) classifications. Only faint clusters in the range $m_V$ = 23.5 to 24.5 mag are included, i.e., fainter than for most of the PHANGS-HST and LEGUS
human classifications. Four boxes representing different ages are included, along with the number of objects in each box. While these figures are similar to Figures \ref{fig:cc_c1}, \ref{fig:cc_c2}, and \ref{fig:cc_c3} for brighter clusters in some regards (e.g., the ability to find old clusters in Box~1), they also show important difference, such as fewer numbers of Class~2 and Class~3 clusters. 
}
\label{fig:cc_faint}
\end{figure}

Figure \ref{fig:cc_faint} shows the  colour-colour plots for RESNET and VGG classifications of Class~1, 2, and 3, with $m_V$ magnitudes between 23.5 and 24.5 mag,
reaching roughly one magnitude fainter than shown in Figures \ref{fig:cc_c1}, \ref{fig:cc_c2}, \ref{fig:cc_c3}, and \ref{fig:cc_c4}  (i.e., the $M_V < -7.8$ mag criteria is equivalent to $m_V$ = 23.44 mag for NGC 1566). 
The magnitude range was chosen to insure that the number of Class~1 and 2 objects combined is roughly the same as in Figure \ref{fig:cc_c1}, so that direct comparisons can be made. 

The results for the fainter sample look similar to the brighter sample for Class~1. However, for Classes 2 and 3 there are some important differences that should be kept in mind. 
First, the fainter sample has a larger scatter in measured  colours.
This is
due to a combination of larger photometric errors, an increase in stochasticity (especially for Class~2 where individual stars are more prevalent), and  a higher degree of classification errors.

In Figure \ref{fig:cc_faint}, the number of Class~1 sources which fall in Box~1 for RESNET and VGG are roughly the same as those in Figure \ref{fig:cc_c1} for RESNET and VGG (i.e., 51 and 47 for the faint clusters compared to 46 and 45 for the brighter clusters). A human examination of objects that were classified by both RESNET and VGG as Class~1 shows that nearly all ($\approx$ 90 \%) of these objects would also be classified as Class~1 in PHANGS-HST, with only three of the objects classified in the pairs and triplets bins instead. If the criterium is relaxed to include {\em either} RESNET or VGG rather than {\em both}, the number drops only slightly ($\approx$ 85 \%).
We conclude that RESNET and VGG are able to correctly identify Class~1 Box~1 objects down to $m_V$ = 24.5 mag, allowing us to approximately double the number of old clusters by using the deeper RESNET and VGG classifications.

A similar result is found for Class~1 sources which fall in Box~2, 
where we find 175 and 169  classified by RESNET and VGG, respectively, in this fainter magnitude range (see Figure \ref{fig:cc_faint}), compared with 157 and 148 for their brighter counterparts (Figure \ref{fig:cc_c1}). 
The comparison does not hold up as well for the faint, 
young Class~1 objects in Box~3, with 51 (RESNET) and 46 (VGG) for the faint objects, compared with 82 and 76 for the brighter ones in Figure \ref{fig:cc_c1}.

Unlike the case for the brighter Box~1 and 2, Class~1 clusters in Figure \ref{fig:cc_faint} for NGC 1566, the distributions for the fainter 
 Class~2 and 3 clusters look  quite different than their brighter counterparts in Figure \ref{fig:cc_c2}, with larger numbers for the older clusters in Box~1 and 2, and smaller numbers for the younger clusters in Box~3 and 4. The most dramatic example of this difference is a decrease from 101 for the brighter clusters in Figure \ref{fig:cc_c2} for VGG  Box~4, to 27 for the fainter clusters in Figure \ref{fig:cc_faint} in the same box.
 This is consistent with what we found in Figure \ref{fig:agree_mag}, with a low agreement fraction for faint Class~3 objects (60 \%), and especially for faint Class~2 (40 \%) objects. 
The fact that RESNET and VGG find fewer faint sources with  colours consistent with young ages is the strongest systematic bias in the current study. 

To summarize this section, the distribution of classified objects in a U-B vs. V-I colour-colour diagram can be used as a figure-of-merit to test which of the classification methods provide the best results. We find that all four methods give comparable results for brighter sources, with the largest systematic differences being due to the presence of fewer Class 3 candidates, as designed into the PHANGS-HST selection criteria, and the smaller number of young Class 2 and 3 objects found by RESNET and VGG. An examination of the fainter objects (i.e., $m_V$ between 23.5 and 24.5 mag) show that while the Class 1 and Class 1 + Class 2 samples are quite robust, care must be taken when using Class 2 alone or when Class 3 is used. We remind people that the treatment of Class 3 (compact associations) has been largely superseded by the multi-scale approach described in K. Larson et al. (in preparation).

\section{Completeness Estimates}
\label{sec:complete}

Attempts to estimate completeness in cluster  catalogues generally follow the approach taken by stellar  catalogues, i.e., adding artificial objects of different magnitudes to the image and then finding what fraction can be recovered. Unfortunately, the situation is more difficult for clusters, making the resulting estimates more uncertain. 
The fundamental problem is that in the stellar case, there is a single point spread function for all objects and these are generally just 
added to an image with little or no background. For clusters, the objects come in a variety of shapes (circular, elongated, wide range of asymmetries) and different sizes. 

In addition, they are often embedded in  variable  and high background regions from the underlying galaxy, or in  crowded regions since stars tend to be born in clusters and associations. All of these issues make the task of estimating completeness for clusters a difficult one. 

In D. Thilker et al. (in preparation)
we follow this standard approach to estimate completeness to the degree possible by adding simulated clusters to images and determining how many we recover.  We also compare results from different studies (e.g., PHANGS-HST and LEGUS) to estimate relative completeness levels based on a comparison of the intersection and the union of the two studies. 
This typically results in an agreement in the number of  objects  at about the 70 \% level down to $m_v$ = 24.0 for the various galaxies.

Below we approach the question from two different directions to provide a sanity check on aspects of completeness that might fall outside the standard practice of adding objects to an image.

\subsection{Comparisons with a Human-Selected Catalogue}
\label{sec:comp_human}

In this subsection we
compare our classifications with a cluster  catalogue that is completely manually selected (i.e., without using a candidate list to start with), in order to assess how many clusters may be missing from the normal candidate selection process described in 
D. Thilker et al. (in preparation),
and to better understand if the different methods are identifying similar objects.
A similar study was done in M83 as reported in \citet{chandar10a}. Agreement between human-selected and hybrid methods (automatically selected followed by human classifications)  
was found to be $\approx$ 60 \% in that study. This is similar to the 70 \% estimate mentioned above when comparing the intersection and the union for the current study. 

\begin{figure}
\begin{center}

\includegraphics[width =3.2in, angle= 0]{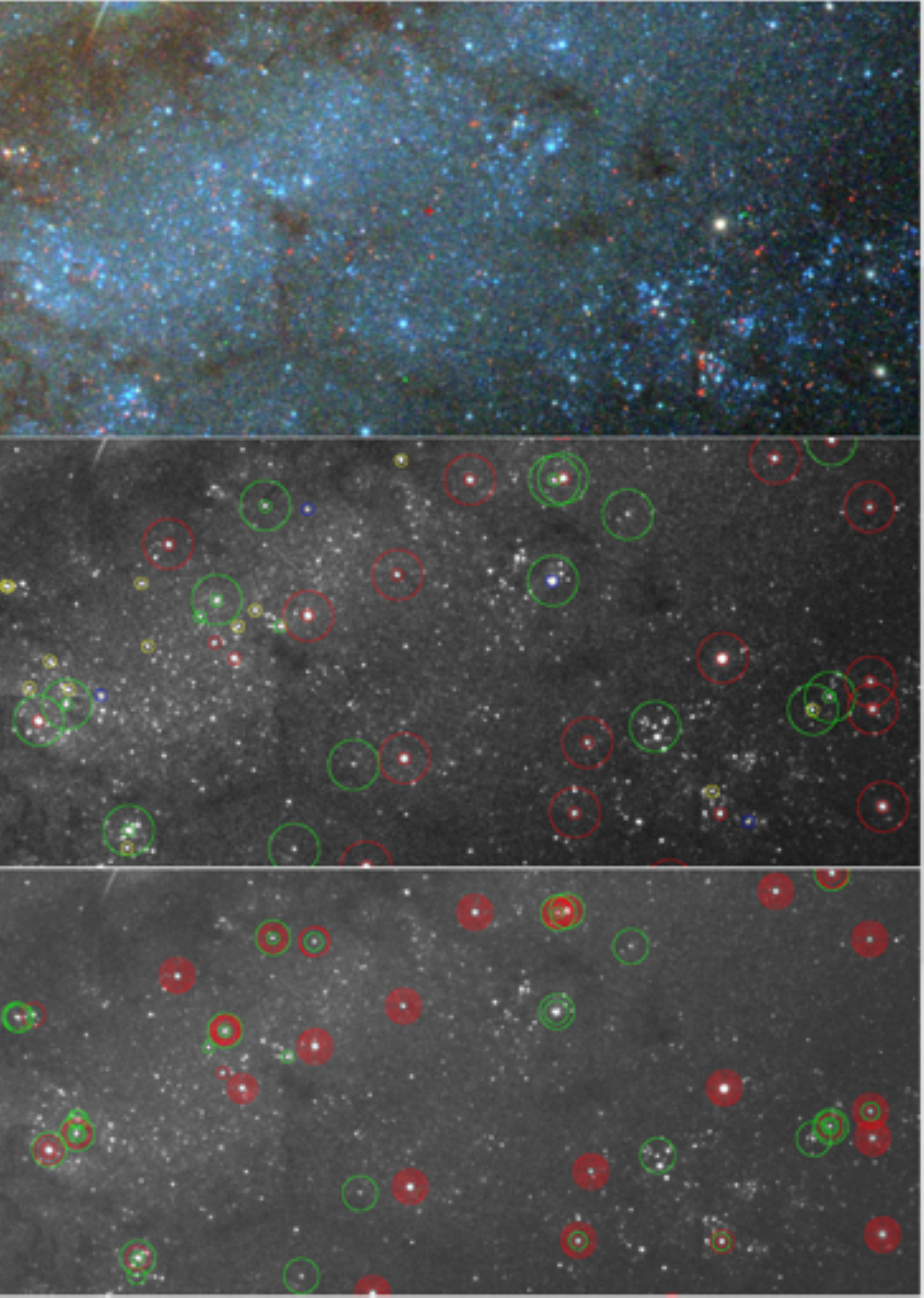}

\end{center}
\caption{A search for ``missing'' clusters and a comparison of the classifications for different methods for a  field in NGC 628. The top panel shows a  colour image from the HLA interactive display tool - 
\url{https://archive.stsci.edu/hlsp/phangs-hst}. We suggest that readers electronically enlarge the image; pick out objects they think are clusters, and then compare with the panels below. The middle panel shows Class~1 (red) and Class~2 (green) clusters from the  human-selected cluster classification (large circles) and the PHANGS-HST  catalogues (small circles). Class~3 (blue - compact associations) and Class~4 (yellow - artifacts) are also included for PHANGS-HST. The bottom panel uses the same  colour scheme but includes all five classifications from PHANGS-HST (smallest circles), to LEGUS, to RESNET, to VGG, to the human-selected classifications (largest circles).
}
\label{fig:img_compl}
\end{figure}

As an illustrative example, we perform this test in a typical field in NGC 628, as shown in Figure \ref{fig:img_compl}. 
No candidate clusters are identified in the top panel to avoid guiding the eye. We suggest that readers electronically enlarge the image and study this field in detail, picking out objects they identify as clusters, then compare their selection with the middle and bottom panels to see if they are included in the various  catalogues. The middle panel shows Class~1 (red) and Class~2 (green) clusters from the human-selected cluster classification (large circles) and the PHANGS-HST  catalogues (small circles). Class~3 (blue) and Class~4 (yellow) are also included for PHANGS-HST. The bottom plot uses the same  colour scheme but includes all five classifications from PHANGS-HST (smallest circles), to LEGUS, to RESNET, to VGG, to the human-selected classifications (largest circles).

The middle panel of Figure \ref{fig:img_compl} shows good general agreement between the human-selected and PHANGS-HST classifications, with 14 of 14 exact matches for Class~1 (red) and 6 of 11 exact matches for Class~2 (green). For Class~2, three of the non--matches were classified as Class~1, one was
classified as Class~3 (compact association), and one was classified as Class~4.

These numbers are  similar to those found in Figures \ref{fig:agree_class} and \ref{fig:agree_mag}, and in \citet{wei20}, although the 14 for 14 Class~1 matches is noteworthy and probably reflects the fact that 
the field is  not as  crowded as some other regions of this and other galaxies in our sample. 

Next, we note that there are four (all Class~2) of 29 clusters in the human-selected  catalogues with no counterpart in PHANGS-HST (i.e., 86 \% match) and conversely, five of 28 (i.e., 82 \% match) of the PHANGS-HST (Class 1 + 2) clusters with no counterparts in the human-selected  catalogue. These numbers are higher (better) than the roughly 60 \% number found in \citet{chandar10b}, probably because we are working at a brighter magnitude cutoff ($m_V$ = 23.0) and only including Class 1 and Class 2 clusters in the comparison. Hence we conclude that the number of ``missing'' Class~1 and Class~2 clusters due to different selection of the original candidates is relatively low, approximately in the 15 \% range down to
$m_V$ = 23 mag. 

Most of the  human-selected  clusters with no counterparts in Figure \ref{fig:img_compl} are near the faint end  of the sample, and  several are just outside the MCI polygon used for selection for the PHANGS-HST sample (see D. Thilker et al., in preparation).

 Hence the mismatch is not due to radical differences in morphology (e.g., very diffuse clusters) but to normal uncertainties such as visually estimating the appropriate brightness limit for the magnitude cutoff, and estimating whether the profile of a faint object can be distinguished from a star.

In the bottom panel of Figure \ref{fig:img_compl} we show sources classified as Classes~1 and 2 from all five methods. It is reassuring to find 12 clusters with Class~1 classifications in all five.  However, we also note that there are no Class 2 objects with this level of agreement. This reflects the lower agreement fractions in Figures \ref{fig:agree_class} and \ref{fig:agree_mag} for Class~2. Following this line of inquiry we find that there are eight objects which agree in three or four out of the five classification methods. At the other end, we find 12 objects that are only identified by one method, and two objects identified by two classification methods. 

A few other trends that can be established by a careful examination of Figure \ref{fig:img_compl} are: 1) for the objects with four or five classifications in common, the machine learning programs (RESNET and VGG) find fewer Class~2 objects than the human-selected classifications (this has already been seen in Figure \ref{fig:cc_c2} and Figure \ref{fig:cc_faint}),
2) over 50~\% of the artifacts in the PHANGS-HST  catalogue (the small yellow circles in the middle panel) 
are pairs and triplets, and 3) these artifacts are generally found in the more crowded regions of the image, as might be expected.

While the conclusions we can draw based on one illustrative example are limited, overall we find the agreement fairly  reassuring and supportive of the general conclusion that the machine learning classifications (RESNET and VGG) are competitive with the human classifications. In addition, we found no clear cases of large diffuse clusters  that are missing in this field.

\subsection{Estimating Completeness for Objects Brighter than $M_V$ of $-10$}
\label{sec:comp_hd}

Another method of estimating completeness for a subset of the data is to determine how many of the objects with magnitudes brighter than that expected for the brightest star (i.e., $M_v$ = -10 mag, the Humphrey-Davidson (H-D)  limit; \citealt{humphreys79}) are classified as clusters. This approach has two potential issues. The first is that  some of these bright objects are foreground stars. In principle, these can be identified using parallax measurements from GAIA, an approach being investigated in 
D. Thilker et al. (in preparation).
The second potential problem is that some of the brightest, youngest clusters are likely to be ultracompact (e.g., \citealt{smith20}), and hence may be difficult to distinguish from individual stars, especially in more distant galaxies. Incompleteness due to limitations of spatial resolution is also relevant for the smaller clusters in general, and is a major component of the completeness testing being performed in 
D. Thilker et al., (in preparation).

We use NGC~628, the closest of our five galaxies at 9.8~Mpc, to make a first check of completeness at the bright end.
Nineteen objects brighter than $M_V = $-10 mag are found, with four classified as saturated stars by PHANGS-HST. On inspection these are all obvious stars with diffraction spikes and Airy rings. 
A fifth object was classified as the nucleus of NGC~628 (Class~4.6).  Three additional sources were  classified as stars (Class~4.1). Of these, two are clearly stars since they show Airy rings, while the third is potentially an ultra-compact cluster, with FWHM = 2.2 pix.
However, it remains possible that this source is a foreground star, and GAIA (DR2) measurements are inconclusive because they have a large uncertainty.
One object classified as Class 4.10 (= too faint to tell) is likely to be a cluster based on its location in an intense star-forming region, even though its visual appearance and FWHM do not clearly distinguish it from a point source.  This object is the only clear case of a bright cluster that is missing from the PHANGS-HST NGC~628  catalogue.
Hence, 10 of 11 objects (or 10 out of 12 if the potential ultra-compact cluster is included) appear to have been correctly classified as clusters. The completeness for the brightest Class~1 and Class~2 clusters therefore appears to be in the 80 to 90 \% range for NGC~628, based on the PHANGS-HST  catalogue.

The classification of H-D objects is slightly worse when the LEGUS,  RESNET, and VGG classifiers are used, with eight, seven, and seven of the 10 clusters, respectively, identified as Class 1 or 2 objects. 
The most common misidentification are Class~4 sources (i.e., objects interpreted as a single star) as might be expected.
The object identified as the nucleus by PHANGS-HST (i.e., Class~4.6), is classified as Class~4 by LEGUS, but as  Class~1 by both RESNET and VGG. Hence, researchers should be careful to check the classification of the brightest objects if their study is sensitive to them. 
All four methods classified the foreground stars correctly. 

\subsection{Issues Related to Completeness and Systematic Differences in Classification}

\subsubsection{Double Counting}
\label{sec:comp_double}

The Dolphot package used in PHANGS-HST for both stellar and cluster detection 
was designed for crowded stellar fields. It uses an iterative approach, finding peaks, fitting the PSF and subtracting it, and then refitting to see if new peaks can be detected. While this works well if all the objects are stellar, in fields where both stars and clusters are present the software sometimes detects additional, false peaks after it subtracts out slightly resolved clusters. 

These false peaks are removed during the human classifications and designated Class 4.9 (= redundant - see Table \ref{tab:table3}). Operationally this is performed by classifying the brighter object and then giving any candidate within a radius of five  pixels a value of Class~4.9. This procedure also avoids double (or more) counting in Class 3 compact associations. LEGUS does not include this redundancy check, hence the much larger number of redundant (multiply detected) Class~3 objects in Figure \ref{fig:img_3}.

If uncorrected, the number of multiple detections of Class~3 objects for the RESNET and VGG classifications would be even higher than for LEGUS.
For this reason, redundant objects are removed from the RESNET and VGG machine learning classifications in a post-processing step of the pipeline (see  
D. Thilker et al., in preparation)
using the same algorithm as employed for the human check. This results in numbers which are roughly 50 lower than for LEGUS  in Figure \ref{fig:cc_c3}. The numbers are still considerably higher than for PHANG-HST due to the removal of pairs and triplets as artifacts (i.e., Classes~4.2 and 4.3), as discussed in Section \ref{sec:cc}.

\subsubsection{Trends in Classification Based on Confusion Matrices}
\label{sec:comp_matrix}

\bigskip

A standard tool used in many machine learning studies is the ``confusion matrix''. \footnote{Our procedure for making confusion matrices is non-standard, due to the fact that we do not define one study as ``ground truth'' but instead average the results for the two studies being compared, as discussed in Section \ref{sec:compare}. This works fine for the diagonal, but for the off-diagonal values it requires a reflection across the diagonal to identify the appropriate box to average. } This graphic provides a concise  way to visualize how often the classifications are in agreement, as well as insight into the most common types of systematic differences.

\begin{figure}
\begin{center}
\includegraphics[width = 3.3in, angle= 0]{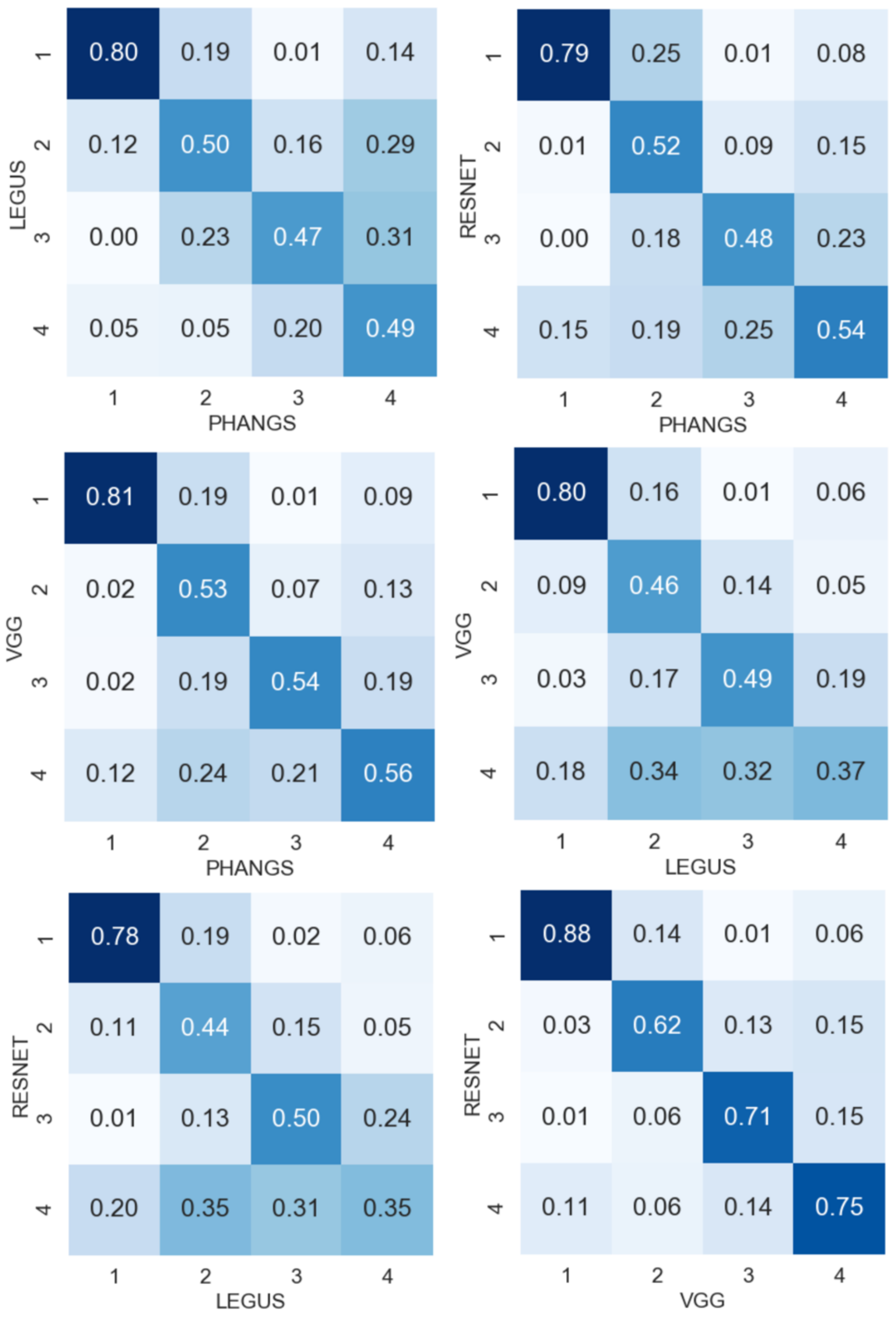}
\end{center}
\caption{The confusion matrices for all six comparisons between PHANGS-HST, LEGUS, RESNET and VGG for  NGC 1566. 
}
\label{fig:confusion_matrix_6plot}
\end{figure}

Figure \ref{fig:confusion_matrix_6plot} shows the confusion matrices for the six comparisons between different methods for the  bright ($m_V < 23$~mag) objects in  NGC 1566.
Ideally, the darkest  colour (highest percentage  of matches) would be found along the diagonal. In all six cases we find the highest matching fraction is for Class~1 objects (approximately 80 \%), as we also did in Figure \ref{fig:agree_class} and Table \ref{tab:table2}.

The off-diagonal values allow us to see where the most common systematic differences in classifications occur. For example, we note that a common difference is for  Class 4~objects from PHANGS-HST to be classified as Class~2  by LEGUS (i.e., 29 \% of the time according to the top left panel in Figure \ref{fig:confusion_matrix_6plot}). This is primarily due to the difference in how pairs and triplets are classified by the two studies, as shown in Figure \ref{fig:cc_c3} and discussed in Sections \ref{sec:procedure} and \ref{sec:cc}.
The converse of this (i.e. a PHANGS-HST Class~2 object being classified as a LEGUS Class~4 object) is very rare, only 5 \% of the time according to the upper left matrix in Figure \ref{fig:confusion_matrix_6plot}. This is the largest systematic difference between the two studies. 

Is the same trend seen between LEGUS and the two machine learning methods? Figure \ref{fig:confusion_matrix_6plot} shows a similar strong trend does exist for LEGUS to classify objects as Class 2 that RESNET and VGG classify as Class 4, (i.e., 35 \% and 34 \%  values vs. 5 \% and 5 \%) in the two  panels with LEGUS on the X-axis. Hence this tendency is not unique to the PHANGS-HST vs. LEGUS comparison.

Also of interest are cases where both an off-diagonal value and its converse are high, indicating a  large random uncertainty (i.e., hard to classify) rather than a systematic difference. The best example is that all the comparisons between Class 3  and Class 4 are between 14 \% and 32 \%, even for RESNET vs. VGG.

On the other end of the scale, Class~1 objects are almost never mistaken for  Class~3 objects in any of the comparisons (i.e., between 0 and  3 \% in all 12 relevant comparisons in Figure \ref{fig:confusion_matrix_6plot}).

An interesting comparison involving the machine learning  classifications is a strong tendency for both RESNET and VGG to classify objects as Class 1 that PHANGS-HST  classifies as Class 2 (i.e.,  19 \% and 25 \% of the time. The converse is only found 1 \% and 2 \% of the time. This same effect was found in Figure \ref{fig:img_compl}. 
A comparison between the two machine learning algorithms alone (bottom-right panel) shows that RESNET has a stronger tendency  to find Class 1 rather than Class 2 objects (i.e., 14 \% compared to 3 \%) when compared to VGG. We note that using a sample including both Class~1 + 2  (i.e., the ``standard" sample) will eliminate this problem.  Hence, while the largest systematic differences are caused by different definitions (e.g., whether pairs and triplets belong in Class 2 or Class 4 - differences of $\sim$25 \%), we also find that the two machine learning classifications can have sizeable differences (up to $\sim$ 10 \%), even though they used the same training sample.

Only NGC 1566 has been used as an illustrative example above. However, we find very similar trends in the confusion matrices for all five galaxies.

\subsubsection{Systematic differences between PHANGS-HST and LEGUS}
\label{sec:phangs_vs_legus}

The primary difference between PHANGS-HST and LEGUS classifications appears to be the inclusion of more pairs by LEGUS (e.g., Sections \ref{sec:comp_human}). 
Here, we make a more systematic comparison by examining the number of objects classified as pairs by PHANGS-HST but identified as Classes~1, 2, or 3 by LEGUS, RESNET or VGG in NGC 1566.
For LEGUS classifications of PHANGS-HST pairs, we find 18 of 478 (4~\%)  identified as Class~1, 43 of 404  (11~\%) identified as Class~2, and 32 of 691 (5~\%) identified as Class~3 objects.
The percentages  are lower for RESNET (3~\%, 5~\% and 3~\%) and for VGG (3 ~\%, 4 \%, 2 \%).
While the percentages of these systematic differences in identifications arising from close pairs of stars are relatively small, we note that the most affected objects are once again Class~2 (asymmetric clusters).

Differences in how studies define a cluster or association are relevant to the discussion of completeness. A specific borderline class that might be considered are triplets. In PHANGS-HST these are considered Class 4 artifacts (as shown in Table \ref{tab:table3}), while LEGUS generally  includes triplets in Class 3. From the numbers in 
Table \ref{tab:table3} we see that including or excluding triplets changes the total number of clusters and compact associations by about 15 \%. Hence, while this is an important contribution to the overall estimate of completeness, it is not a dominant component. One of the reasons we have classified the triplets separately is to
allow users to include them in their definition of clusters and associations, if they choose.

To summarize this section, the fact that clusters come in many shapes and sizes, and that different studies use somewhat different definitions for clusters and associations, makes it challenging to estimate completeness. Completeness can be as high as 90 \% or better for Class~1 + 2 clusters brighter than $m_v$ = 23.5 mag found in uncrowded regions or those with low background (e.g., Section \ref{sec:comp_human}), and as low as 10 \%  for Class~3 sources when compared with stellar associations defined using a watershed approach 
(K. Larson et al., in preparation). Various sanity checks performed in this section suggest typical completeness  numbers in the 70 \% to 80 \% range when considering Class 1 + 2 clusters over the full ensemble of environments. A more detailed look at the question of completeness as a function of magnitude, crowding, background and other properties will be included in a future study
(D. Thilker et al., in preparation).

\section{Dependence of science results on cluster classification method}
\label{sec:sci}

How much do science results, such as the shape of the  cluster mass and age distributions, depend on the classes and classification algorithm used?
Some studies (e.g., \citealt{bastian12a}, \citealt{krumholz19}, \citealt{adamo21}) have concluded that source selection is the major factor which has led different groups to reach different conclusions.
Other studies (e.g., \citealt{chandar14}) find 
that the mass and age functions are, within the errors, similar when  catalogues using different selection criteria are used.

\begin{figure}
\begin{center}
\includegraphics[width = 3.3in, angle= 0]{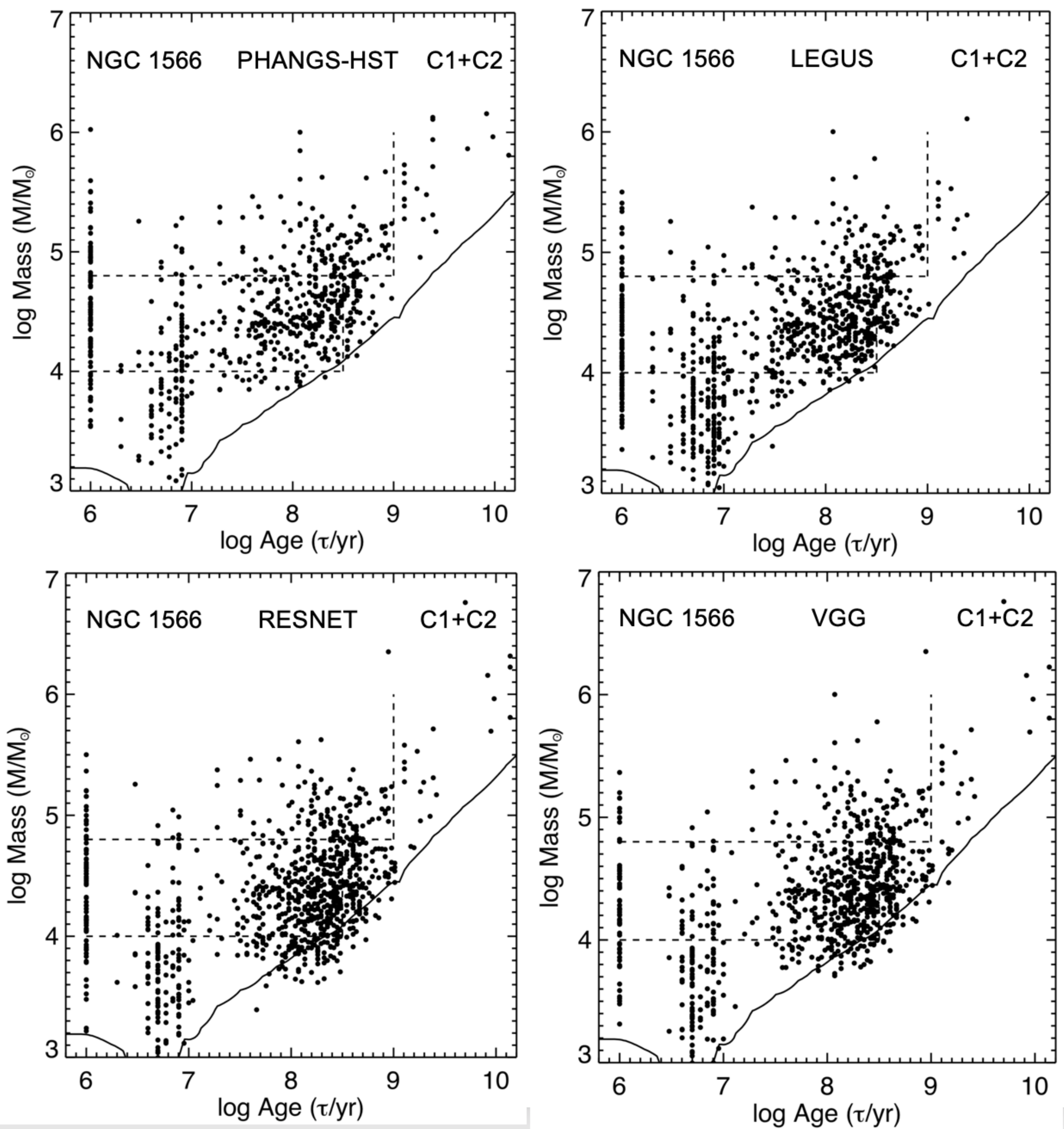}

\end{center}
\caption{Log mass vs. log age diagrams for NGC 1566 for the four classification methods using the Class 1 + 2 classifications. The dotted lines show the various cuts used in the determination of mass and age functions.  }
\label{fig:mass_age}
\end{figure}

In this section we address the question using the four main classification methods for a single galaxy, NGC 1566. 
A. Mok et al. (in preparation) will examine  a larger sample of PHANGS-HST galaxies in the future.
Figure \ref{fig:mass_age} shows the mass-age diagrams of Class~$1+2$ clusters in NGC 1566 separately based on all four classification methods. Ages are taken from the  non-stochastic $\chi^2$ SED fitting rather than the Bayesian analysis (see \citealt{turner21}). The mass-age distributions look similar, supporting the idea that the machine learning classifications provide results similar to those based on human classifications. There are fewer clusters in the PHANGS-HST and LEGUS samples since they have brighter magnitude cutoffs than the RESNET and VGG samples. 

\begin{figure}
\begin{center}
\includegraphics[width = 2.8in, angle= 0]{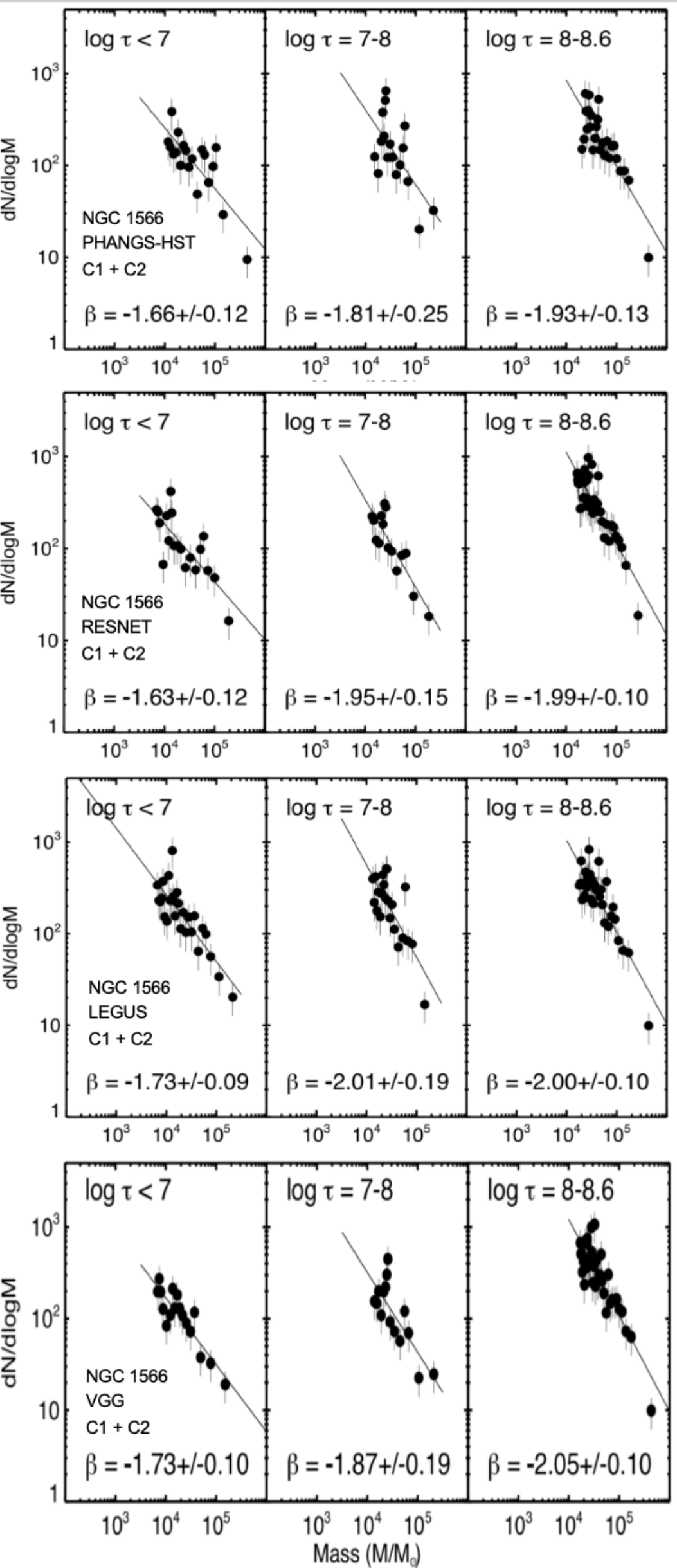}
\end{center}
\caption{Mass functions for NGC 1566 for each of the four classification methods using the Class~1 + 2 classifications. Fits are made for three age ranges for each method. The overall mean value is $\beta$ = 1.86 with an uncertainty in the mean of 0.04. The mean values for each method are the same within 2 sigma limits.  }
\label{fig:mass_func}
\end{figure}

Cluster mass functions can be described, to first order, by a simple power law, $dN/dM \propto M^{\beta}$.
Figure \ref{fig:mass_func} shows the Class $1+2$ cluster mass functions in the three indicated intervals of age, again for the four different classification methods. Here, the mass functions are plotted with an equal number of objects in each bin, and the power law index $\beta$ is determined from the best linear fit.
The best fit values of $\beta$ are compiled in
Table \ref{tab:table4}.
We find that the mean and standard deviation of all 12 determinations are $-1.86 \pm 0.04$.
The mean values of $\beta$, found by averaging the three different age ranges together for each classification method, 
are $-1.80$, $-1.91$, $-1.86$, $-1.88$ 
for PHANGS-HST, LEGUS, RESNET, and VGG,
respectively. All four are within the $2\sigma$ error estimates, and are consistent with being drawn from the same distribution. Repeating the exercise using 
a sample that includes only the Class~1 clusters 
yields values of $\beta=-1.90 \pm 0.03$ for the mean of all 12 estimates, and $-1.85$, $-1.86$, $-1.95$, and $-1.93$ for the four methods separately. These are again all within $2\sigma$ uncertainties, and indicates that {\em the mass function is insensitive to the specific classification algorithm and whether or not Class~$1+2$ or Class~1 clusters are used}.

\begin{figure}
\begin{center}
\includegraphics[width = 3.3in, angle= 0]{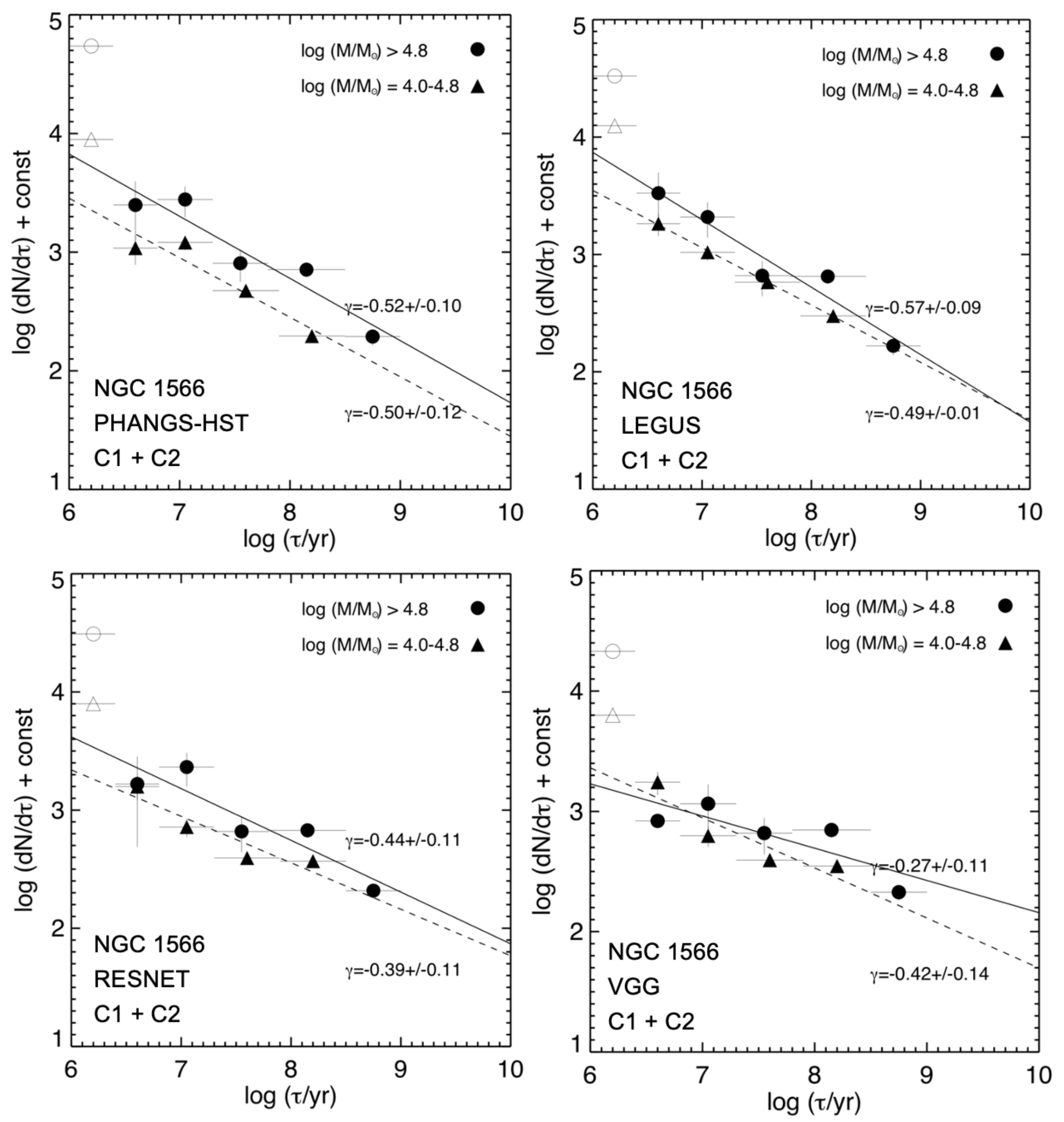}
\end{center}
\caption{Age distributions for NGC 1566 for each of the four classification methods using the Class~$1+2$ sample. Fits are made for two mass ranges for each method. The youngest points are shown as open symbols and are not included in the fits since they appear to be systematically high. 
The overall mean value is $\gamma$ = -0.45 with an uncertainty in the mean of 0.03. 
}
\label{fig:age_dis}
\end{figure}

\begin{figure}
\begin{center}
\includegraphics[width = 3.3in, angle= 0]{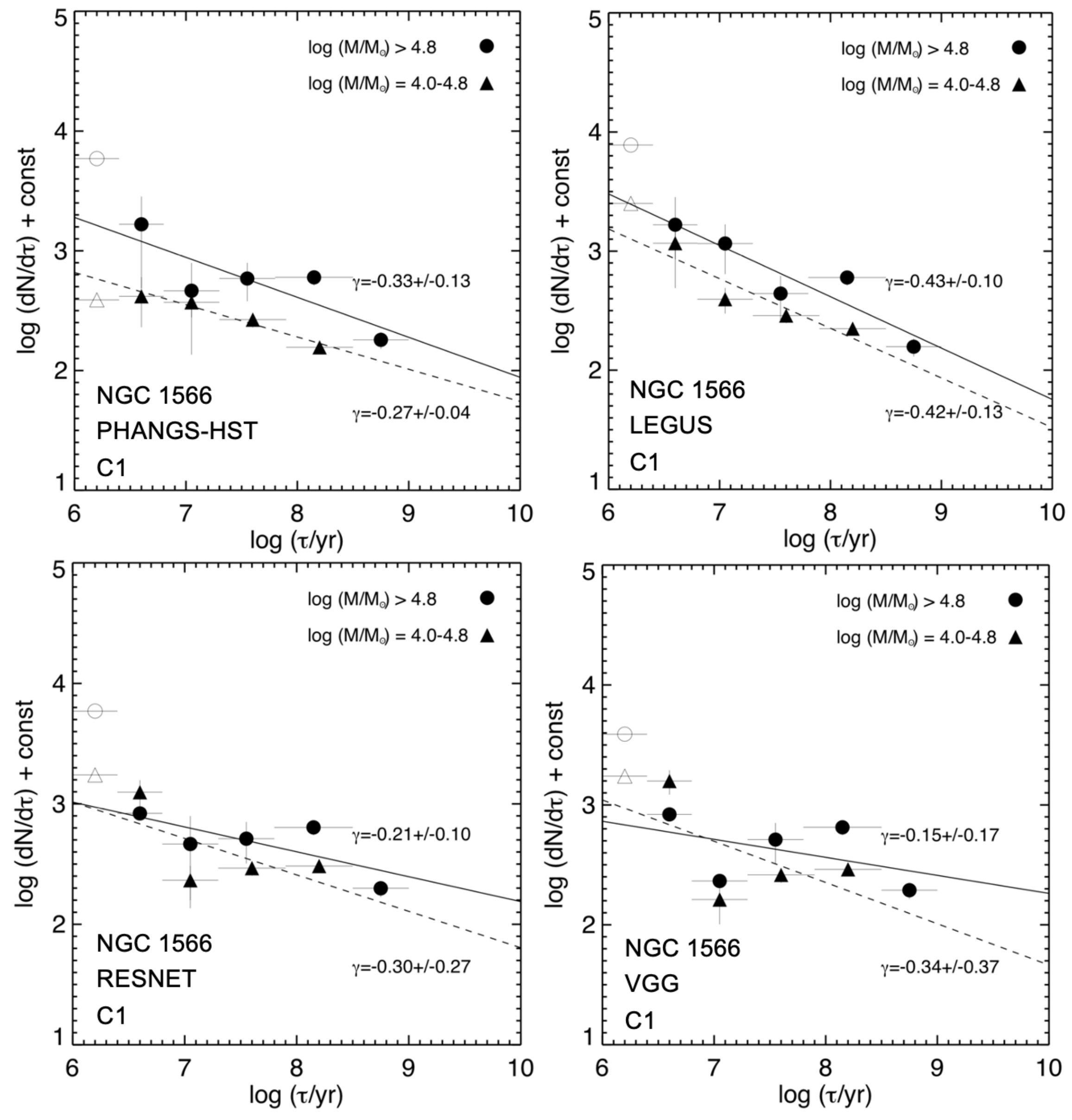}
\end{center}
\caption{Age distributions for NGC 1566 for each of the four classification methods using only the Class~1 sample. Fits are made for two mass ranges for each method. The youngest points are shown as open symbols and are not included in the fits since they appear to be systematically high. The overall mean value is $\gamma$ = -0.31 with an uncertainty in the mean of 0.03. 
}
\label{fig:age_dis_c1}
\end{figure}

Figure~\ref{fig:age_dis} shows the age distributions based on Class~$1+2$ clusters in NGC~1566, and Figure~\ref{fig:age_dis_c1} shows the results if only the Class~1 clusters are included.
The age distribution can be fit by a single power law, $dN/d\tau \propto \tau^{\gamma}$.  In these figures, the inner (youngest) data points, shown with open symbols, are excluded from the fits because they are systematically high compared with the rest of the distributions.
In Table~\ref{tab:table5}, we present fit results for the power-law index $\gamma$ both with and without the youngest data point included in the fit.

For the Class~$1+2$ sample in NGC~1566, the mean and standard deviation of all eight fits (excluding the youngest age bin) gives $\gamma= -0.45 \pm 0.03$.
The mean values of $\gamma$ (averaging the two mass ranges together) are 
$-0.51$, $-0.53$, $-0.42$, $-0.34$
for PHANGS-HST, LEGUS, RESNET, and VGG,
respectively. We note there is tentative evidence that the machine learning classifications give slightly flatter values for $\gamma$ at about the $3\sigma$ level. 
If we repeat this exercise but now {\em include} the youngest age bin, the results are: $\gamma= -0.68 \pm 0.03$ from all eight fits, and $-0.75$, $-0.74$, $-0.66$, and $-0.58$.  Including the data point representing the youngest age range in the fits therefore results in a power law index $\gamma$ for the age distribution which is steeper by $\approx-0.2$ than if the youngest point is excluded. 

If we repeat this exercise for the Class~1 sample alone (i.e., Figure \ref{fig:age_dis_c1}), excluding the youngest age bin from the fits, we find the mean from all eight fits $\gamma = -0.31 \pm 0.03$, and the mean values of $\gamma$ from the different methods (averaging the two mass ranges together) are  
$-0.43$ (PHANGS-HST), $-0.30$ (LEGUS), $-0.24$ (RESNET), and $-0.26$ (VGG).
Not surprisingly, the age distributions when only Class~1 sources are included are somewhat flatter, by $\approx0.1-0.2$, than those which include Class~$1+2$.

To summarize, the cluster mass functions are   similar, with a power-law index of $\beta=-1.9 \pm 0.1$, regardless of the age interval or specific classification method that is used, or whether Class~$1+2$ or Class~1 are included.  The age distribution is more sensitive to the specific selection method, age interval for the fit, and object class.  We find a range of values for the power-law index $\gamma$, with the steepest value $\gamma=-0.68 \pm 0.03$ found for Class~$1+2$ clusters fit over the full age range (log~$(\tau/\mbox{yr})=6.0-9.0$), and the shallowest $\gamma=-0.31 \pm 0.03$ found for Class~1 clusters where the youngest age data point is excluded (log~$(\tau/\mbox{yr})=6.5-9.0$). 
These small differences are similar to what was found in
 \citet{krumholz19}.
We remind the reader that these specific results are for the sample with ages taken from the non-stochastic $\chi^2$ SED fitting.
Similar, but slightly different results are found using the Bayesian analysis (see \citealt{turner21}).

\section{Future Work}
\label{sec:improve}

It is reassuring that the agreement fractions for RESNET and VGG classifications based on the LEGUS-BCW training sets (see \citealt{wei20})
are roughly as good as or slightly better than those from human classifications.  Based on the results presented in this work, particularly through the figures of merit, we conclude that both human and machine learning classifications do a fairly good job.
In the future, we plan to continue improving the quality of the machine learning classifications so they are eventually superior to the human classifications.

\subsection{Using Different Training Sets} 
\label{sec:improv_train}

As described in \citet{wei20}, besides the LEGUS-BCW training sets a second training set  called the  LEGUS 3-person-consensus classifications was also used. The results were similar, as reported in Figure 3 of \citet{wei20}, with the mean value of the four classes equal to 66 \% for  LEGUS-BCW galaxies and 64 \% for the LEGUS 3-person-consensus. Due to this similarity, our first test will be to combine the two training sets to determine if the resulting classifications are improved by having larger numbers in the training set.

One trade-off to be considered is whether it is better to have a more specific (i.e., training objects taken from the same galaxy) or the largest possible training set. 
To test this, one of our experiments will be to use only a set of the new PHANGS-HST human classifications rather than including the previous LEGUS related training sets. 

Another experiment will be to use the larger
set of artifacts than just Class 4 (e.g., single dominant stars, pairs, triplets, saturated stars, diffraction spikes, cosmic rays, background galaxies, etc. -- see Table \ref{tab:table3}). The hypothesis is that with the wide variety of types of artifacts all mixed  together, the  algorithm might not be able to train as optimally as it might with similar object types in each class.

\subsection{Quality of Training Samples}
\label{sec:impro_quality}

The quality of the training set also plays an important role in the classification accuracy of the algorithm. Some parameters that may limit the quality of the training set are crowding, magnitude, background, and distance. We plan to examine the role of each of these parameters in the future by using different subsets as training sets.

Another experiment is to identify a high quality training set by only including objects that have the same estimated classification in at least seven of the ten trials used by both the RESNET and VGG algorithms (see \citealt{wei20} and Appendix A).

In the future, and as outlined in \citet{wei20}, we plan to organize a cluster classification challenge using roughly a dozen different people to classify  clusters in a few different galaxies. The resulting
sample could be used as a training set to see if the classification agreement improves.

\subsection{Objective Approaches - Simulated and Colour-Colour Selected  Training Sets}
\label{sec:improv_obj}

In this paper we explore the development of automatically selected cluster  catalogues. However, it would not be fair to call these ``objectively selected''  catalogues since they are based on ``subjectively selected'' training sets. In this subsection we discuss two approaches to more objectively
determined  catalogues.

The first approach would be to use simulations, as described in 
D. Thilker et al. (in preparation)
as part of our future training and testing. We are currently using these simulations to test our completeness levels and will extend this approach to see how well the RESNET and VGG algorithms recover the input classifications.
A similar study is \citet{bia2019}, who used neural network classifcations based on simulations for resolved star clusters in M31.

The use of simulations has a number of pros and cons. On the positive side, it provides a better estimate of ``truth", since we know what kinds of clusters are inserted. It also provides a more rigorous method of tracking dependencies on crowding, background, luminosities and distance. On the negative side, the simulations do not capture the full range of morphologies, especially for Class 2 clusters that are by definition asymmetric and irregular. 

A related approach would be to use actual clusters scaled to different magnitudes.
These objects would be drawn from within the four boxes in the  colour-colour diagrams introduced in Section  \ref{sec:cc} to select the objects that go into the training sets, rather than using morphology. We would explore the resulting success fractions (i.e., how many of the objects  have the  colours appropriate for the box they were trained for). 
One potential short-coming of this approach is the effect of reddening, hence we will experiment with galaxies with both high and low overall reddening.

\section{Summary and Conclusions}
\label{sec:summary}

Our goal in this project is to develop automated and repeatable classification methods for making star cluster  catalogues that are at least as good as, 
and  in the future hopefully better than, human classifications. 
Detailed comparisons have been made between  catalogues of star clusters developed using human classifiers (i.e., PHANGS-HST and LEGUS) and convolutional neural network models (RESNET and VGG) trained using deep transfer learning techniques, as described in \citet{wei20}. In the current paper we focus on the results for five PHANGS-HST (and LEGUS) galaxies (NGC 628, NGC 1433, NGC 1566, NGC 3351, NGC 3627).  
The primary results are outlined below.

1. We describe in detail how human classifications are made for PHANGS-HST. This includes the automated identification of cluster candidates using the MCI approach described in 
D. Thilker et al. (in preparation), human inspection of both the V-band image and a  colour image, examination of surface brightness profiles and measurement of the FWHM,  examination of contours, and stretching of the contrast to see if the object grows like a star or a cluster.

2. We examine agreement fractions between pairs of methods for each of the classes. We find that results produced by the convolutional neural network models (RESNET and VGG), 
using models trained with the BCW-only classifications as described in \cite{wei20}, 
 are comparable in  quality to the PHANGS-HST and LEGUS human classifications with
 typical agreement for the four methods around 70 to 80  \% for Class~1 clusters (symmetric, centrally concentrated),  
 40 to 70 \% for Class~2 clusters (asymmetric, centrally concentrated), 50 to 70 \% for Class~3 (compact associations), 
 and 50 to 80 \% for Class~4 (artifacts such as stars and pairs of stars). Our focus is on the Class~1 and 2 clusters; work on the Class~3 associations has been largely superseded by the watershed approach developed in K. Larson et al. (in preparation). 

3. The dependence of agreement fractions on several properties, including magnitudes, crowding,  surface brightness background, and distance are explored. While the dependence on magnitude for Class~1 clusters is quite flat, the dependence of Class~2 is much steeper, with only 30 \% agreement found at magnitudes fainter than $m_V$ = 24 mag. Using a sample of  Class~1 + 2 clusters together alleviates much of this problem, resulting in agreement fractions essentially the same as Class~1 alone. Crowding results in  the strongest effect, showing relatively steep dependencies for all four classes.

4.  The distribution of data points in  colour-colour diagrams is used as a ``figure of merit" to test the absolute performances of the different methods, supporting the finding that the automated classifications are comparable in quality to human classifications for $M_V$ brighter than -7.8 mag. 
The agreement is somewhat poorer for fainter magnitudes,
especially for Class~2 and 3 were the machine learning methods find more old clusters, but fewer young objects than the human classifications.

5. Issues related to completeness are explored including comparisons with a completely human-selected  catalogue, identification of common trends in classification based on confusion matrices, and an examination of whether objects too bright to be individual stars (i.e., the Humphrey-Davidson limit) are being found and included in the cluster  catalogues. The most  important common trend is related to differences in definitions, with LEGUS including more pairs and triplets as Class 3 (compact associations) while PHANGS-HST classifies them as artifacts (Class 4).

6. Mass and age distributions are examined for NGC 1566 as a function of the four classification methods. We find essentially the same mass functions in all four cases, and relatively weak dependencies on the classification methods for the age distributions. 
Using the  Class~1  catalogue alone and dropping the youngest data point results in a slope in the age distribution which is a slightly shallower than using the Class~1 +2 sample or including the youngest data point.

Based on these results we conclude that the human classifications and the machine learning
classifications are of comparable quality, although there are various caveats to be aware of, as discussed in the text. We recommend that researchers experiment with both the human and machine learning cluster  catalogues, and use this experimentation to  provide a measure of how it affects their science results, if at all. 
Similarly, we recommend that  researchers experiment with samples using different combinations of classes, as in Section \ref{sec:sci}, although the Class~1 + 2 catalog  can be considered a standard in many cases. 

Finally, we include an appendix which provides  an overview on how to apply the \citet{wei20} neural network models to classify star cluster candidates in other galaxies.
If  comparable data sets to the PHANGS-HST or LEGUS datasets are available (i.e., five-band HST observations including the F555W band for galaxies within a similar distance range), the \citet{wei20} models can be used. If not, researchers can develop training sets from their own data and train the algorithms themselves. We have also discussed future work to improve the models, including re-training the RESNET and VGG neural networks with our new BCW human classifications for PHANGS-HST star clusters. We will make available all of our neural network models at the same website as discussed in Appendix A, along with Jupyter notebooks that illustrate their application.

\bigskip

Acknowledgements:
We thank the referee for a number of insightful comments that we feel have greatly improved the paper. 
This study is based on observations made with the NASA/ESA Hubble Space Telescope, obtained from the data archive at the Space Telescope Science Institute. STScI is operated by the Association of Universities for Research in Astronomy, Inc. under NASA contract NAS 5-26555.  Support for Program number 15654 was provided through a grant from the STScI under NASA contract NAS5-26555.
JMDK and MC gratefully acknowledge funding from the Deutsche Forschungsgemeinschaft (DFG, German Research Foundation) through an Emmy Noether Research Group (grant number KR4801/1-1) and the DFG Sachbeihilfe (grant number KR4801/2-1). JMDK gratefully acknowledges funding from the European Research Council (ERC) under the European Union's Horizon 2020 research and innovation programme via the ERC Starting Grant MUSTANG (grant agreement number 714907). TGW acknowledges funding from the European Research Council (ERC) under the European Union\'s Horizon 2020 research and innovation programme (grant agreement No. 694343).
EAH and WW gratefully acknowledge National Science 
Foundation (NSF) awards OAC-1931561 and OAC-1934757.
FB acknowledges funding from the European Research Council (ERC) under the European Unions Horizon 2020 research and innovation programme (grant agreement No.726384/Empire).

Data Availability:
The data underlying this article are available at the Mikulski Archive for Space Telescopes at \url{https://archive.stsci.edu/hst/search_retrieve.html} under proposal GO-15654.  High level science products associated with HST GO-15654 are provided at \url{https://archive.stsci.edu/hlsp/phangs-hst}.

\clearpage

\begin{table}
 \caption{Agreement Fractions for the Five Program Galaxies}
 \label{tab:table1}
 \begin{tabular}{lcccccc}
  \hline
  Classification & Class~1 & Class~2 & Class~3 & Class~4 & mean & Class~1+2\\
  Comparison \\
  \hline
  
  NGC 628 (9.8 Mpc) \\
  PHANGS vs. LEGUS & 0.81 & 0.54 & 0.58 & 0.43 &  0.59 & 0.85 \\
  PHANGS vs. RESNET & 0.79 & 0.40 & 0.60 & 0.56  & 0.59 & 0.84\\
  PHANGS vs. VGG & 0.82 & 0.49 & 0.58 & 0.57 & 0.61 & 0.84 \\
  LEGUS vs. RESNET & 0.78 & 0.38 & 0.49 & 0.54  & 0.55 & 0.81\\
  LEGUS vs. VGG & 0.78 & 0.35 & 0.46 & 0.52  & 0.53 & 0.81\\
  mean for above values & 0.80 (0.02)$^a$ & 0.43 (0.08) & 0.54 (0.06) & 0.52 (0.06) & 0.57 (0.04) & 0.83 (0.02) \\
  RESNET vs. VGG & 0.90 & 0.58 & 0.69 & 0.76 & 0.73 & 0.89 \\
  \bigskip \\

  NGC 1433 (18.6 Mpc) \\
  PHANGS vs. LEGUS & 0.74 & 0.54 & 0.54 & 0.59 & 0.60 & 0.80 \\
  PHANGS vs. RESNET & 0.76 & 0.56 & 0.64 & 0.71 & 0.67 & 0.76 \\
  PHANGS vs. VGG & 0.81 & 0.54 & 0.58 & 0.64 & 0.64 & 0.71 \\
  LEGUS vs. RESNET & 0.68 & 0.56 & 0.53 & 0.56 & 0.58 & 0.68 \\
  LEGUS vs. VGG & 0.68 & 0.56 & 0.53 & 0.56 & 0.58 & 0.68  \\
  mean for above values & 0.73 (0.06) & 0.55 (0.01) & 0.56 (0.05) & 0.62 (0.06) & 0.62 (0.04) & 0.74 (0.04) \\
  RESNET vs. VGG & 0.88 & 0.69 & 0.76 & 0.86 & 0.80 & 0.86 \\
  \bigskip \\
  
  NGC 1566 (17.7 Mpc) \\
  PHANGS vs. LEGUS & 0.80 & 0.50 & 0.47 & 0.49 & 0.56 & 0.82 \\
  PHANGS vs. RESNET & 0.79 & 0.52 & 0.49 & 0.66 & 0.62 & 0.82 \\
  PHANGS vs. VGG & 0.82 & 0.53 & 0.55 & 0.68 & 0.64 & 0.82 \\
  LEGUS vs. RESNET & 0.78 & 0.44 & 0.51 & 0.48 & 0.55 & 0.80 \\
  LEGUS vs. VGG & 0.80 & 0.47 & 0.50 & 0.50 & 0.57 & 0.80 \\
   mean for above values & 0.80 (.01) & 0.49 (.04) & .50 (0.03) & 0.56 (0.10) & 0.59 (0.04) & 0.81 (0.01) \\
  RESNET vs. VGG & 0.88 & 0.62 & 0.71 & 0.89 & 0.77 & 0.88 \\
  \bigskip \\

  NGC 3351$^{b,c}$ (10.0 Mpc)\\
  PHANGS vs. LEGUS & 0.82 & 0.53 & 0.46 & 0.73 & 0.64 & 0.81 \\
  PHANGS vs. RESNET & 0.81 & 0.54 & 0.49 & 0.74 & 0.64 & 0.81 \\
  PHANGS vs. VGG & 0.82 & 0.54 & 0.51 & 0.72 & 0.65 & 0.78 \\
  LEGUS vs. RESNET & 0.83 & 0.62 & 0.63 & 0.74 & 0.70 & 0.86 \\
  LEGUS vs. VGG & 0.85 & 0.60 & 0.56 & 0.69 & 0.67 & 0.81 \\
  mean for above values & 0.82 (0.01) & 0.57 (0.04) & 0.53 (0.07) & 0.72 (0.02) & 0.66 (0.03) & 0.82 (0.03) \\
  RESNET vs. VGG & 0.90 & 0.72 & 0.79 & 0.90 & 0.83 & 0.89 \\
  \bigskip \\

  NGC 3627$^b$ (11.3 Mpc)\\
  PHANGS vs. LEGUS & 0.82 & 0.68 & 0.63 & 0.73 & 0.72 & 0.89 \\
  PHANGS vs. RESNET & 0.81 & 0.60 & 0.60 & 0.72 & 0.68 & 0.90 \\
  PHANGS vs. VGG & 0.82 & 0.63 & 0.66 & 0.72 & 0.71 & 0.90\\
  LEGUS vs. RESNET & 0.88 & 0.74 &0.74 & 0.84 & 0.80 & 0.93 \\
  LEGUS vs. VGG & 0.86 & 0.72 & 0.76 & 0.83 & 0.79 & 0.92\\
  mean for above values & 0.84 (.03)  & 0.67 (.06)  & 0.68 (0.07)  & 0.77 (0.06) & 0.74 (0.05)  & 0.91 (0.02)  \\
  
  RESNET vs. VGG & 0.91 & 0.75 & 0.80 & 0.84 & 0.82 & 0.94 \\
  \\
  mean of means (n = 5) & 0.80 (0.02) & 0.54 (0.04) & 0.56 (0.03) & 0.64 (0.05) & 0.64 (0.03) & 0.82 (0.03)\\

  \hline
 \end{tabular}
\end{table}
NOTE:  Agreement fractions obtained by requiring one-to-one matches in position with a 2 pixel radius (i.e., using the intersection rather than union of the two studies being compared),  separately using both studies as the denominator, and taking the mean of the two determinations.

$^a$ Values in parenthesis are uncertainties in the mean (i.e., the stddev divided by the sqrt of N).

\clearpage

$^b$ Classifications for NGC 3351 and NGC 3627 were done for LEGUS by BCW, and were therefore part of the BCW-only classifications used in the development of the \citet{wei20} model. This may have resulted in the slightly better agreement fractions for these two galaxies in this table.

$^c$ The NGC 3351 LEGUS field of view was roughly 30~\% smaller than the fields of view for the other studies. Only the overlapping fields of view were used to calculate the agreement fractions.

\bigskip

\clearpage

\begin{table}
 \caption{Colour-Colour Statistics for Four Classification Methods }
 \label{tab:table2}
 \begin{tabular}{lccccc}
  \hline
Galaxy   & NGC 628 (9.8 Mpc)  & NGC 1433 (18.6 Mpc)  & NGC 1566 (17.7 Mpc)  & NGC 3351 (10.0 Mpc) & NGC 3627 (11.3 Mpc)\\
 Approach & PH/LG/RS/VG$^a$  & PH/LG/RS/VG$^a$  & PH/LG/RS/VG$^a$  & PH/LG/RS/VG$^{a,b}$ & PH/LG/RS/VG$^a$  \\

Class / Box \\
  \hline

 $M_V$ < -7.8 mag\\
  Class 1/Box 1 & \textcolor{blue}{28/27/26/28-0.2$^c$} & \textcolor{blue}{24/21/24/22-0.3} & \textcolor{red}{52/34/46/45-1.1} & \textcolor{blue}{16/14/17/18-0.4} & \textcolor{blue}{38/37/46/46-0.8} \\
  Class 1/Box 2 & \textcolor{blue}{15/13/18/18-0.6} & 5/3/5/4 & \textcolor{blue}{132/135/157/148-1.0} & 0/0/1/0 & \textcolor{blue}{98/104/113/114-0.7}\\
  Class 1/Box 3 & \textcolor{blue}{29/32/30/30-0.2} & 7/11/11/8 & \textcolor{blue}{84/84/82/76-0.4} & 2/2/10/8 & \textcolor{blue}{86/80/99/80-1.0}\\
  Class 1/Box 4 & 12/8/10/8 & 7/13/4/3 & \textcolor{blue}{23/25/18/16-0.9} & 2/6/5/3 & 9/3/4/4\\

   \bigskip\\

  Class 2/Box 1 & 0/0/0/0 & 2/1/0/0 & 11/11/0/4 & 1/0/0/0 & 12/7/3/6 \\
  Class 2/Box 2 & 4/3/0/0 & 1/2/0/0 & 25/26/6/9 & 0/0/0/0 & 24/16/11/11\\
  Class 2/Box 3 & 15/13/6/10 & 10/9/2/4 & \textcolor{red}{73/85/34/45-3.1} & 12/5/6/9 & \textcolor{red}{66/51/42/55-1.4}\\
  Class 2/Box 4 & \textcolor{red}{34/35/20/21-1.5} & \textcolor{blue}{40/34/39/41-0.5} & \textcolor{red}{110/127/89/101-1.5} & \textcolor{red}{21/12/22/17-1.1} & \textcolor{blue}{16/19/18/13-0.7}\\

     \bigskip\\

  Class 3/Box 1 & 0/0/0/0 & 0/0/0/1 & 0/6/1/1 & 2/1/3/2 & 2/0/1/1 \\
  Class 3/Box 2 & 1/0/0/1 & 1/1/1/3 & 6/8/6/9 & 0/1/2/1 & 3/1/1/3 \\
  Class 3/Box 3 & 6/3/2/8 & 5/1/6/4 & \textcolor{red}{30/69/73/89-3.1} & 4/2/7/8 & \textcolor{blue}{27/26/36/37- 1.0}  \\
  Class 3/Box 4 & \textcolor{red}{30/23/41/54-2.2} & \textcolor{red}{36/37/58/62-2.0} & \textcolor{red}{78/239/257/327-7.0} & 7/5/9/19 & \textcolor{red}{19/17/24/28-1.1} \\

     \bigskip\\

  Class 4$^d$/Box 1 & 1/1/3/2 & 7/7/26/28 & 10/5/37/34 & 6/2/12/12 & 10/22/16/13 \\
  Class 4/Box 2 & 15/2/25/24 & 11/25/61/60 & 35/20/94/95 & 7/1/9/11 & 21/36/32/30 \\
  Class 4/Box 3 & 18/3/53/49  & 29/24/58/61 & 122/79/274/266 & 18/6/33/31 & 55/105/85/91 \\
  Class 4/Box 4 & 44/17/100/97 & 53/30/93/98 & 145/116/400/363 & 15/6/18/21 & 30/54/47/51 \\

     \bigskip\\

  Class 1+2/Box 1 & \textcolor{blue}{28/27/26/28-0.2} & \textcolor{blue}{26/23/24/22-0.4} & \textcolor{red}{63/45/46/50-1.2} & \textcolor{blue}{17/14/17/18-0.4} & \textcolor{blue}{50/44/49/52-0.5} \\
  Class 1+2/Box 2 & \textcolor{blue}{19/16/18/18-0.3} & 6/5/5/4 & \textcolor{blue}{157/161/163/157-0.2} & 0/1/0/0 & \textcolor{blue}{122/120/124/125-0.2} \\
  Class 1+2/Box 3 & \textcolor{blue}{44/40/36/40-0.5} & \textcolor{blue}{17/18/13/12-0.8} & \textcolor{red}{157/138/116/121-1.6} & \textcolor{blue}{14/12/16/17-0.6} & \textcolor{blue}{152/131/141/136-0.8} \\
  Class 1+2/Box 4 & \textcolor{red}{46/43/30/29-1.4} & \textcolor{blue}{47/47/43/44-0.3} & \textcolor{red}{133/152/107/117-1.7} & \textcolor{blue}{23/18/27/20-0.8} & \textcolor{blue}{25/22/22/17-0.7} \\
   \hline
 \end{tabular}
\end{table}

NOTES: The boxes are defined as:

Box 1 - Old: 0.95<V-I<1.5 and -0.4<U-B<1.0

Box 2 - Intermediate: 0.2<V-I<0.95 and -0.4<U-B<1.0

Box 3 - Young: 0.1<V-I<0.95 and -1.1<U-B<-0.4

Box 4 - Very young: -0.6<V-I<0.95 and -1.8<U-B<-1.1

$^a$ PH = PHANGS-HST, LG = LEGUS, RS = RESNET, VG = VGG.

$^b$ The NGC 3351 LEGUS field of view was roughly 30~\% smaller than the fields of view for the other studies. Only the overlapping fields of view were used to calculate the colour-colour statistics.

$^c$ The values after the hyphens are the Quality Ratios (QR),  defined as the standard deviation of the four measurements divided by the square root of the mean number of objects in the box. Only values with all four columns having more than 12 measurements are included. 
Numbers in \textcolor{blue}{blue} have ``good'' agreement (i.e., QR $<$ 1). Numbers in \textcolor{red}{red} have ``poor'' agreement (i.e., QR $>$ 1).

$^d$ Class 4 was not used in the analysis since the machine learning methods used a larger empirical selection region of the MCI plane, resulting in much larger numbers of artifacts. This would skew the value of QR if included.

\clearpage

\begin{table}
 \caption{Breakdown of Classes and Subclasses for the Five Program Galaxies using PHANGS-HST Classifications}
 \label{tab:table3}
 \begin{tabular}{lccccc}
  \hline
 Galaxy (mag cutoff) & N 628 ($m_V$ $<$23.0) & N 1433 ($m_V$ $<$24.1) & N 1566 ($m_V$ $<$23.5) & N 3351 ($m_V$ $<$24.0) & N 3627 ($m_V$ $<$22.5) \\
  \hline
  Primary Classes\\
  Total & 1129 & 578 & 1541 & 1029 & 637\\
  C1 - symmetric cluster & 262 & 85 & 387 & 132 & 266 \\
  C2 - asymmetric cluster & 214 & 103 & 285 & 154 & 144 \\
  C3 - compact association & 185 & 99 & 163 & 136 & 70 \\
  C4 - artifacts (i.e., C4.1 - C4.12) & 468 & 291 & 706 & 607 & 157 \\
  \\subclasses included in C4:\\
  
  C4.1 - star & 116 & 35 & 112 & 119 & 56\\
  C4.2 - pair & 163 & 80 & 302 & 181 & 59\\
  C4.3 - triplet  & 83 & 44 & 132 & 101 & 25\\
  C4.4 - saturated star & 3 & 1 & 10 & 1 & 1 \\
   C4.5 - diffraction spike & 0 & 0 & 1 & 0 & 0\\
   C4.6 - nucleus of galaxy & 1 & 1 & 1 & 1 & 0  \\
  C4.7 - background galaxy & 0 & 3 & 5 & 1 & 2\\
  C4.8 - fluff (no peak)  & 0 & 7 & 26 & 2 & 1 \\
  C4.9 - redundant & 92  & 113 & 65 & 92 & 11\\
  C4.10 - too faint to tell & 0 & 0 & 28 & 5 & 1\\
   C4.11 - edge & 1 & 7 & 13 & 12 & 1 \\
  C4.12 - bad pixel & 0 & 1 & 11 & 0 & 0\\

 \hline
 \end{tabular}
\end{table}

NOTE: 1. For the NGC 3627 mosaic, only the southern field which overlaps with LEGUS is used. The full human  PHANGS-HST catalogue has 1368 objects in it. 

2. For NGC 1566, the whole field has been humanly classified to at least $m_V$ = 23.5 mag, but parts of the image have been spot checked to fainter magnitudes. 
\clearpage

\clearpage

\begin{table}
 \caption{Mass Function Fits for NGC 1566}
 \label{tab:table4}
 \begin{tabular}{lcccccccccccc}
  \hline
  Sample & PHANGS & & & LEGUS & & & RESNET & & & VGG & &  \\
  log Age range & 6 - 7 & 7 - 8 & 8 - 8.6 & 6 - 7 & 7 - 8 & 8 - 8.6 & 6 - 7 & 7 - 8 & 8 - 8.6 & 6 - 7 & 7 - 8 & 8 - 8.6 \\
  Sample \\
  \hline

  C1+C2  & -1.66  & -1.81  & -1.93  & -1.73 & -2.01 & -2.00  & -1.63 & -1.95  & -1.99 & -1.73  & -1.87  & -2.05 \\
  & (0.12) & (0.25) & (0.13)  & (0.09) & (0.19) & (0.10) & (0.00) & (0.00) & (0.00) & (0.10) & (0.19) & (0.10) \\

mean (n=3) & -1.80 (.08) &  &  & -1.91 (.09) & & & -1.86 (.11) & & & -1.88  (.09)& &\\

mean (n=12) & -1.86 (.04) &  &  & & & & & & & & &\\

\\
  C1  & -1.64  & -1.97  & -1.95  & -1.80 & -1.84 & -1.94  & -1.92 & -1.98  & -1.96 & -1.79  & -2.01  & -2.00 \\
  & (0.14) & (0.37) & (0.13)  & (0.09) & (0.26) & (0.10) & (0.20) & (0.17) & (0.11) & (0.18) & (0.36) & (0.11) \\

mean (n=3) & -1.85 (.11) &  &  & -1.86  (.04) & & & -1.95 (.02) & & & -1.93 (.07) & &\\

mean (n=12) & 1.90 (.03) &  &  & & & & & & & & &\\

  \hline
 \end{tabular}
\end{table}

\begin{table}
 \caption{Age Distribution Fits for NGC 1566}
 \label{tab:table5}
 \begin{tabular}{lcccccccccccc}
  \hline
  Sample & PHANGS-HST & &  LEGUS & &  RESNET & &  VGG &   \\
 Mass range  & log Mass 4-4.8 &  $>$ 4.8  & log Mass 4-4.8 & $>$ 4.8 & log Mass 4-4.8 & $>$ 4.8 & log Mass 4-4.8 & $>$ 4.8\\
 Sample \\
\hline

  C1+C2 (log Age = 6.5 - 9)  & -0.50  & -0.52 & -0.49 & -0.57  & -0.39 & -0.44 & -0.42  & -0.27 \\
  & (0.12) & (0.10) & (0.01)  & (0.09) & (0.11) & (0.11) & (0.14) & (0.11) \\
mean (n = 2) & -0.51 (.01) &  & -0.53 (.04) & & -0.42 (.02) &  &  -0.34 (.07) & \\
mean (n = 8) & -0.45 (.03) &  \\
 \\

  C1+C2 (log Age = 6 - 9) & -0.72  & -0.78  & -0.72  & -0.77 & -0.63 & -0.68  & -0.60 & -0.56 \\
  & (0.16) & (0.18) & (0.15)  & (0.14) & (0.17) & (0.18) & (0.15) & (0.21) \\

mean (n = 2) & -0.75 (.03) &  & -0.74 (.02) & & -0.66 (.02) & &  -0.58 (.02) & \\
 mean (n = 8) & -0.68 (.03)&  \\

\\
   C1 (log Age = 6.5 - 9) & -0.43  & -0.43  & -0.27  & -0.33 & -0.34 & -0.15  & -0.30 & -0.21 \\
  & (0.13) & (0.10) & (0.11)  & (0.11) & (0.27) & (0.10) & (0.37) & (0.17) \\

mean (n = 2) & -0.43 (.00) &  & -0.30 (.03) & & -0.24 (.09) & &  -0.26 (.04) & \\
mean (n = 8) & -0.31 (.03) &  \\
 \\

  C1 (log Age = 6 - 9) & -0.21  & -0.47  & -0.53  & -0.56 & -0.41 & -0.41  & -0.44  & -0.35\\
  & (0.05) & (0.13) & (0.11)  & (0.11) & (0.18) & (0.15) & (0.23) & (0.17) \\

mean (n = 2) & -0.34 (.13) &  & -0.54 (.01) & & -0.41 (.00) & &  -0.40 (.05) & \\
mean (n = 8) & -0.42 (.04) &  \\

  \hline
 \end{tabular}
\end{table}

\clearpage 

\bibliographystyle{mnras}  
\bibliography{draft}

\section{APPENDIX A - A General Introduction to  Deep Learning Software  }
 \label{sec:appendix}

\subsection{Using the Existing \citealt{wei20} Deep Transfer Learning Models}
\label{sec:app_exist}

To encourage broad usage of our trained deep transfer learning models we provide the trained networks, associated python scripts, and a step-by-step tutorial guide at \url{https://archive.stsci.edu/hlsp/phangs-hst} with digital object identifier \doi{10.17909/t9-r08f-dq31}.

Many aspects of the description are specific to our PHANGS-HST dataset and cluster  catalogues, but we attempt to provide sufficiently  generalised discussion so that anyone with comparable data can classify their own sources. A short description of the main steps for classification and for new training (if desired) is provided below. 

The process is reliant upon a few prerequisites.  Our models have been trained using five band HST imaging in the F275W, F336W, F438W, F555W, and F814W filters.  
A cluster candidate  catalogue containing source positions is 
required.  The trained models from \citet{wei20} and our python scripts
 (available at the website noted above) must also be downloaded.  In terms of computing resources, our PHANGS-HST classifications were accomplished using a GPU instance on Amazon Web Services (AWS).  
This is not required, and the work could be undertaken even on a personal desktop machine having the proper python packages (e.g. pytorch) installed.  To facilitate the process, we provide a Jupyter Notebook.

We begin our classification procedure with the production of FITS subimages (299$\times$299 pixel) centered on the position of each candidate, one subimage per band per source. These subimages are assembled into a single multi-extension FITS (MEF) file per candidate.  As described in \citet{wei20}, we generate a total of twenty classifications for each object using ten ResNet18- and ten VGG19-based models.  This is accomplished with a wrapper script that makes the individual calls specific to each of the networks.  Finally, the resulting classifications are consolidated into a single file in which we tabulate the per-candidate classification mean, median, mode, standard deviation, and a quality factor defined as the fraction of outcomes matching the mode, alongside the individual classifications for each of the ResNet18 and VGG19 architectures.

\subsection{Creation of New Models with Deep Transfer Learning}
\label{sec:app_train}

In the event that one does not wish to use the cluster classification models published in \citet{wei20}, our scripts also provide a guide for creating new independently-trained deep learning models. 

For training, the machine learning algorithm requires pre-classified sets of images with the same format as described above. 
These must be split into a training sample and a test sample to allow the model to iteratively test its accuracy. 

Depending on the processor used, batch size, and the number of batches run, the training process may take a few hours to a few days to complete. Thus, it is generally advisable to train with the greater processing power of a GPU. For example, training the ResNet18 architecture using 10000 batches with 32 objects per batch takes about 4 hours with the p3.2xlarge GPU instance on AWS, while training the VGG19 architecture with the same parameters will take about 50$\%$ longer. These training times roughly scale with the number of batches and batch size.

Batch size, learning rate, and number of batches all influence the effectiveness of the trained model. 
To create a batch, a random object is first chosen, and then a random object 
from the
training 
file is selected. Next, the selected 
image is rotated between 0 and 360 degrees (by 90 degree intervals) and also has a 50$\%$ chance of being flipped. This augmentation is designed to manufacture a larger sample of objects for training and makes it very rare for the model to train on the exact same image multiple times. 

Once all of the object images in a batch are collected, the training procedure performs matrix multiplication on the original, generic model (e.g. ResNet18 pre-trained on $>1$ million images from the ImageNet database) and then compares its predicted classes (essentially a random guess to begin with) for the test objects with their ground truth classes. 
The degree to which the matrix multiplication changes per iteration during training is determined by the learning rate; a faster learning rate will make larger modifications, which may train a model faster, but may also result in a less accurate final model. 
These steps are then repeated for the desired number of batches; thus 10000 batches correspond to 10000 modifications to the initial model. The user should note, however, that it is possible to over-train the model, where the model becomes overly adept at classifying the objects on which it trains rather than accurately classifying a completely new set of objects. 

All the scripts necessary to do what is outlined in this appendix, a fully commented Jupyter notebook, and other more detailed documentation are  available at the website listed above.

\end{document}